\documentclass[preprint,floatfix,superscriptaddress,showpacs,nofootinbib,amssymb,amsmath,aps,prc,eqsecnum]{revtex4-1}

\usepackage{references}
\usepackage[dvips]{graphicx}
\usepackage{latexsym,epsfig,bm,psfrag,subfigure}
\usepackage{color}         
\usepackage[bookmarksnumbered,bookmarksopen,colorlinks,citecolor=blue,linkcolor=blue]{hyperref} %

\newcommand{\ugamma}[1]{\gamma^{#1}}

\newcommand{\dgammafive}{\gamma_{5}}

\newcommand{\psibar}[1]{\overline{#1}}
\newcommand{\half}{\ensuremath{\frac{1}{2}}}

\newcommand{\slashed}[1]{\not\!#1}
\newcommand{\lag}{\mathcal{L}}

\newcommand{\ucpartial}[1]{\partial^{#1}}
\newcommand{\dcpartial}[1]{\partial_{#1}}
\newcommand{\mn}{\mu\nu}

\newcommand{\den}[1]{\rho_{#1}}
\newcommand{\dvecp}[1]{\frac{d^{3}\vec{#1}}{(2\pi)^{3}}}
\newcommand{\espec}[2]{\epsilon_{#1}(#2)}
\newcommand{\eq}[1]{Eq.~(\ref{#1})}
\newcommand{\eqs}[2]{Eqs.~(\ref{#1})-(\ref{#2})}
\newcommand{\fig}[1]{Fig.~\ref{#1}}

\def\Tr{\mathop{\rm Tr}\nolimits}
\def\grho{g_{\rho}}
\def\gs{g_{s}}
\def\gv{g_{v}}
\def\mnstar{M^{\ast}}
\def\mland{m^{\ast}}
\def\msstar{m_{s}^{\ast}}
\def\vp{\mathcal{V}}
\def\mup{\mu_{p}}
\def\mun{\mu_{n}}
\def\nup{\nu_{p}}
\def\nun{\nu_{n}}
\def\phib{\psibar{\phi}}
\def\vb{\psibar{V}}
\def\rhob{\psibar{b}}
\def\estar{E^{\ast}}
\def\ed{\mathcal{E}}
\def\fd{\mathcal{F}}
\def\sd{\mathcal{S}}
\renewcommand{\vec}{\boldsymbol}
\def\denunit{\mathrm{fm}^{-3}}

\begin{document}


\title{Hot and dense matter beyond relativistic mean field theory}

\author{Xilin Zhang} 
\email{xilinz@uw.edu}
\affiliation{Physics Department, University of Washington, 
Seattle, WA 98195, USA} 
\author{Madappa Prakash}
\email{prakash@ohio.edu}
\affiliation{Department of Physics and Astronomy, Ohio University, Athens, OH  45701, USA}

\date{\today}

\begin{abstract}
Properties of hot and dense matter are calculated in the framework of quantum hadro-dynamics by including contributions from two-loop (TL) diagrams arising from the exchange of iso-scalar and iso-vector mesons between nucleons. Our extension of mean-field theory (MFT) employs the same five density-independent coupling strengths which are calibrated using the empirical properties at the equilibrium density of iso-spin symmetric matter.  Results of calculations from the MFT and TL approximations are compared for conditions of density, temperature, and proton fraction encountered  in the study of core-collapse supernovae, young and old neutron stars, and mergers of compact binary stars.  The TL results for the equation of state (EOS) of cold pure neutron matter at sub- and near-nuclear densities agree well with those of modern quantum Monte Carlo and effective field-theoretical approaches.  Although the high-density EOS in the TL approximation for cold and beta-equilibrated neutron-star matter is substantially softer than its MFT counterpart, it is able to support a $2M_\odot$ neutron star required by recent precise determinations. In addition, radii of $1.4M_\odot$ stars are smaller by $\sim 1$ km than obtained in MFT and lie in the range  indicated by analysis of astronomical data.   In contrast to MFT, the TL results also give a better account of the single-particle or optical potentials extracted from analyses of medium-energy proton-nucleus and heavy-ion experiments. In degenerate conditions, the thermal variables are well reproduced by results of Landau's Fermi-Liquid theory in which density-dependent effective masses feature prominently. The ratio of the thermal components of pressure and energy density expressed as $\Gamma_{th}=1+(P_{th}/\epsilon_{th})$, often used  in astrophysical simulations, exhibits a stronger dependence on density than on proton fraction and temperature in both MFT and TL calculations. The prominent peak of $\Gamma_{th}$ at supra-nuclear density found in  MFT is, however, suppressed in TL calculations. This outcome is analogous to  results of non-relativistic models when exchange contributions from finite-range interactions are included in addition to those of contact interactions. \\


\end{abstract}




\maketitle

 
 \newpage

\section{Introduction} 
\label{sec:intro}
The equation of state  of dense matter plays a central role in describing the collective properties of laboratory nuclei, medium-energy heavy-ion collisions, and astrophysical phenomena involving core-collapse supernovae,  neutron stars from their birth to old age, and mergers of compact binary stars.   In hydrodynamic simulations of compact objects, the equation of state (EOS) is required for baryon densities $\rho_B$ ranging from $10^{-8}~\denunit$ to several times the nuclear saturation density of $\den{0}=0.16\,\denunit$, temperatures $T$ up to 100 MeV and beyond, and electron fractions $Y_e=\rho_e/\rho_B$ from 0  to 0.5.  The physical state of matter, determined by the minimization of the free energy, depends on the ambient conditions characterized  by $\rho_B,~T$ and $Y_e$. At sub-nuclear densities and moderate temperatures ($T$ up to $\sim20$ MeV), the preferred phase is inhomogeneous containing nucleons, light nuclear clusters (e.g., $d, t,$ and $He$) as well as heavier neutron-rich nuclei. At near-nuclear and supra-nuclear densities,  a homogeneous phase of bulk matter comprised of nucleons prevails.  Baryons beyond nucleons (e.g., hyperons), phase transitions to Bose condensates, and sub-hadronic degrees of freedom (quarks) may also enter in the description of the EOS as $\rho_B$ increases well beyond $\rho_0$. In astrophysical settings, contributions from charge neutralizing leptons (electrons, positrons and muons),  neutrinos of all flavors, and photons must also be considered when appropriate. Examples of EOS's  based on non-relativistic and relativistic field-theoretical descriptions of  nucleonic matter at the mean-field level that are currently  used in astrophysical applications can be found in Refs. \cite{Lattimer91,SLM94,Shen1,sys07,Hempel10,Shen2b,shen2011new,Oconnor,Hempel12,Steiner}.

Attempts to constrain the EOS to be consistent with the empirical properties of nuclei and bulk nuclear matter with varying isospin content, data from medium energy-heavy-ion collisions and astrophysical observations are growing in number (see the recent review in Ref. \cite{LattimerPrakash16}, and references therein).  On the nuclear front, experimental data on nuclear masses, symmetry energy and its density dependence, neutron skin thicknesses, dipole polarizabilities, level densities, etc., have been used  to pin down the EOS at near- and sub-nuclear densities. Collective flow observables, such as the mean transverse momentum vs rapidity, elliptic flow, etc., in medium heavy-ion collisions have shed light on the single-particle potential felt by nucleons (used in the construction of the EOS) for $\rho_B$ up to $\sim 3\rho_0$ for nearly isospin-symmetric bulk matter. On the astrophysical front, precise determinations of neutron star masses up to 2$M_\odot$ \cite {Demorest10,Antoniadis13} have put stringent constraints on the high-density EOS. Estimates of neutron star radii, which further constrain the EOS at supra-nuclear densities, are beginning to become available (see Ref. \cite{LattimerPrakash16} for a summary and relevant references). These developments have provided the impetus for studies of hot and dense matter beyond the mean-field level for the conditions encountered  in astrophysical applications.

In this paper, we apply the theory of quantum hadro-dynamics  (QHD) - commonly known as the Walecka model (see Refs.~\cite{Serot:1984ey,Serot:1997xg} for reviews) - beyond the mean-field level to study the nuclear EOS at both zero and finite temperatures for proton fractions $x=\rho_p/\rho_B$ ranging from zero for pure neutron matter (PNM) to 0.5 for symmetric nuclear matter (SNM).  QHD is a relativistic field theory in which nucleons (protons and neutrons) interact via the exchange of iso-scalar scalar $\phi$ and vector $V^{\mu}$, and isovector pseduo-scalar $\vec{\pi}$ and vector $\vec{\rho}^{\mu}$ mesons.  The theory respects the internal discrete symmetries of parity, charge symmetry, and time invariance, and continuous $SU(2)\otimes SU(2)$ chiral symmetry and its spontaneous breaking that are required by the underlying theory of quantum chromodynamics (QCD) \cite{Serot:1997xg,Furnstahl:1996zm}. A recent systematic  study of chiral symmetry in QHD can be in found Refs.~\cite{Serot:2011yx,Serot:2012rd}. The mean-field approximation of QHD has been widely used to study the EOS of bulk matter and the properties of finite nuclei \cite{Serot:1984ey,Serot:1997xg,Bender:2003jk}. Recent EOS tables based on relativistic mean-field theory (MFT) constructed for use in astrophysical simulations can be found in \cite{Shen2b,shen2011new}.

A systematic improvement beyond the MFT approximation was carried out in Refs.~\cite{Furnstahl:1989wp,Furnstahl:1996zm} in terms of a loop expansion for computing the ground state energy of nuclear matter. However,  two-loop (TL) contributions  beyond MFT were found to be very large and more significant than contributions from MFT so that the loop expansion could not be regarded as perturbatively convergent.  The ensuing unphysical predictions for the bulk properties at the nuclear equilibrium density resulted in abandoning loop expansion as a viable scheme for some time. The method was revived in Ref. \cite{Prakash:1991fx} where form factors at the vertices of the loops were used to regulate the high-momentum behavior of the loop contributions from the Lamb shift and vacuum fluctuation pieces with the result that the empirical properties of nuclear matter were recovered.

An effective loop-renormalization scheme was not available until  recent studies \cite{Hu:2007na,McIntire:2007ud} treated QHD as an effective field theory  (see Refs.~\cite{Weinbergprl1967, Weinbergpr1968, Weinbergphysica1979,Weinbergbookv2} for accounts on EFT, and Refs.~\cite{Kubis:2000zd, Scherer:2009bt, Fuchs:2003sh, Schindler:2005ke} for chiral EFT including heavy mesons) and adapted infrared loop regularization \cite{Tang:1996ca, Ellis:1997kc, Becher:1999he}.  In  chiral EFT, a proper power counting in the single nucleon sector \cite{Gasser:1987rb} was made feasible either by using a non-relativistic theory or by using infrared regularization which expands anti-nucleon pole contributions as contact terms \cite{Tang:1996ca, Ellis:1997kc, Becher:1999he}. In the scheme for many-body systems, when treating the TL contribution to the ground state energy, only the terms having manifest density dependence are kept; the pieces with anti-nucleon (or negative energy state) contributions, e.g., ``Lamb shift'' and ``vacuum fluctuation'' terms, as well as the pure meson loops \cite{Serot:2011yx,Serot:2012rd}, are considered as due to short range physics and renormalized by terms already present in the QHD Lagrangian. This procedure had been partially employed in an early TL study \cite{Chin:1977iz}, and in various Hatree-Fock (HF) calculations,  e.g., \cite{Horowitz:1981zg,Horowitz:1982kp}, but without ample justification. It should be pointed out that field-theoretical studies are still incomplete, the current three-loop calculation being in the exploratory stage \cite{McIntire:2007pv}.  

Our objectives in this work are to (1) extend the TL calculations by including  iso-vector meson ($\vec{\pi}$, and $\vec{\rho}^{\mu}$) exchanges in the loops with non-linear scalar meson self-interactions in the Lagrangian; (2) fit the coupling strengths utilizing the available nuclear properties and study the phenomenology in detail; (3) develop the finite temperature formalism; and (4) study the thermal properties relevant to astrophysical phenomena such as core-collapse supernovae, proto-neutron stars, and mergers of compact binary stars.  These issues have not been addressed together in  previous studies \cite{Chin:1977iz, Hu:2007na,McIntire:2007ud}. This work is also motivated by the observation in Refs.~\cite{Prakash:1988zz,Welke:1988zz} that exchange contributions are needed to reconcile the single-particle potential (or the real part of the optical potential) with the collective flow observables in medium energy heavy-ion collision experiments. Our results for the single-particle potential provide a contrast to other relativistic versions  in Refs.~\cite{Weber:1992qc,Gaitanos:2012hg,Antic:2015tga}. 

As noted in Refs.~\cite{Horowitz:1981zg,Horowitz:1982kp}, the relativistic HF calculation is similar to the TL one in \cite{Chin:1977iz}, but the TL formalism is much simpler than the HF which requires self-consistency. This advantage can benefit  studies of finite nuclei. Different versions of  HF calculations  \cite{Long:2005ne, Long:2007dw, Sun:2008zzk, Miyatsu:2011bc, Katayama:2012ge} exist, some of which include vertex form factors and/or density dependent couplings. From the EFT perspective, these procedures introduce uncertainties that are degenerate with the non-linear couplings in the Lagrangian. Thus far, finite temperature calculations have not been addressed in relativistic HF calculations. 

There exist numerous non-relativistic EOS calculations. Recently, chiral EFT's have been applied to study two- and  three-nucleon potentials, with couplings determined from low-energy nucleon-nucleon scattering data and properties of light nuclei (see reviews in Refs. \cite{Epelbaum:2008ga,Machleidt:2011zz}). Different microscopic perturbative schemes are employed, including the Hamiltonian framework using the EFT potential with or without similarity renormalization group transformation \cite{ Kruger:2013kua, Drischler:2013iza, Bogner:2009bt}, as well as the EFT Lagrangian framework \cite{Fritsch:2002hp, Kaiser:2001jx, Fiorilla:2011sr,  Holt:2013fwa}. The numerically intensive non-perturbative methods, based on either an empirical or an EFT potential, have been used mainly to study light nuclei and neutron matter \cite{Akmal:1997ft, Gezerlis:2013ipa, Gandolfi:2009fj, Epelbaum:2008vj}. Thermal properties have been addressed in Refs~\cite{Fritsch:2002hp, Kaiser:2001jx, Fiorilla:2011sr, Holt:2013fwa}  below and around the nuclear saturation density.  Extensions to supra-nuclear densities well beyond $\rho_0$ have been hampered owing both to the non-relativistic treatment and the relatively small high-energy scale required in the EFT approach.  

In contrast to the above  calculations, the QHD  couplings are calibrated at the saturation density using nuclear bulk properties with sub- and supra- nuclear properties emerging as predictions. According to the density functional theory \cite{Bender:2003jk}, such fitted couplings implicitly include some effects of higher-order, many-body correlations that are not included in the approximate energy density functional. Including TL contributions improves the density functional by adding non-analytic density dependences to the MFT density functional.  There are also non-relativistic density functionals (e.g., Skyrme \cite{Skyrme1958}), which are based on contact nucleon-nucleon interactions. These functionals have been  widely used to study properties of nuclei (see review~\cite{Bender:2003jk}) as well as of high-density matter. Recently, a detailed study of thermal properties using non-relativistic density functionals has been carried out in Ref.~\cite{Constantinou2014}.

The organization of the paper is as follows. In Sec. \ref{sec:lag}, the Lagrangian density of QHD featuring  interactions between nucleons and mesons is presented.  Section \ref{sec:ftft} is devoted to a discussion of the finite temperature formalism at the MFT and TL levels. Formulas to calculate zero temperature properties at the MFT and TL levels are given in Sec. \ref{sec:zeroT}. A self-consistent procedure to calculate the thermal properties is described in Sec. \ref{sec:SC} along with a comparison to a perturbative approach. Results at zero temperature for isospin symmetric and asymmetric matter, and for structural attributes of neutron stars, are presented in Sec. \ref{sec:results0}. This section also includes comparisons with results of modern non-relativistic approaches as well a discussion of the single-particle potential of relevance to heavy-ion collisions. Thermal properties are studied in Sec. \ref{sec:resultst} where results of relevance to astrophysical simulations of core-collapse supernovae, proto-neutron stars and mergers of compact binary stars are discussed. This section also contains a comparison of   the exact numerical calculations with those in the degenerate limit to illustrate how the Landau effective mass captures the behavior of the thermal state variables as a function of density to leading order effects in the temperature. Our summary and conclusion are contained in Sec. \ref{sec:sumup}.  Working formulas for the TL contributions are collected in Appendix A. In Appendix B, non-relativistic limit expressions for the TL contributions are given. Appendix C contains expressions that facilitate the evaluation of the single-particle spectrum.  Degenerate and non-degenerate limit expressions to examine the thermal properties  are summarized in Appendix D.

\section{The Lagrangian density of QHD} 
\label{sec:lag}

The Lagrangian density of QHD was proposed in Ref.~\cite{Furnstahl:1996wv}, in which the non-linear realization of $SU(2) \otimes SU(2)$ chiral symmetry was included. Such symmetry realization was systematically studied in Refs.~\cite{Serot:2011yx,Serot:2012rd}. Here we only mention the relevant interaction terms. The interactions between nucleons and mesons are delineated  in the Lagrangian density
\begin{eqnarray}
\lag_{N}&=&
        \psibar{N}\left[i\ugamma{\mu}\left(\dcpartial{\mu}
        +ig_{\rho}\rho_{\mu}+ig_{v}V_{\mu}\right)+\frac{g_{A}}{f_{\pi}}\ugamma{\mu}\dgammafive\, \dcpartial{\mu}\pi-M+g_{s}\phi \right]N  \ .   \label{eqn:Nlaglowest}
\end{eqnarray}
Here, $N=(p,n)^{T}$ is the isospin-multiplet of proton and nucleon Dirac spinor fields; $\ugamma{\mu}$ and $\dgammafive$ are the Dirac matrics, and $\partial_{\mu}\equiv \partial / \partial x^{\mu}$; $\rho_{\mu}\equiv \rho_{\mu}^{i}\, \tau_{i}/2$ and $\pi\equiv\pi^{i}\,\tau_{i}/2$ are iso-vector vector and pseduo-scalar (Goldstone) fields, with  $\tau^i$ being isospin Pauli matrices and ${i}=0,\pm 1$ as isospin indices; $\phi$ and $V^{\mu}$ are iso-scalar scalar and vector fields. The pion decay constant $f_{\pi}=93$ MeV, and nucleon axial charge $g_A=1.26$ \cite{Serot:2011yx,Serot:2012rd}. The pseudo-vector nucleon-$\pi$ interaction is a result of chiral symmetry breaking \cite{Furnstahl:1996wv,Serot:2011yx,Serot:2012rd}. 

The Lagrangian density describing meson interactions is 
\begin{eqnarray}
\lag_{\mathrm{meson}} &=& \half \,
       \partial_{\mu}\phi\,\partial^{\mu}\phi -\left(\half+\frac{\kappa_{3}}{3!}\frac{g_{s}\phi}{M}
+\frac{\kappa_{4}}{4!}\frac{g_{s}^{2}\phi^{2}}{M^{2}}\right)m_{s}^{2}\phi^{2} 
       + \half \, \ucpartial{\mu}\pi^{i} \dcpartial{\mu}\pi_{i}
       -\frac{1}{2}  m^{2}_{\pi}\pi^{i} \pi_{i}  \notag
       \\[5pt]
&& {}-\frac{1}{4} \, V^{\mu\nu}V_{\mu\nu}+\half m^{2}_{v}\, V_{\mu}V^{\mu}-\frac{1}{4}\rho_{\mu\nu}^{i}\rho^{\mu\nu}_{i} +\half m^{2}_{\rho} \rho^{i}_{\mu}\rho^{\mu}_{i} \,,
  \label{eqn:lmeson}
\end{eqnarray}
where 
\begin{eqnarray}
 V_{\mu\nu} &\equiv&
\partial_{\mu} V_{\nu} - \partial_{\nu} V_{\mu} \quad {\rm and} \nonumber \\ 
 \rho_{\mn} &\equiv& \partial_{\mu}\rho_{\nu}-\partial_{\nu}\rho_{\mu}+ig_{\rho}[\rho_{\mu}\,, \, \rho_{\nu}]  \,.
 \label{eqn:rhomunudef}
\end{eqnarray}
 are the field tensors. The coupling constant $\grho$ in \eq{eqn:rhomunudef} indicates $\rho^i_\mu$ couples to the iso-vector vector current including its own contributions, which is also known as universal vector meson dominance (UVMD) \cite{Furnstahl:1996wv}. The same coupling is assigned to $\pi$-$\pi$-$\rho$ interaction \cite{Serot:2011yx,Serot:2012rd} without being shown here. The masses of the different fields are:  $M=939$ MeV (nucleon), $m_s=550$ MeV ($\phi$ meson), $m_{v}=783$ MeV ($V^{\mu}$), $m_{\rho}=770$ MeV  ($\rho^{i}_\mu$),  and $m_{\pi}=138$ MeV ($\pi^i$). According to the expansion scheme proposed in Refs.~\cite{Furnstahl:1996wv,Hu:2007na,McIntire:2007ud}, the meson masses, $m_s$, $m_v$, $m_{\rho}$ are on the order of the nucleon mass $M$, and the non-linear couplings, $\kappa_3$ and $\kappa_4$ are of order  $1$.  The five coupling strengths, $g_{s}$, $g_{v}$, $g_{\rho}$, $\kappa_3$, and $\kappa_4$ are calibrated against empirical properties of nuclear matter at its equilibrium density. 

\section{Finite temperature formalism}
\label{sec:ftft}

Although the finite temperature field theory formalism can be found in various text books (e.g., \cite{Kapusta:2006pm}), we summarize the relevant formulas in the QHD context to see how TL effects are manifested.  (Chiral EFT studies of thermodynamics can be found in e.g. Ref.~\cite{Holt:2013fwa}, but no mean field minimization is needed in these calculations. The finite temperature formalism for  MFT calculations in QHD can be found in \cite{Furnstahl:1990zza, Furnstahl:1990xy, Furnstahl:1991vk}.) The grand canonical partion function is~\cite{Kapusta:2006pm,Serot:1984ey}
\begin{eqnarray}
Z &\equiv&{\rm Tr\ exp}\left[{-\beta\left(H-\mup N_p -\mun N_n\right)} \right] \equiv {\rm exp} \left[-\beta \Omega\left(T,V,\mu_{p,n}; \phib,\vb,\rhob \right)\right] \,.
\end{eqnarray}
Here, $\beta\equiv 1/T$ with $T$ being temperature; $V$, $\mu_p$, $\mu_n$ are the volume and nucleon chemical potentials; $\phib$, $\vb$, and $\rhob$ are the ensemble average values of $\phi$, $V^{0}$, and $\rho^{0 	,0}$ fields at a given temperature. (Only the zeroth components of $V^{\mu}$ and neutral $\rho^{0,\mu}$ develop non-zero expectation values due to rotation and isospin symmetry.) In the following, $bg$ denotes the collection of these expectation values. The relations between $bg$ and $(T,\mu_i)$ can be obtained by extremizing the grand canonical chemical potential $\Omega$:
\begin{eqnarray} 
\left.\frac{\partial \Omega}{ \partial \phib}\right\vert_{T,\mu_{p,n}}&=&\left.\frac{\partial \Omega}{ \partial \vb}\right\vert_{T,\mu_{p,n}}=\left.\frac{\partial \Omega}{ \partial \rhob}\right\vert_{T,\mu_{p,n}}=0 \ .  \label{eqn:fieldeqnOmega}
\end{eqnarray}
Other state variables, such as the particle number, entropy, and pressure that depend on $T$ and $\mu_i$ can be computed via
\begin{eqnarray}
N_i = -\left.\frac{\partial \Omega}{\partial \mu_i} \right\vert_{T,V,\mu_j,bg}\,, \quad 
S=-\left.\frac{\partial \Omega}{\partial T}\right\vert_{V,\mu_{p,n},bg} \quad {\rm and} \quad
\Omega = - PV=E-TS-\sum_{i=p,n} \mu_i N_i 
\end{eqnarray}
We do not need to differentiate $\Omega$ with respect to the $bg$ variables, because their values  extremize $\Omega$. In homogenous matter, the volume dependence of the state variables can be factored out.  From now on, we define densities of $\Omega$, $E$, and $S$ as $\omega$, $\ed$, and $\sd$:
\begin{eqnarray}
\omega(T,\mu_{p,n}; bg)&=&-P(T,\mu_{p,n}; bg)=\ed-T\sd-\sum_i \mu_i \den{i}  \label{eqn:thermalconsistency} \\
\den{i} &=& -\left.\frac{\partial \omega}{\partial \mu_i}\right\vert_{T,\mu_j,bg} \quad {\rm and} \quad
\sd = -\left.\frac{\partial \omega}{\partial T}\right\vert_{\mu_{p,n},bg} 
\label{eqn:rhodpdmu}
\end{eqnarray}
In this study, we use $\rho_{p,n}$ and $T$ as the independent variables. The free energy density is
\begin{eqnarray}
\fd(\den{p,n},T;bg)\equiv \ed-T\sd &=&\omega+\sum_i \mu_i \den{i} \ , 
\end{eqnarray}
with $\mu_i$ being a function of temperature, density, and $bg$ through \eq{eqn:rhodpdmu}. Based on the free energy density $\fd(\den{p,n},T;bg)$, the other state variables are computed using
\begin{eqnarray}
\sd =  - \left.\frac{\partial \fd}{\partial T}\right\vert_{\den{p,n},bg} \quad {\rm and} \quad
\mu_i = \left.\frac{\partial \fd}{\partial \den{i}}\right\vert_{T,\den{j},bg} \ ,  \label{eqn:solvemufreeE} 
\end{eqnarray}
and $P(T,\den{p,n};bg)$ through \eq{eqn:thermalconsistency}. In the derivatives above, $bg$ are held fixed because of the  identities [see \eq{eqn:fieldeqnOmega}]:
\begin{eqnarray}
\left.\frac{\partial \fd(T,\den{p,n};bg)}{ \partial \phib}\right\vert_{T,\den{p,n}}&=& \left.\frac{\partial \omega(T,\mu_{p,n};bg)}{ \partial \phib}\right\vert_{T,\mu_{p,n}}\\
\left.\frac{\partial \fd(T,\den{p,n};bg)}{ \partial \vb}\right\vert_{T,\den{p,n}}&=& \left.\frac{\partial \omega(T,\mu_{p,n};bg)}{ \partial \vb}\right\vert_{T,\mu_{p,n}} \\
\left.\frac{\partial \fd(T,\den{p,n};bg)}{ \partial \rhob}\right\vert_{T,\den{p,n}}&=& \left.\frac{\partial \omega(T,\mu_{p,n};bg)}{ \partial \rhob}\right\vert_{T,\mu_{p,n}} \label{eqn:solvebg} \ , 
\end{eqnarray}

\subsection{Formalism at the MFT level} 
\label{subsec:MFTgeneral}

The grand canonical chemical potential $\omega_{(0)}(T,\mu_{p,n};bg)$ at the mean-field level (MFT) is \cite{Furnstahl:1990zza, Furnstahl:1990xy, Furnstahl:1991vk}:
\begin{eqnarray}
\omega_{(0)}=\vp(\phib)-\half m_{v}^{2} \vb^{2}-\half m_{\rho}^{2} \rhob^{2}-\gamma_{s} T \sum_{i} \int \dvecp{k}\ln\left[1+e^{-\beta \left(\estar(k)-\nu_{i}\right)}\right]   \,.
\end{eqnarray}
Here,   
\begin{eqnarray}
\vp(\phib)\equiv \left(\half+\frac{\kappa_{3}}{3!}\frac{g_{s}\phib}{M}
+\frac{\kappa_{4}}{4!}\frac{g_{s}^{2}\phib^{2}}{M^{2}}\right)m_{s}^{2}\phib^{2} \ .
\end{eqnarray}
For densities and temperatures of interest here, contributions from  anti-nucleons and mesons are negligible and are not included. The various symbols are: $\gamma_{s}=2$ is the spin-degeneracy, $\estar(k)\equiv \sqrt{{\mnstar}^{2}+k^{2}}$  with $\mnstar\equiv M-\gs\phib$ and $k\equiv |\vec{k}|$, $\nup\equiv\mup-\gv\vb-\half \grho \rhob$, and $\nun\equiv\mun-\gv\vb+\half \grho \rhob$. From now on, unless explicitly stated, the momentum variable, e.g., $k$, is the norm of the space component, $|\vec{k}|$.  

For completeness, we collect the MFT results in the following. The Fermi-Dirac distribution is denoted as $n_{i}(k) \equiv \left\{\mathrm{exp}\left[\beta\left(\estar(k)-\nu_{i}\right)\right]+1\right\}^{-1}$. By using  \eq{eqn:rhodpdmu}, we can compute state variables as functions of $T$ and $\mu_{p,n}$. 
\begin{eqnarray}
\den{i}(T,\nu_{p,n};\phib)&=&\gamma_{s} \int \dvecp{k} n_{i}(k) \ , \label{eqn:solverhofrommu} \\
\ed_{(0)}(T,\mu_{p,n};bg)&=& \vp(\phib)-\half m_{v}^{2} \vb^{2}-\half m_{\rho}^{2} \rhob^{2}+\den{p}\left(\gv \vb +\half \grho \rhob\right)+\den{n} \left(\gv \vb -\half \grho \rhob\right) \notag \\
&&{}+\gamma_{s}\sum_{i} \int \dvecp{k} \estar(k)n_{i}(k) \ , \label{eqn:edmftdef}\\
\sd_{(0)}(T,\mu_{p,n};bg) &=&-\gamma_{s}\sum_{i} \int \dvecp{k} \bigg[\big(1-n_{i}(k)\big)\ln\big(1-n_{i}(k)\big)+n_{i}(k)\ln\left(n_{i}(k)\right)\bigg] \label{eqn:entmftdef} \\ 
\fd_{(0)}(T,\mu_{p,n};bg) &=& \omega_{(0)}+ \sum_i \mu_i \den{i} =\ed_{(0)}-T\sd_{(0)} \, \label{eqn:fdmftdef} \\
P_{(0)} (T,\mu_{p,n};bg) &=& - \omega_{(0)} 
\end{eqnarray}

To solve $bg$ for  a given $T$, the conditions to be met are: 
\begin{eqnarray}
 \left.\frac{\partial \omega_{(0)}}{\partial \phib} \right\vert_{\mu_{p,n},T}&=&\frac{d \vp(\phib)}{d \phib} -\gs \gamma_{s}\sum_{i} \int \dvecp{k} \frac{\mnstar}{\estar(k)} n_{i}(k)   = 0\ , \label{eqn:solvephiMFT} \\
\left.\frac{\partial \omega_{(0)}}{\partial \vb} \right\vert_{\mu_{p,n},T} &=&-m_{v}^{2}\vb +\gv \den{B}  = 0\ , \\
\left.\frac{\partial \omega_{(0)}}{\partial \rhob} \right\vert_{\mu_{p,n},T} &=&
-m_{\rho}^{2}\rhob +\half \grho \left(\den{p}-\den{n}\right)  = 0\ .
\end{eqnarray}
The relations in \eq{eqn:solverhofrommu} and \eq{eqn:solvephiMFT} can be combined to solve $\mu_{p,n}$ and $bg$ to yield $\den{p,n}$ and $T$, based on which all the other state variables become functions of $\den{p,n}$ and $T$.

\subsection{Formalism including two-loop contributions} 
\label{sec:TLgeneral}

\begin{figure}
	\centering
		\includegraphics[width=4cm]{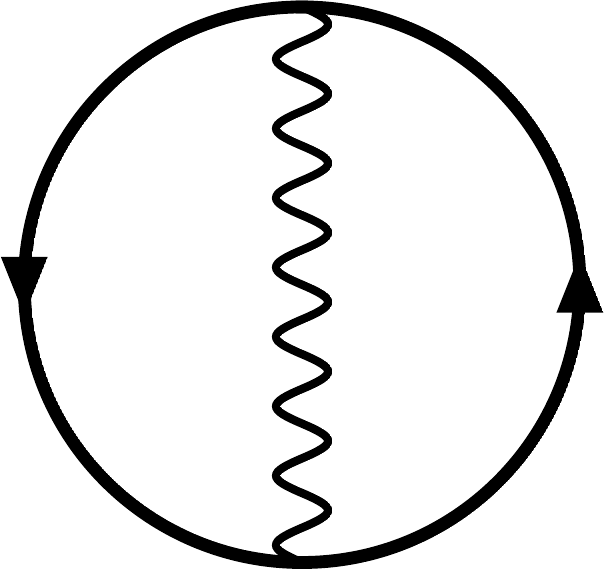}
     \caption{Feynman diagram for the two-loop contribution to the grand canonical potential density $\omega$.  
     The exchange of mesons  $\phi$, $V^{\mu}$, $\rho^{\mu}$ and $\pi$ is indicated by the wavy line.} \label{fig:two-loop}
\end{figure} 

The two-loop contribution to $\omega$ at finite temperature can be computed in the imaginary-time formalism \cite{Furnstahl:1990zza, Furnstahl:1990xy, Furnstahl:1991vk,Kapusta:2006pm}. The conclusion in Refs.~\cite{Hu:2007na,McIntire:2007ud} is that the properly regularized loop contribution arises only from the $G_{D}$ propagator, which is the density-dependent piece of the full baryon's propagator written as 
\begin{eqnarray}
G_{i}(p)=(\slashed{p}^{\ast}+\mnstar) \left[\frac{1}{p^{\ast 2}-{\mnstar}^{2}+i\epsilon}-i 2\pi n_{i}(p) \delta (p^{\ast 2}-{\mnstar}^{2})\right] \equiv G_{F}(p)+G_{D}(p) \ . 
\end{eqnarray}
The Fermi-Dirac distribution $n_{i}(p)$ is as in Section~\ref{subsec:MFTgeneral}, and  $p^{\ast 0}=p^{0}-\gv \vb - \grho \rhob\, t_{3}/2$ ($t_{3}=\pm 1 $ for proton and nucleon). This assertion, although only proved for zero temperature can be generalized to finite temperature.  Terms depending on either the anti-nucleon or meson density, as well as the vacuum polarization terms independent of densities, can be subsumed in the meson couplings of the Lagrangian and need not  be kept here. The real-time formalism \cite{Furnstahl:1990zza, Furnstahl:1990xy, Furnstahl:1991vk} can also be used, but at the expense of extra $2\times 2$ matrix structures for the vertices and propagators. 

The general meson exchange two-loop contribution to the grand chemical potential density  $\delta\omega_{(1)}$ has the following general structure \cite{Kapusta:2006pm}:
\begin{eqnarray}
\delta\omega_{(1)} &=& \frac{\gamma_{s}}{4} g^{2} \int \frac{d^{4}p}{(2\pi)^{4}} \frac{d^{4}q}{(2\pi)^{4}} \Tr\left[G_{D}(q)\Gamma(p-q) G_{D}(p) \Gamma(q-p) \right] D(q-p)  \,,
\end{eqnarray}
where the interaction vertex $\Gamma$ involves isospin. We first introduce the following definitions considering the trace in the above equation: 
\begin{eqnarray}
 f_{s}(p^{\ast},q^{\ast}) &\equiv& 4\left(p^{\ast}\cdot q^{\ast}+{\mnstar}^{2}\right) \ , \notag \\
f_{v}(p^{\ast},q^{\ast})&\equiv& 8\left(p^{\ast}\cdot q^{\ast}-2{\mnstar}^{2}\right) \ , \notag \\
 f_{pv}(p^{\ast},q^{\ast}) 
&\equiv & 16 \left(p^{\ast}\cdot q^{\ast}-{\mnstar}^{2}\right) \,, 
\end{eqnarray}
where $f_{s}$, $f_{v}$, and $f_{pv}$ are due to scalar, vector, and psedovector coupings; $p^{\ast}\cdot q^{\ast} \equiv p^{\ast \mu} q^{\ast}_{\mu}=\sqrt{{\mnstar}^2+|\vec{p}|^2}\sqrt{{\mnstar}^2+|\vec{q}|^2}-\vec{p} \cdot \vec{q}$.
Moreover, the meson propagators can be expressed in terms of 
\begin{eqnarray}
D(k;m)\equiv \frac{1}{k\cdot k -m^{2}} \ .
\end{eqnarray}
With these notations, the TL contributions from the four meson exchanges are:
\begin{eqnarray}
\delta\omega_{(1,\phi)} &=& -\frac{\gamma_{s}}{4} \gs^{2} \int d\tau_{\vec{p}} d\tau_{\vec{q}}~ f_{s}(p^{\ast},q^{\ast}) D(k;\msstar) \left[n_{p}(p)n_{p}(q)+n_{n}(p)n_{n}(q)\right] \ , \label{eqn:chempot1phi} \\
\delta\omega_{(1,v)} &=& -\frac{\gamma_{s}}{4} \gv^{2} \int d\tau_{\vec{p}} d\tau_{\vec{q}}~ f_{v}(p^{\ast},q^{\ast}) D(k;m_{v}) \left[n_{p}(p)n_{p}(q)+n_{n}(p)n_{n}(q)\right] \ , \label{eqn:chempot1v} \\
\delta\omega_{(1,\rho)} 
&=& -\frac{\gamma_{s}}{16} \grho^{2} \int d\tau_{\vec{p}} d\tau_{\vec{q}}~ f_{v}(p^{\ast},q^{\ast}) D(k;m_{\rho}) \notag \\ 
&&  \hskip8em\relax \times \left[n_{p}(p)n_{p}(q)+n_{n}(p)n_{n}(q)+4n_{p}(p)n_{n}(q)\right] \ , \label{eqn:chempot1rho} \\
\delta\omega_{(1,\pi)}
&=& -\frac{\gamma_{s}}{16} \left(\frac{g_{A}\mnstar}{f_{\pi}}\right)^{2}  \int d\tau_{\vec{p}} d\tau_{\vec{q}}~ f_{pv}(p^{\ast},q^{\ast}) D(k;m_{\pi})  \notag \\ 
 && \hskip8em\relax \times \left[n_{p}(p)n_{p}(q)+n_{n}(p)n_{n}(q)+4n_{p}(p)n_{n}(q)\right] \ .  \label{eqn:chempot1pi} 
\end{eqnarray}
In the above expressions, $d\tau_{\vec{p}}\equiv d^3 \vec{p}/\left[(2\pi)^3{2\estar(p)}\right]$; $p^{\ast}$ and $q^{\ast}$ are all ``on shell'', i.e., $p^{\ast 2}=q^{\ast 2}={\mnstar}^{2}$; $k=p^{\ast}-q^{\ast}$.  The scalar meson $\phi$ mass used in the two-loop is modified to 
\begin{eqnarray}
{\msstar}^{2}\equiv {m_{s}}^{2} \left(1+\kappa_{3}\frac{\gs \phib}{M}+\frac{\kappa_{4}}{2}\frac{(\gs \phib)^{2}}{M^{2}}\right) \ . \label{eqn:msstardef}
\end{eqnarray}
owing to its non-linear self-interactions.

The choice of $k=p^{\ast}-q^{\ast}$ implicitly assumes UVMD as mentioned in section~\ref{sec:lag}. From energy momentum conservation, in $D(k;m_{\pi})$, $k=p-q$, whose time component $k^0$ could be different from $\estar(p)-\estar(q)$ if $\vec{\rho}_\mu$ does not couple to the iso-vector vector current from $\vec{\pi}$. This would lead to a time-like $\vec{\pi}$ propagator in the loop, signaling $\vec{\pi}$ production. The same argument applies to the $\vec{\rho}_\mu$ meson. Without further knowledge of $\vec{\pi}$ and $\vec{\rho}_\mu$ propagators in dense medium, we assume UVMD in this study. 
 
Integrations of the angular dependences in \eqs{eqn:chempot1phi}{eqn:chempot1pi} are easily performed. The resulting expressions are presented  in Appendix~\ref{app:simplifiedtwoloop}. In the non-relativistic limit, i.e., in the low-density region, these expressions become physically transparent, and are shown in Appendix~\ref{subsec:TLhnlimit}. The importance of relativistic kinematics and retardation effects will be discussed in the section on results.

\subsection*{Perturbative analysis} \label{subsec:finiteTpert}

Here, the approximation scheme proposed in Refs.~\cite{Drut:2009ce, Holt:2013fwa} is applied whereby the relation between the density $\den{i}$ and the chemical potential $\mu_i$ is kept the same as in MFT, i.e., \eq{eqn:solverhofrommu}. As a result, $\mu_i$ is not the physical chemical potential, and will be labeled as $\mu_i^{(0)}$ in the following. As outlined in Refs.~\cite{Drut:2009ce, Holt:2013fwa}, this approximation provides a systematic expansion of the free energy density functional at finite temperature. The so-called anomalous diagrams arising in the zero-temperature limit \cite{Waleckamanybodybook,Kohn1960,Luttinger1960} start at the three-loop level, and are not considered in this calculation. To get $\omega_{(1)}$, we sum up the MFT and TL contributions to get
\begin{eqnarray}
\omega_{(1)}(T,\mu_{p,n}^{(0)};bg)=\omega_{(0)}(T,\mu_{p,n}^{(0)};bg)+\delta\omega_{(1)}(T,\nu_{p,n};\phib) \ ,  \label{eqn:omega1def}
\end{eqnarray}
with $\nup\equiv\mu_p^{(0)}-\gv\vb-\half \grho \rhob$, and $\nun\equiv\mu_n^{(0)}-\gv\vb+\half \grho \rhob$. 
Here $\delta\omega_{(1)}$ depends on $\nu_{p,n}$ and $\phib$ but not on $\vb$ and $\rhob$. By using \eq{eqn:solverhofrommu}, we compute $\den{i}$ for given  $\mu_i^{(0)}$,  $T$, and $bg$. Note that in this approximation, $\mu_i^{(0)}$ is not the physical chemical potential which motivates the notation used. Based on 
\begin{eqnarray}
\fd_{(1)} &\equiv&\omega_{(1)}(T,\mu_{p,n}^{(0)};bg)+\sum_i \mu_i^{(0)} \den{i} \ ,  
\end{eqnarray}
and $\fd_{(0)}=\omega_{(0)}+\mu_i^{(0)}\den{i}$ (\eq{eqn:fdmftdef}), we get 
\begin{eqnarray}
\fd_{(1)}=\fd_{(0)}(T,\den{p,n};bg)+\delta\omega_{(1)}(T,\nu_{p,n};\phib)  \label{eqn:fd1def}
\end{eqnarray}
To solve for $\phib$, $\vb$ and $\rhob$ for given $T$ and $\den{p,n}$, we invoke \eq{eqn:solvebg} (note that $\delta\omega_{(1)}$ does not have manifest $\vb$ and $\rhob$ dependences) and require the following conditions to be satisfied: 
\begin{eqnarray}
\left.\frac{\partial \fd_{(1)}}{\partial \vb}\right\vert_{\den{p,n},T}&=& \left.\frac{\partial \fd_{(0)}}{\partial \vb}\right\vert_{\den{p,n},T} =-m_{v}^{2}\vb +\gv \den{B} = 0 \notag \\
\left.\frac{\partial \fd_{(1)}}{\partial \rhob}\right\vert_{\den{p,n},T}&=& \left.\frac{\partial \fd_{(0)}}{\partial \rhob}\right\vert_{\den{p,n},T}=
-m_{\rho}^{2}\rhob +\half \grho \left(\den{p}-\den{n}\right)  = 0 \notag \\ 
\left.\frac{\partial \fd_{(1)}}{\partial \phib}\right\vert_{\den{p,n},T}&=&\left.\frac{\partial \fd_{(0)}}{\partial \phib}\right\vert_{\den{p,n},T} + \left.\frac{\partial \delta\omega_{(1)}\left(T,\nu_{p,n}(T,\den{p,n},\phib);\phib\right)}{\partial \phib}\right\vert_{\den{p,n},T} = 0 \notag \\
&=&\frac{d \vp(\phib)}{d \phib} -\gs \gamma_{s}\sum_{i} \int \dvecp{k} \frac{\mnstar}{\estar(k)} n_{i}(k) \notag \\
 &&+\left.\frac{\partial \delta\omega_{(1)}\left(T,\nu_{p,n};\phib\right)}{\partial \phib}\right\vert_{\nu_{p,n},T} + \sum_i \left.\frac{\partial \delta\omega_{(1)}\left(T,\nu_{p,n};\phib\right)}{\partial \nu_i} \right\vert_{\nu_{j\neq i},T,\phib} \left.\frac{\partial \nu_i(T,\den{p,n}; \phib)}{\partial \phib}\right\vert_{T,\den{p,n}} \ .
\end{eqnarray}
From the above mean field equations and \eq{eqn:fd1def}, we can compute the entropy density $\sd_{(1)}$, the physical chemical potential $\mu_i$, the energy density $\ed_{(1)}=\fd_{(1)}+T\sd_{(1)}$, and the pressure $P_{(1)}=-\fd_{(1)} +\sum_i \mu_i \den{i}$ using \eq{eqn:solvemufreeE}.  Note that in general $\mu_i \neq \mu_i^{(0)}$ and  $P_{(1)}\neq -\omega_{(1)}$.

\section{Zero temperature properties} 
\label{sec:zeroT}

\begin{table}
   \caption{Empirical properties used to constrain the coupling strengths in the Lagrangian for  the MFT and two-loop calculations. The various symbols are: $\rho_0$, the equilibrium density, $B_0$, the binding energy, $K_{v,0}$, the compression modulus, 
   $m_0^*/M$, the Landau effective mass scaled with the vacuum mass, all for isospin symmetric nuclear matter; $S_{2,0}$ is the nuclear symmetry energy.} 
    \begin{ruledtabular} 
   \begin{tabular}{c c c c c } 
           $\rho_{0}$ ($\mathrm{fm}^{-3}$)
           & $B_0$ (MeV) 
           & $S_{2,0}$ (MeV)  
           & $K_{v,0}$ (MeV) 
           & $\mland_{0}/M$ \\  \hline
           $0.16$
           & $16.0$
           & {$35.0$} 
           &  {$250$} 
           & {$0.73$} 
   \end{tabular}      
   \end{ruledtabular}
   \label{tab:saturationpoint}
\end{table}

As mentioned in section~\ref{sec:intro}, the coupling constants in the Lagrangian are fixed using the empirical properties of nuclear matter. The properties used for calibrating the model are listed in Table~\ref{tab:saturationpoint}, where the lower index $``0"$ denotes quantities evaluated at 
$\den{0}=2 k_{F0}^2/(3\pi^2)$ where $P(\den{0})=0$ and $\ed(\den{0})/\den{0}-M=-B_0$. The symmetry energy and incompressibility are defined as 
\begin{eqnarray}
K_v \equiv 9 \left.\frac{\partial P}{\partial \den{B}}\right\vert_{x} \quad {\rm and} \quad
S_{2} \equiv  \left.\frac{1}{8\den{B}} \frac{\partial^2 \ed }{\partial x^2}\right \vert_{x=0.5,\den{B}}  \ ,
\end{eqnarray}
where $x=\rho_p/\rho_B$ is the proton fraction. The single particle spectrum is given by
\begin{eqnarray}
\gamma_s\, \espec{i}{k}=\frac{\partial \ed \left[n_{p,n};\phib[n_{p,n}], \vb[n_{p,n}], \rhob[n_{p,n}] \right] }{\partial n_{i}(k)} =\left. \frac{\partial \ed\left[n_{p,n};\phib, \vb, \rhob \right] }{\partial n_{i}(k)} \right\vert_{bg}
\label{eqn:singlespec}
\end{eqnarray}
as $\ed$ is a functional of the nucleon distribution function $n_i(q)$, and $bg$; the latter is also a functional of $n_i(q)$. The calculation is greatly simplified owing to $bg$ extremizing $\ed$. The Landau effective mass, proportional to the density of states at the Fermi surface, defined by 
\begin{eqnarray}
\mland_i(\den{B},x) \equiv k_{F,i} \left.\left(\frac{\partial \espec{i}{k}}{\partial k} \right)^{-1}  \right\vert_{k=k_{F,i}}  \label{eqn:mlanddef} 
\end{eqnarray}
helps to examine the thermodynamics in the degenerate limit (see Appendix~\ref{sec:degnondeglimits}). When convenient, we use the variables $(\den{B},\,x)$ or $(\den{p},\, \den{n})$ to indicate isospin asymmetry.

\subsection{Mean field theory (MFT)} 
\label{sec:MFT}

At zero temperature, the MFT energy density is   
\begin{eqnarray}
\ed_{(0)}(\den{p,n};bg)&=&\vp(\phib)-\half m_{v}^{2} \vb^{2}-\half m_{\rho}^{2} \rhob^{2}+\rho_p \left(\gv \vb +\half \grho \rhob\right)+\rho_n \left(\gv \vb -\half \grho \rhob\right) \notag \\
&& + \gamma_{s}\sum_{i=n,p} \int \dvecp{k} \estar(k)n_{i}(k) \,,
\label{eqn:MFTed}
\end{eqnarray}
where $n_i(k)=\theta(k_{Fi}-k)$ and  $\den{i}={\gamma_s k_{Fi}^3}/(6\pi^2)$.   The expectation values of the meson fields satisfy  
\begin{eqnarray}
\gv \vb  &=& \frac{\gv^2}{m_v^2} \rho_B \label{eqn:MFTvEOM} \quad {\rm and} \quad
\grho \rhob = \frac{\grho^2}{m_{\rho}^2 }\frac{\rho_p-\rho_n}{2} \notag \\
\rho_{s,p}+\rho_{s,n} &\equiv& \gamma_{s}\sum_{i} \int \dvecp{k} \frac{\mnstar}{\estar(k)} n_{i}(k) =-\frac{\partial \vp(\phib)}{\partial M^{\ast}} \,,
\label{eqn:MFTphiEOM}
\end{eqnarray}
where $\den{s,i}$ are the nucleon scalar densities.  Based on the above equations,
\begin{eqnarray}
\left.\frac{\partial M^{\ast}}{ \partial \rho_B}\right\vert_{x} &=& (-) \frac{x \frac{M^{\ast}}{E^{\ast}_{F,p}}+(1-x) \frac{M^{\ast}}{E^{\ast}_{F,n}}}{-3 \frac{\rho_p}{E^{\ast}_{F,p}}-3 \frac{\rho_n}{E^{\ast}_{F,n}}+\left(\frac{\partial^2}{\partial {M^{\ast}}^2}-\frac{3}{M^{\ast}} \frac{\partial}{\partial M^{\ast}}\right)\vp(\phib)} \label{eqn:dMstardrhoB} \, , \\
\left.\frac{\partial M^{\ast}}{ \partial x}\right\vert_{\rho_B} &=& (-) \frac{ M^{\ast} \rho_B \left( \frac{1}{E^{\ast}_{F,p}}-\frac{1}{E^{\ast}_{F,n}}\right)}{-3 \frac{\rho_p}{E^{\ast}_{F,p}}-3 \frac{\rho_n}{E^{\ast}_{F,n}}+\left(\frac{\partial^2}{\partial {M^{\ast}}^2}-\frac{3}{M^{\ast}} \frac{\partial}{\partial M^{\ast}}\right)\vp(\phib)} \, .
\label{eqn:MstarxprhoB}
\end{eqnarray}
Here $E_{F,i}^{\ast}\equiv\sqrt{k_{F,i}^{2}+{M^{\ast}}^2}$. The first equation above agrees with the  result obtained in Ref.~\cite{Steiner:2004fi} for $x=0.5$. The second order partial derivative  vanishes owing to isospin symmetry around $x=0.5$.

The chemical potentials are given by
\begin{eqnarray}
\mu_{i}&=&\ \partial \ed /\partial \den{i} = \gv \vb + t_3 \frac{1}{2} \grho \rhob +E_{F,i}^\ast  \ , \label{eqn:MFTmu}
\end{eqnarray}
with $t_3=+1$ and $-1$ for proton and neutron. The incompressibility $K_v$ and symmetry energy $S_{2}$ at arbitrary $\den{B}$ and $x$  are~\cite{Steiner:2004fi} 
\begin{eqnarray}
K_{v, (0)} &=& 9 \rho_B \bigg[
\frac{\gv^2}{m_v^2} +\frac{\grho^2}{m_{\rho}^2} \left(x-\frac{1}{2}\right)^2 +\pi^2 \left(\frac{x^2}{k_{F,p} E^{\ast}_{F,p}}+\frac{(1-x)^2}{k_{F,n} E^{\ast}_{F,n}}\right) \label{eqn:MFTkv} \\
&& +\left(x\frac{M^{\ast}}{E^{\ast}_{F,p}}+(1-x)\frac{M^{\ast}}{E^{\ast}_{F,n}}\right) \left.\frac{\partial M^{\ast}}{ \partial \rho_B}\right\vert_{x} \bigg] \notag \\
S_{2,(0)}&=& \frac{\rho_B}{8}\left[\frac{\grho^2}{m_{\rho}^2}+ \pi^2  \left(\frac{1}{k_{F,p}E_{F,p}^{\ast}}+\frac{1}{k_{F,n}E_{F,n}^{\ast}} \right)+ \frac{1}{\rho_B} \left(\frac{M^{\ast}}{E_{F,p}^{\ast}}-\frac{M^{\ast}}{E_{F,n}^{\ast}}\right)\left.\frac{\partial M^{\ast}}{ \partial x}\right\vert_{\rho_B}\right] \,,
\label{eqn:MFTs2}
\end{eqnarray}
where the lower index $(``0")$ stands for results at the MFT level. 
The single-particle spectrum from \eq{eqn:singlespec} reads as
\begin{eqnarray}
\espec{(0),i}{k}&=&\gv \vb + t_3 \frac{1}{2} \grho \rhob +\sqrt{k^{2}+{M^{\ast}}^2} \ .
\end{eqnarray}
As a result, $\mland_i=E_{F,i}^{\ast}$. At the equilibrium density $\rho_0$, the  Fermi momentum $k_{F0}=1.333\,\mathrm{fm}^{-1}$  and $\mland_{0}=0.73 M$
(c.f. Table~\ref{tab:saturationpoint}), whence the value of $\phib$ or equivalently $\mnstar$ at $\rho_0$ is  $\mnstar_0=\sqrt{{\mland_0}^2-k_{F0}^2}=0.674 M$ and  $g_s \phib_0 = M-\mnstar_0$. 

The five couplings, $\gs,\, \gv,\, \grho,\, \kappa_{3}$ and $\kappa_4$ are determined as follows.  
At $\rho_0$, the thermodynamic identity simplifies to $\ed -\sum_i \mu_i \den{i}=0$ because $P=0$. Thus,  the proton and nucleon chemical potentials in isospin-symmetric matter are $\mu_{0}=\ed/\den{0}=M-B_0$. Using Eqs.~(\ref{eqn:MFTmu}) and~(\ref{eqn:MFTvEOM}), 
\begin{eqnarray}
\mu_0-E^{\ast}_{F0}&=&  \frac{\gv^2}{m_v^2} \rho_0 \quad {\rm with} \quad E^{\ast}_{F0}=\sqrt{k_{F0}^2+{\mnstar_0}^2} 
\ . \label{eqn:5paraMFT1}
\end{eqnarray}
From the symmetry energy constraint in \eq{eqn:MFTs2}, we have 
\begin{eqnarray} 
S_{2,0}-\frac{k_{F0}^2}{6E_{F0}^{\ast}}&=&\frac{\rho_0}{8}\frac{\grho^2}{m_{\rho}^2} 
\ . \label{eqn:5paraMFT2}
\end{eqnarray}
From the binding energy constraint,  
\begin{eqnarray}
M-B_0-\frac{\ed_{kin0}}{\rho_0} &=&\frac{\vp(\phib_0)}{\rho_0}+\frac{1}{2} \frac{\gv^2}{m_v^2} \rho_0 \,,   \label{eqn:5paraMFT3}
\end{eqnarray}
where $\ed_{kin0}$ is the last term in \eq{eqn:MFTed} at $\rho_0$. 
From  the self-consistent equation for $\phi$ in \eq{eqn:MFTphiEOM}, we have
\begin{eqnarray}
\frac{1}{\gs \phib_0} \frac{\rho_{s,0}}{m_s^2}=
\frac{1}{\gs^2}+\frac{\kappa_3}{\gs^2} \frac{1}{2} \frac{\gs \phib_0}{M}+\frac{\kappa_4}{\gs^2} \frac{1}{6}  \left(\frac{\gs \phib_0}{M}\right)^2 \ ,
\label{eqn:5paraMFT4}
\end{eqnarray}
where $\den{s,0}$ is the nucleon scalar density at $\rho_0$. 

From Eq.~(\ref{eqn:MFTkv}) for the incompressibility  (for the second term on the right,  see \eq{eqn:dMstardrhoB}), 
\begin{eqnarray}
&& K_{v,0}-3 \frac{k_{F0}^2}{E^{\ast}_{F0}} = 9 \rho_0 \bigg[
\frac{\gv^2}{m_v^2}  +\frac{M^{\ast}_0}{E^{\ast}_{F0}} \left.\frac{\partial M^{\ast}}{ \partial \rho_B}\right\vert_{x=0.5} \bigg]  \ . \label{eqn:5paraMFT5}
\end{eqnarray}
These five equations can be readily solved; the so determined coupling strengths  are shown in Table~\ref{tab:two-loop_mine}.
The symmetry energy stiffness parameter evaluated at the saturation density is 
\begin{eqnarray}
L =  3 \den{B} \left.\frac{\partial S_2 }{\partial \den{B}}\right\vert_{x=0.5} 
&=& 2S_2^{(kin)}\left[ 1 - 18 \left( \frac {S_2^{(kin)}}{k_{F}} \right)^2 \left\{1 + 3 \left(\frac{M^*}{k_{F}} \right)^2 
\left. \frac {\partial \ln M^*}{\partial \ln \rho_B}\right|_{x=0.5} \right\} \right] \notag \\
&& + \frac 38 \frac {g_\rho^2}{m_\rho^2}~\rho_B \,, \label{eqn:LMFTdecomp}
\end{eqnarray}
where $S_2^{(kin)} = {\displaystyle {\frac 16 \frac {k_{F}^2}{E_{F}^*} }}$.

\begin{table}
 \caption{Coupling strengths for the MFT and TL calculations. Values for the masses are $M=939$ MeV, $m_{s}=550$ MeV,  $m_{v}=783$ MeV, $m_{\rho}=770$ MeV, and $m_{\pi}=138$ MeV, whereas $f_{\pi}=93$ MeV and  $g_A=1.26$. 
 The last column shows the resulting symmetry energy stiffness parameter $L$. 
 The parameter sets below give the same saturation density, binding energy, and symmetry energy  as 
  in Table~\ref{tab:saturationpoint}. The sets labelled TL(235), TL(250), and TL(270) give $K_v=235$, 250 and $270$ MeV, with Landau effective masses $m^\ast=0.74,~0.73$, and $0.72 M$, respectively.} 
    \begin{ruledtabular} 
   \begin{tabular}{c c c c c c c } 
           & $\gs^{2}$ 
           & $\gv^{2}$  
           & $\grho^{2}$ 
           & $\kappa_3$
           & $\kappa_4$
           &$L$ (MeV)\\  \hline
					 MFT
           & $96.36$
           & $118.45$ 
           & $70.13$
           & $2.08$
           &$-6.77$
           & $103.62$  \\  
          
            TL(235)
           & $71.26$
           & $49.58$ 
           & $60.72$
           & $5.94$
           & $-2.48$
           & $83.66$   \\    
	
					  TL(250)
           & $74.03$
           & $56.58$ 
           & $57.97$
           & $4.84$
           & $-4.47$
           & $85.09$   \\    
            TL(270)
           & $74.65$
           & $61.45$ 
           & $58.06$
           & $3.70$
           & $1.97$
           & $84.51$   \\     													
   \end{tabular}    
   		\end{ruledtabular}
 \label{tab:two-loop_mine}   
\end{table}

\subsection{Two-loop contributions } 
\label{subsec:TLfull}

At zero temperature, 
\begin{eqnarray}
\ed_{(1)}(\den{p,n};bg)=\ed_{(0)}(\den{p,n};bg) + \delta\omega_{(1)}\left(T\rightarrow 0 \right) \equiv \ed_{(0)}(\den{p,n};bg) + \delta\ed_{(1)}(\den{p,n};\phib) \ ,
\end{eqnarray}
where the second term  is evaluated with the Fermi distribution $n_i(k) =  \theta\left(k_{Fi}-k\right)$ in $\delta\omega_{(1)}(T,\nu_{p,n};\phi)$. The single particle spectrum, $\espec{(1),i}{k}=\epsilon_{(0),i}(k)+ \delta\epsilon_{(1),i}(k)$ is discussed in Appendix~\ref{sec:degnondeglimits}. At the two-loop level, the five coupling constants are determined from 
\begin{eqnarray}
\mu_0-E^{\ast}_{F0}&=&  \frac{\gv^2}{m_v^2} \rho_0 +\left.\frac{\partial \delta\ed_{(1)}}{\partial \rho_B}\right\vert_{\phib_0,x=0.5} \,  \label{eqn:muconstraintTL} \\
S_{2,0}-\frac{k_{F0}^2}{6E_{F0}^{\ast}}&=&\frac{\rho_0}{8}\frac{\grho^2}{m_{\rho}^2}+\delta S_{2,(1)}\,  \label{eqn:s2constraintTL} \\
M-B_0-\frac{\ed_{kin0}}{\rho_0} &=&\frac{\vp(\phib_0)+\delta\ed_{(1)}}{\rho_0}+\frac{1}{2} \frac{\gv^2}{m_v^2} \rho_0  \,  \label{eqn:beconstraintTL} \\
K_{v,0}-3 \frac{k_{F0}^2}{E^{\ast}_{F0}}  &=& 9 \rho_0 \bigg[
\frac{\gv^2}{m_v^2}  +\frac{M^{\ast}_0}{E^{\ast}_{F0}} \left.\frac{\partial M^{\ast}}{ \partial \rho_B}\right\vert_{x=0.5} \bigg]  + \delta K_{v,(1)}\,  \label{eqn:kvconstraintTL} \\ 
\frac{E^{\ast}_{F0}}{m^{\ast}_0}&=& 1+\frac{E^{\ast}_{F0}}{k_{F0}} \left. \frac{\partial \delta \epsilon_{(1)}(k) }{\partial k }\right\vert_{k=k_{F0}} \ , \label{eqn:mlconstraintTL}
\end{eqnarray}  
where $\ed_{kin0}$ is the same as in the MFT calculation. 
The relation between $\mnstar$ and $\den{B}$ at $x=0.5$, and hence $\partial \mnstar / \partial \den{B}$, is provided by
\begin{eqnarray}
\rho_{s} + \left.\frac{\partial \delta\ed_{(1)}}{\partial M^{\ast} } \right\vert_{\rho_B, x=0.5}+\frac{d \vp(\phib)}{ d M^{\ast}}=0 \ .
\end{eqnarray}
In \eq{eqn:muconstraintTL},
  the TL contribution is $\left.\frac{\partial \delta\ed_{(1)}}{\partial \rho_B}\right\vert_{\phib_0,x=0.5}$. 
 In \eq{eqn:s2constraintTL}, the TL contribution to symmetry energy   (note $\left.\frac{\partial M^{\ast}}{ \partial x}\right\vert_{x=0.5}=0$) is
\begin{eqnarray}
\delta S_{2,(1)} &=&\frac{1}{8 \rho_B}\left.\frac{\partial^2  \delta\ed_{(1)}}{\partial x^2}\right\vert_{\rho_B,bg,x=0.5}   \ . \label{eqn:deltaS2}
\end{eqnarray}
In \eq{eqn:beconstraintTL}, the extra energy due to TL is $\ed_{(1)}/\rho_0$. In \eq{eqn:kvconstraintTL}, the TL contribution is 
\begin{eqnarray}
\delta K_{v,(1)}&=& 9\rho_B\left[\left.\frac{\partial^2 \delta\ed_{(1)}}{\partial \rho_B^2} \right\vert_{bg, x=0.5}+\left.\frac{\partial M^{\ast}}{\partial \rho_B}\right\vert_{x=0.5}\left.\frac{\partial^2 \delta\ed_{(1)}}{\partial M^{\ast} \partial \rho_B}\right\vert_{x=0.5} \right]   \ .
\end{eqnarray} 
These equations for fixing the five parameters are highly nonlinear. The fitted couplings and the predicted stiffness parameter 
$L$ are shown in Table~\ref{tab:two-loop_mine}.

\section{Self-consistent calculation of thermal effects} \label{sec:SC}

For given values of the chemical potential $\mu_{p,n}$ and temperature $T$, the single-particle spectrum can be decomposed as
\begin{eqnarray}
\epsilon_i \big[p;bg,n_{p,n}\left(k\right)\big] &=& \epsilon_{(0),i}(p;bg) + \delta\epsilon_{(1),i}\big[p;bg,n_{p,n}\left(k\right)\big]  \ , \label{eqn:specT}\\ 
n_i\left(k\right) &=& \frac{1}{e^{\beta\left(\epsilon_i\left(k\right)-\mu_i \right)} + 1}\,, \quad
\rho_i = \gamma_s \int \dvecp{k} n_{i}\left(k\right)  \ .  
 \label{eqn:nksc}
\end{eqnarray}
The first term in Eq.~(\ref{eqn:specT})$, \epsilon_{(0),i}$, corresponds to the spectrum of MFT defined in Eq.~(\ref{eqn:especMFT}). The second term,  $\delta\epsilon_{(1),i}$,  arises from TL contributions the $T=0$ analog of which is given in 
Eq.~(\ref{eqn:especTL}) with $n_i(k)=\theta\left(k_{Fi}-k\right)$. Through its dependence on $n_i(k)$ in Eq.~(\ref{eqn:nksc}) at finite $T$, $\delta\epsilon_{(1),i}$  depends on the full spectrum $\epsilon_i$ itself. This feature necessitates a self-consistent procedure for the determination of $\epsilon_i$ similar to that encountered in Hartree-Fock calculations (see also, Refs.~\cite{Welke:1988zz,2015PhRvC..92b5801C} in the context of non-relativistic models with finite-range interactions).  
At given values of $\rho_B$ and $T$, the chemical potentials $\mu_{p,n}$, the meson fields $bg$, the spectra $\epsilon_i(k)$ and the distribution functions $n_i(k)$ can be determined by an iterative process starting with the $T=0$ spectra as  guesses and updating the results with each iteration until convergence is achieved.  
Based on the self-consistent spectra and distribution functions, the entropy density is given by 
\begin{eqnarray}
\sd\big[n_{p,n}(k)\big]&=&-\gamma_{s}\sum_{i} \int \dvecp{k} \bigg[\big(1-n_{i}(k)\big)\ln\big(1-n_{i}(k)\big)+n_{i}(k)\ln\left(n_{i}(k)\right)\bigg] \ , \label{eqn:scSthdef}
\end{eqnarray}
and the energy density by 
\begin{eqnarray}
\ed\big[bg,n_{p,n}(k)\big]= \ed_{(0)}\big[bg,n_{p,n}(k)\big]+ \delta\ed_{(1)}\big[bg,n_{p,n}(k)\big] \ .
\end{eqnarray}
The expressions for $ \ed_{(0)}$ and $\delta\ed_{(1)}$ can be derived  from $ \ed_{(0)}$ in Eq.~(\ref{eqn:edmftdef}) and $\delta\omega_{(1)}$ in Eqs.~(\ref{eqn:chempot1phi})--(\ref{eqn:chempot1pi}), but with the use of the self-consistent $n(k)$. 
The pressure is then obtained from $P = - \ed + T\sd + \sum_i \mu_i \den{i}$.
 The free energy density ensues from
\begin{eqnarray}
\fd\big[T,bg,n_{p,n}(k)\big] &=&\ed\big[bg,n_{p,n}(k)\big]-T \sd\big[bg,n_{p,n}(k)\big] \ .
\end{eqnarray}
Two points are worth noting here: (1) the $T\rightarrow 0$ limit of the self-consistent calculation agrees with the $T=0$ calculation in section~(\ref{sec:zeroT}), and (2) the self-consistent calculation at the MFT level is the same as the one discussed in section~\ref{subsec:MFTgeneral}.

In order to express the $\fd$ as a function of $T$ and $\den{p,n}$, we first need to determine $\mu_{p,n}$ in terms of $bg$, $\den{p,n}$, and $T$. 
The second step is to solve for $bg$ for given $T$ and $\den{p,n}$. As $bg$ should minimize the free energy at fixed $T$ and $\den{p,n}$, the first derivative of $\fd$ wrt $bg$ vanishes. This derivative can be expressed as 
\begin{eqnarray}
 \left.\frac{\partial \fd\big[T,bg,n_{p,n}\big]}{\partial\, bg} \right\vert_{T,\den{p,n}} &=& 
\left.\frac{\partial \ed\big[bg,n_{p,n}\big]}{\partial\, bg} \right\vert_{T,n_{p,n}} \notag \\
&+& \sum_i \int \dvecp{k} \left.\left[\frac{\partial \ed\big[bg,n_{p,n}\big]}{\partial n_i(k)}- T \frac{\partial \sd\big[n_{p,n}\big]}{\partial n_i(k)} \right] \right\vert_{T,bg, n_{j\neq i}} \left. \frac{\partial n_i\big[k;T,\den{p,n},bg\big]}{\partial\,  bg}\right\vert_{T,\den{p,n}} \label{eqn:dfddbg1} .
\end{eqnarray}
(In order to simplify notation, the $k$-dependence in $n_{p,n}(k)$ is suppressed when used in arguments of functions.) Above, $\partial / \partial n_i(k)$ are functional derivatives, which is why the $\int \dvecp{k}$ is involved.  
Further simplification of Eq.~(\ref{eqn:dfddbg1}) occurs with use of the relations
\begin{eqnarray}
\label{eqn:sdnder}
 \left.\frac{\partial \sd\big[n_{p,n}\big]}{\partial n_i(k)} \right\vert_{T,bg, n_{j\neq i}} &=& \gamma_s  \beta\left(\epsilon_i\big[k;bg,n_{p,n}\big]-\mu_i\right) \ , \\ 
 \left. \frac{\partial \ed\big[bg,n_{p,n}\big]}{\partial n_i(k)} \right\vert_{T,bg, n_{j\neq i}} &=& \gamma_s \epsilon_i\big[k;bg,n_{p,n}\big] \ . \label{eqn:ednder}
\end{eqnarray}
which renders the term involving integrals in Eq.~(\ref{eqn:dfddbg1}) to vanish with the result  
\begin{eqnarray}
\left.\frac{\partial \fd\big[T,bg,n_{p,n}\big]}{\partial\, bg} \right\vert_{T,\den{p,n}}  
  &=& \left. \frac{\partial \ed\big[bg,n_{p,n}\big]}{\partial\, bg} \right\vert_{T,n_{p,n}} \label{eqn:bgeqsSC}
\end{eqnarray}
This indicates that all the meson fields satisfy the same equations as those in the $T=0$ case, except with the theta function substituted with the self-consistent one for $n_i(k)$. The $bg$ can then be determined in terms of $\den{p,n}$, and $T$, based on which $n_i\big[k;T,\den{p,n},bg\big]$, $\ed\big[bg,n_{p,n}\big]$, $\sd\big[n_{p,n}\big]$, and $\fd\big[T,bg,n_{p,n}\big]$ are now functions  of $T$ and $\den{p,n}$.

From Eqs.~(\ref{eqn:sdnder}), (\ref{eqn:ednder}) and the expression in Eq.~(\ref{eqn:bgeqsSC}) set to 0, we can check that
\begin{eqnarray}
\left.\frac{\partial \fd\big[T,bg,n_{p,n}\big]}{\partial\, T} \right\vert_{\den{p,n}} 
= - \sd  \quad {\rm and} \quad
\left.\frac{\partial \fd\big[T,bg,n_{p,n}\big]}{\partial\, \den{i}} \right\vert_{T,\den{j\neq i}} 
=  \mu_i \ .
\end{eqnarray}

\subsection*{Self-consistent vs perturbative calculations}

Here the relationship between the self-consistent and perturbative calculations of thermal effects is examined. We restrict ourselves to the degenerate situation when $T/T_{F_i} \ll1$, where $T_{F_i}$ is the Fermi temperature. The FLT result for the entropy density in Eq.~(\ref{eqn:scSthdef}) is~\cite{flt} 
\begin{eqnarray}
\sd = \frac{\pi^2}{3} T \sum_i N_i(0) \ ,  \quad
N_i(0) = \gamma_s \int \dvecp{k} \delta\left(\epsilon_i\left(k\right)-\mu_i \right) \ \,, 
\end{eqnarray}
where  $N_i(0)$ is the density of states at the Fermi surface of species $i$, and 
 $\epsilon_i(k)$ and $\mu_i$ are the $T=0$ single-particle spectrum and chemical potential, respectively. 
 The Landau effective mass $\mland_i$  in Eq.~(\ref{eqn:mlanddef}) is proportional to $N_i(0)$. 
We begin by rewriting  Eq.~(\ref{eqn:fd1def}) as 
\begin{eqnarray}
\fd_{(1)}\big[T,bg,n_{(0)p,n}(k)\big]&=& \ed_{(1)}\big[bg,n_{(0)p,n}(k)\big]- T \sd_{(0)}\big[n_{(0)p,n}(k)\big] \ . 
\end{eqnarray}
In order to differentiate the perturbative calculation from the self-consistent calculation, the order indices ${(0)}$ and ${(1)}$ in subscripts are kept manifest.  Explicitly, $bg$ in the above equation is solved perturbatively, which is different from the $bg$ in the self-consistent calculation. The distribution functions $n_{(0)p,n}(k)$ are defined at MFT level, i.e., 
\begin{eqnarray}
n_{(0)i}\left(k\right) &=& \frac{1}{e^{\beta\left(\epsilon_{(0)i}\left(k;bg\right)-\mu^{(0)}_i \right)} + 1} \ ,
\end{eqnarray}
where 
$\mu^{(0)}_i$ has been discussed in relation to Eq.~(\ref{eqn:omega1def}) and can be determined in terms of  $bg$  and $T$ through 
\begin{eqnarray}
\rho_i = \gamma_s \int \dvecp{k} n_{(0)i}\left(k\right)  \ .  
\end{eqnarray}
Similar to the derviation in Eq.~(\ref{eqn:dfddbg1}), we can compute the entropy density utilizing  ($bg$ being held fixed due as it extremizes $\fd_{(1)}$ for given $T$ and $\den{p,n}$ )
\begin{eqnarray}
&& \left.\frac{\partial \fd_{(1)}\big[T,bg,n_{(0)p,n}\big]}{\partial\, T} \right\vert_{\den{p,n},bg} = \notag \\
&&-\sd_{(0)}\big[n_{(0)p,n}\big] +  \left.\frac{\partial \ed_{(1)}\big[bg,n_{(0)p,n}\big]}{ \partial\, T} \right\vert_{\den{p,n},bg} - T \left.\frac{ \partial \sd_{(0)}\big[n_{(0)p,n}\big]}{\partial T} \right\vert_{\den{p,n},bg} \notag \\ 
&& = -\sd_{(0)}\big[n_{(0)p,n}\big] +  \gamma_s  
\sum_i \int \dvecp{k} \left[ \epsilon_{(1)i}\big[k; bg,n_{(0)p,n}\big] - \left(\epsilon_{(0)i}\big[k; bg\big] - \mu^{(0)}_i\right)\right]  \left. \frac{\partial n_{(0)i}\big[k;T,\den{p,n},bg\big]}{\partial\,  T}\right\vert_{\den{p,n},bg} \notag \\ 
&& = -\sd_{(0)}\big[n_{(0)p,n}\big] +  \gamma_s  
\sum_i \int \dvecp{k} \delta\epsilon_{(1)i}\big[k; bg,n_{(0)p,n}\big]  \left. \frac{\partial n_{(0)i}\big[k;T,\den{p,n},bg\big]}{\partial\,  T}\right\vert_{\den{p,n},bg} \ . 
\end{eqnarray}
In the degenerate limit, the above reduces to 
\begin{eqnarray}
-\left.\frac{\partial \fd_{(1)}\big[T,bg,n_{(0)p,n}\big]}{\partial\, T} \right\vert_{\den{p,n},bg} 
& = &
T \frac{\pi^2}{3} \sum_i N_{(0)i}(0) \left[1 - \left.\frac{\partial \delta\epsilon_{(1)i}} {\partial \epsilon_{(0) i} } \right\vert_{\epsilon_{(0)i}=\mu^{(0)}_i }  \right]  \ , \\
N_{(0)i}(0) &\equiv & \gamma_s \int \dvecp{k} \delta\left( \epsilon_{(0)i}-\mu^{(0)}_i \right) \ . 
\end{eqnarray}
This suggests that the perturbative results approach the FLT limit, but are controlled by effective masses that are different from the Landau effective masses.  Specifically, 
\begin{eqnarray} 
m^{\ast '}_{i} = E^{\ast}_{F,i} \left[1- \left. \frac{\partial \delta\epsilon_{(1)i}}{\partial\epsilon_{(0)i} } \right\vert_{k=k_{F,i}} \right] \ , 
\end{eqnarray}
whereas the Landau effective masses are (cf. Appendix~\ref{sec:spectrum})
\begin{eqnarray} 
\mland_{i} =E^{\ast}_{F,i}\left[1+ \left. \frac{\partial \delta\epsilon_{(1)i}}{\partial\epsilon_{(0)i} } \right\vert_{k=k_{F,i}} \right]^{-1} \ .
\end{eqnarray}
To first order in the derivative term, the two results agree.
Eliminating the derivative term, 
\begin{eqnarray}
m^{\ast '}_i = E^{\ast}_{F,i}\left[2 - \frac{E^{\ast}_{F,i}}{ \mland_i} \right] \ . \label{eqn:revisedMland}
\end{eqnarray}
In our calculations, the perturbative results indeed approach $m^{\ast '}_i$ instead of $\mland_i$, differences between the two being apparent only at very high densities.

\section{Results at zero temperature} 
\label{sec:results0} 

\begin{figure}
	\centering
		\includegraphics[angle=0,width=0.5\textwidth]{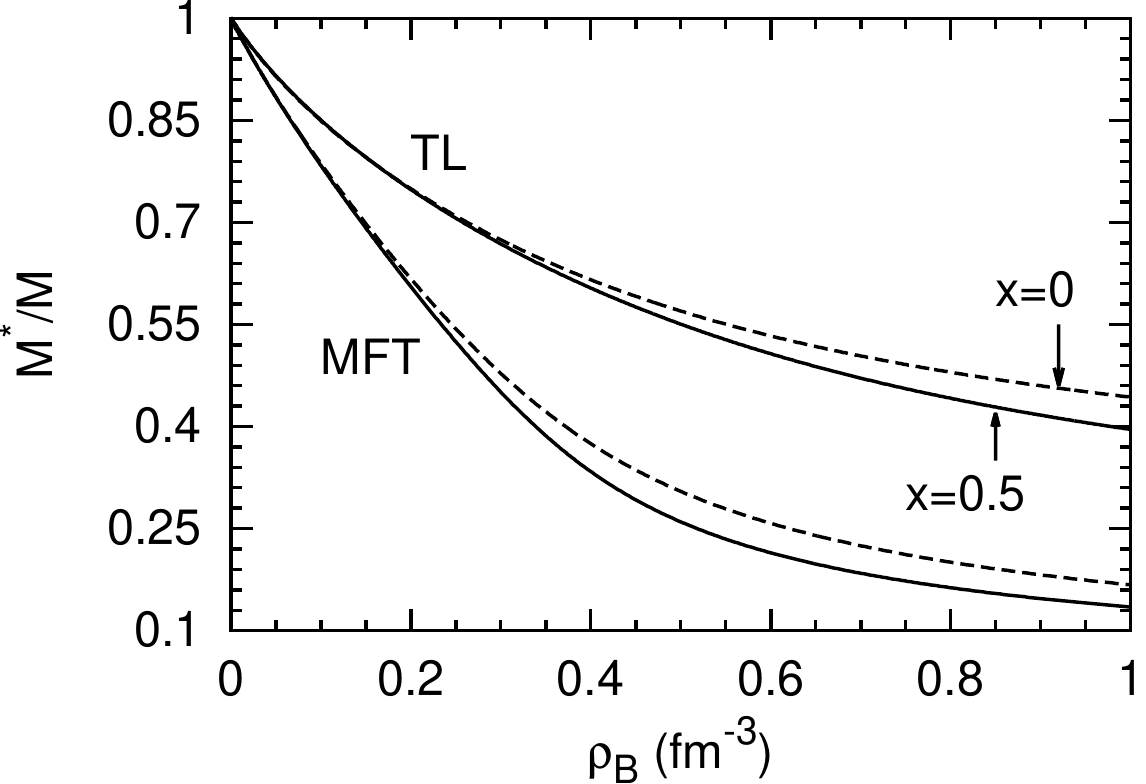}
     \caption{Dirac effective masses $\mnstar/M$ vs. baryon density $\den{B}$ in SNM ($x=0.5$) and PNM ($x=0$) for  the MFT and TL calculations. 
     } \label{fig:Mstarmine}
\end{figure}

For the most part, results of TL numerical calculations in this paper  employ the parameter set TL(250) in Table~\ref{tab:two-loop_mine}, 
which is labeled TL in figures and their associated discussions. The starting point for both MFT and TL calculations is the determination of the Dirac effective masses, $M^*$'s, which feature prominently in the expressions for the energy and pressure. The  scalar couplings  in Table~\ref{tab:two-loop_mine},  
which determine $M^*$'s,  ensure that at $\rho_0$ the Landau effective masses, $m^*$'s, in the two calculations  are the same.  
The density dependences of $M^*$'s in symmetric nuclear matter (SNM)  with proton fraction $x=0.5$ and pure neutron matter (PNM) with $x=0$ are  shown in \fig{fig:Mstarmine}. 
 In TL calculations the decrease of  $\mnstar$  with density is much slower than in MFT. At $1\,\denunit$, $\mnstar/M$ for TL 
is $\sim 0.4~(0.5)$  in SNM (PNM), whereas the corresponding values for  MFT   is $\sim 0.1~(0.2)$. 
The larger values of $\mnstar/M$ in TL calculations arise from the repulsive TL contributions to the single particle energy from the exchange of the scalar meson $\phi$ (see discussion below). 

\begin{figure}
	\centering
		\includegraphics[width=0.5\textwidth]{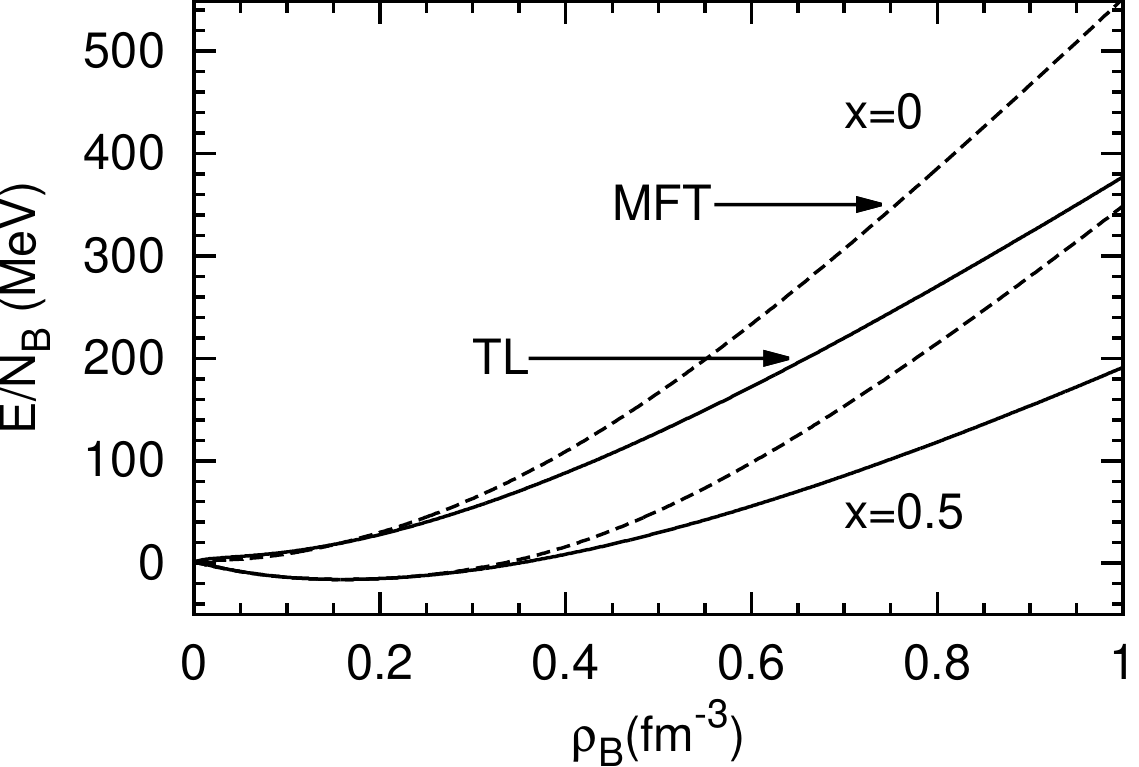}
     \caption{Energy vs. $\den{B}$  in SNM  and PNM  
     for the MFT and TL calculations.} \label{fig:zeroT_BE_mine}
\end{figure}

Figure \ref{fig:zeroT_BE_mine} shows the energy per baryon $E\equiv \ed/\den{B}-M$ from  MFT and TL calculations 
in SNM and PNM. 
For $\den{B} \leq 0.4$ $\denunit$ in SNM, MFT and TL calculations give nearly the same result chiefly because the energy and curvature at the equilibrium density $\rho_0$ are fixed to the same values in obtaining the coupling strengths of the two models. 
But for  densities $\den{B} \geq 0.4$ $\denunit$, the TL energy is much smaller than that for MFT.

\begin{figure}
	\centering
		\includegraphics[angle=0,width=0.5\textwidth]{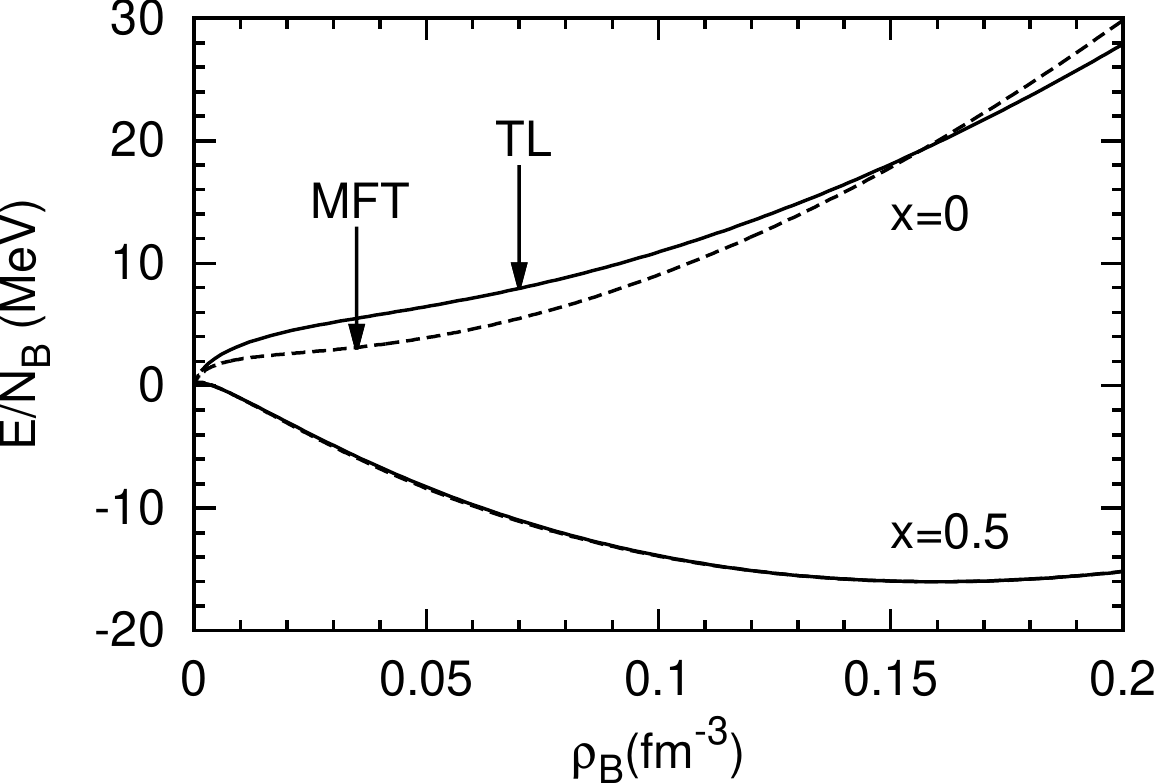}
     \caption{Same as Fig. \ref{fig:zeroT_BE_mine}, but for the low density region.} 
     \label{fig:zeroT_BE_minev2}
\end{figure}

In PNM, results of MFT and TL calculations differ significantly both in the low density region, $\den{B} \leq 0.1\, \denunit$, and in the high density region,  $\den{B} \geq 0.4\,\denunit$. 
In the low density region (\fig{fig:zeroT_BE_minev2}), the TL energy is larger than that of MFT, noted earlier in Refs.~\cite{Chin:1977iz, Horowitz:1981zg,Horowitz:1982kp}.  When $k_{F,i}/m \ll 1$ where $m$ is the mass of the exchanged meson, the leading terms of the exchange energies  vary as   $k_{F,i}^3$ and  $k_{F,i}^5/m^2$ (as can be ascertained by Taylor expanding the non-relativistic exchange integrals in Appendix \ref{subsec:TLhnlimit}), and with overall signs that oppose contributions from the direct (Hartree) terms. The net exchange energies in SNM and PNM thus acquire different density dependences  because of the different coupling strengths associated with the different mesons being exchanged,  as well as $k_F$ in PNM being greater than  that in SNM at the same density $\rho_B$.  It is reassuring that
exchange contributions bring the QHD energy for PNM in the low density region close to  results of non-relativistic 
microscopic calculations, see also Refs.~\cite{Typel:1999yq,Bender:2003jk}.

\begin{figure}
	\centering
		\includegraphics[angle=0,width=0.5\textwidth]{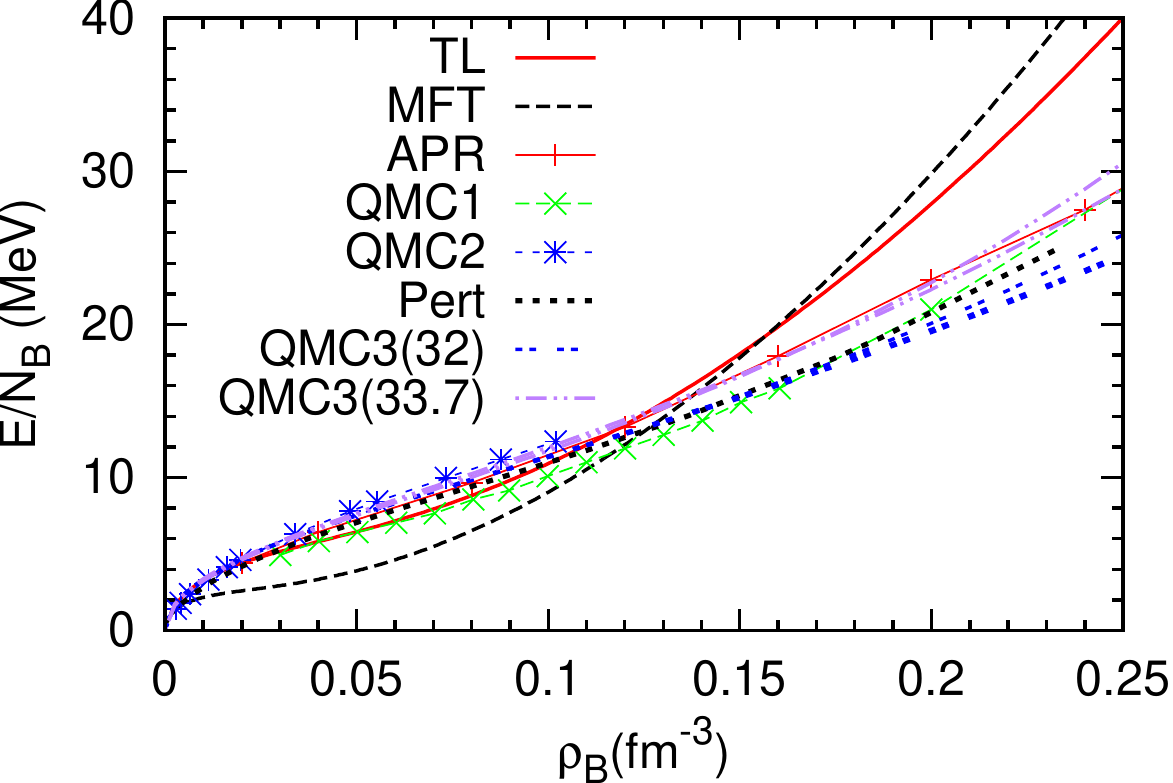}
     \caption{Comparison of the MFT and TL results for PNM  energy vs $\rho_B$ with those of 
     a variational calculation ``APR''~\cite{Akmal:1998cf},  quantum Monte Carlo (QMC) calculations, ``QMC1''~\cite{Gandolfi:2009fj,Armani:2011mn}, ``QMC2'' based on a N3LO chiral potential with momentum cut-off $414$ MeV~\cite{Wlazlowski:2014jna}, and  ``QMC3''~\cite{Gandolfi:2011xu},  and the N3LO chiral perturbation theory calculation (``Pert'') with a momentum cut-off of 500 MeV~\cite{Coraggio:2012ca}. For ``QMC3'', the two curves shown are with $S_2=32$ and 33.7 MeV, respectively.
     } 
\label{fig:zeroT_BE_minevsabinitio}
\end{figure}

\begin{figure}
	\centering
		\includegraphics[width=\textwidth]{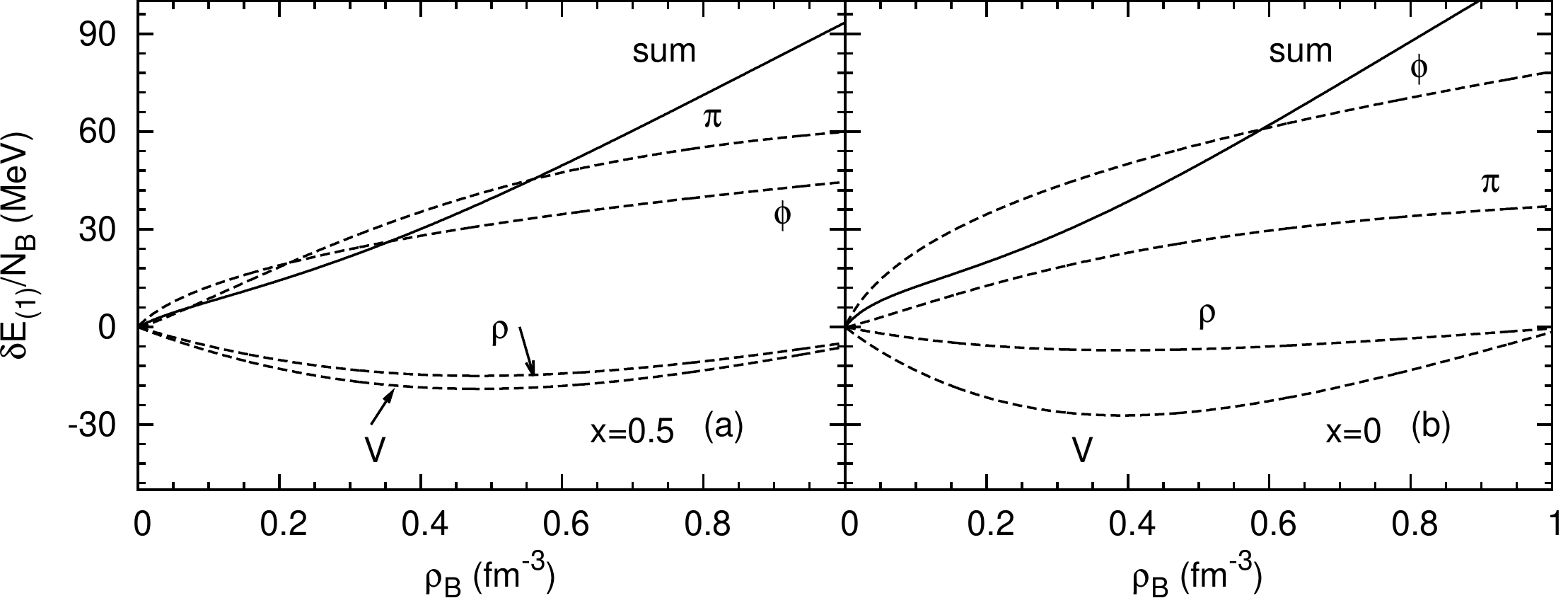} 
     \caption{Two-loop contributions to the energy vs. $\den{B}$ in SNM and PNM. } \label{fig:zeroT_E2nd_mine}
\end{figure}

In Fig.~\ref{fig:zeroT_BE_minevsabinitio}, we compare our MFT and TL results for PNM energy to those of modern microscopic 
calculations~\cite{Akmal:1998cf,Gandolfi:2009fj,Armani:2011mn,Wlazlowski:2014jna,Coraggio:2012ca,Gandolfi:2011xu}, in which nucleon-nucleon interactions that reproduce scattering data and binding energies of light nuclei were used.
In some cases (QMC1, QMC2, and QMC3), the role of different 
three-nucleon interactions were also explored.     
The figure caption provides some details of these calculations. The agreement of the TL results with those of potential model calculations up to  $\rho_B \leq 0.13 \, \denunit$ is much better than for the MFT results. The total energy per nucleon
 in MFT at low densities receives contributions from terms proportional to $\rho_B^{2/3}/M^*(\rho_B)$ from the kinetic energy and $\rho_B$ from direct (Hartree) terms involving $\omega$ and $\rho$ meson interactions.  Additional non-trivial density dependences arise from  the exchange of mesons with different masses, and hence ranges, when TL (Fock) contributions are added to the MFT parts (see \fig{fig:zeroT_E2nd_mine} with its associated discussion below for more details). The agreement with the results of potential model calculations can thus be attributed to the inclusion of TL or exchange diagrams at low densities.
For nuclear densities and beyond, 
the TL energies are smaller than those of MFT, but significantly larger than the results of non-relativistic treatments. Nevertheless, the EOS remains causal owing to the relativistic structure of QHD.   

In order to explore the sensitivity of the TL results to the values of the compression modulus  and symmetry energy chosen to determine the various coupling strengths, additional calculations with $K_{v,0} = (235,270)$ MeV and    $S_{2,0}=(33,35)$ MeV, which yielded $L=(75.6,80.6)$ MeV were performed (see Table~\ref{tab:two-loop_mine}) 
for the parameter values).  In both of these cases, the agreement with the results of non-relativistic approaches for $\rho_B \leq 0.13 \, \denunit$
was satisfactory, but not as good  as for the set  with $K_{v,0} = 250$ MeV, $S_{2,0}=35$ MeV, and $L=85.09$ MeV.  Unless specified otherwise, subsequent results will be for the coupling strengths associated with this latter set.

Individual contributions to the total TL (exchange) energy from the exchange of the various mesons are shown in  \fig{fig:zeroT_E2nd_mine} for both SNM and PNM.  For interactions between nucleons mediated by the  pseudo-scalar $\pi^i$ and scalar $\phi$  mesons,  the TL terms provide positive contributions to the energy, whereas the corresponding direct (Hartree) terms are zero and negative, respectively.  For interactions mediated by the vector mesons $V^{\mu}$ and $\rho^{i,\mu}$ for which the Hartree terms are positive,  contributions from the TL terms are negative except at high densities for which positive contributions ensue  owing to relativistic effects.  Because the $\rho^{i,\mu}$ and $V^\mu$ meson masses are similar, the turnover densities for these two cases are close. In SNM, the turnover density is $1.15\,\denunit$; the corresponding $M^\ast=0.37\,M$, leading to $k_F/M^\ast=1.46$. In PNM, the turnover density is $1.03\,\denunit$, $M^\ast=0.44\,M$, and  $k_F/M^\ast=1.49$. These values may be contrasted with $k_F/M \simeq 2.533$ \cite{Salpeter61}  for the case of photon or gluon exchange between particles of mass $M$. For both SNM and PNM, the net TL contribution to the total energy is positive with a density dependence reflecting contributions from sources with differing masses of the exchanged mesons.

\begin{figure}
	\centering
		\includegraphics[angle=0,width=0.5\textwidth]{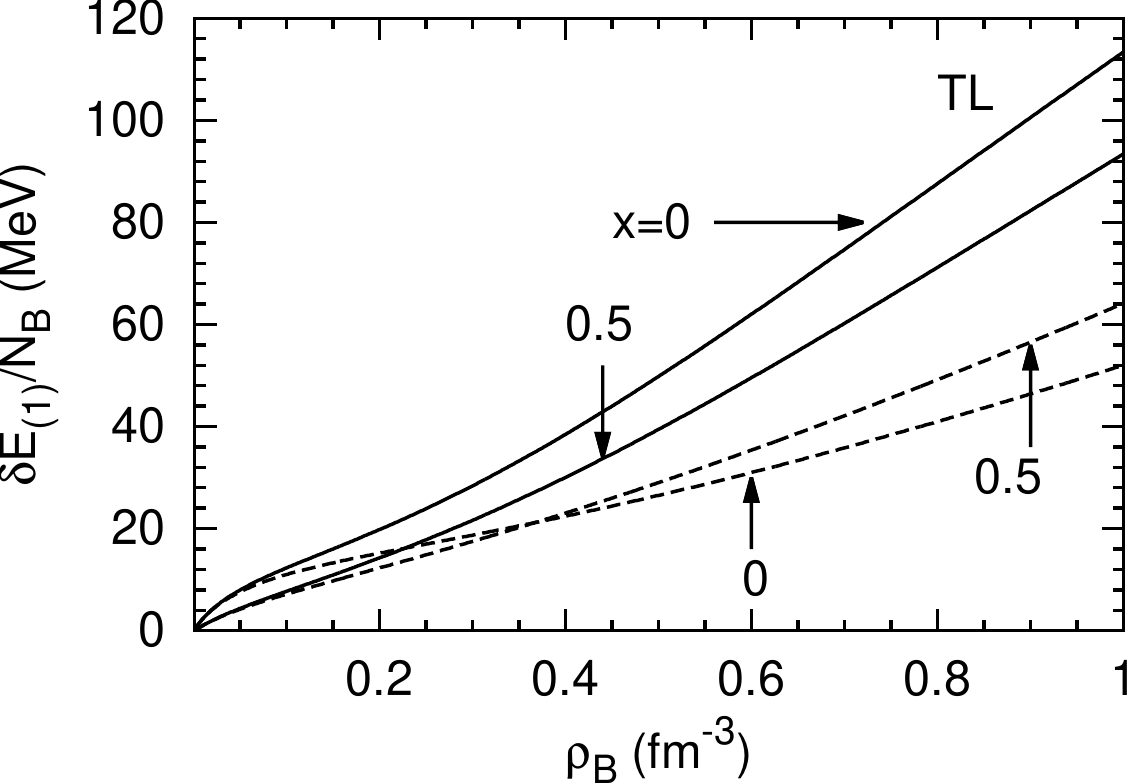} 
     \caption{The total two-loop energy  and its non-relativistic approximation in SNM and PNM. } 
     \label{fig:zeroT_E2nd_fullvsNR}
\end{figure}

In the low density region when $k_F/\mnstar \ll 1$, the non-relativistic expressions for the TL energy given in Appendix~\ref{subsec:TLhnlimit} can be used to determine the density at which relativistic effects begin to become important.  
 In \fig{fig:zeroT_E2nd_fullvsNR}, results for the total exchange energies from the relativistic and 
non-relativistic approximation in both SNM and PNM are shown. 
In SNM, agreement between the two schemes extends up to  $~0.2\, \denunit$, but begins to fail thereafter. 
The  deviation in PNM starts at a lower density,   $~0.1\, \denunit$, than for SNM because of its higher Fermi momentum at the same density. At supra-nuclear densities, relativistic effects are clearly important in both cases. 


\begin{figure}
	\centering
			\includegraphics[angle=0,width=0.5\textwidth]{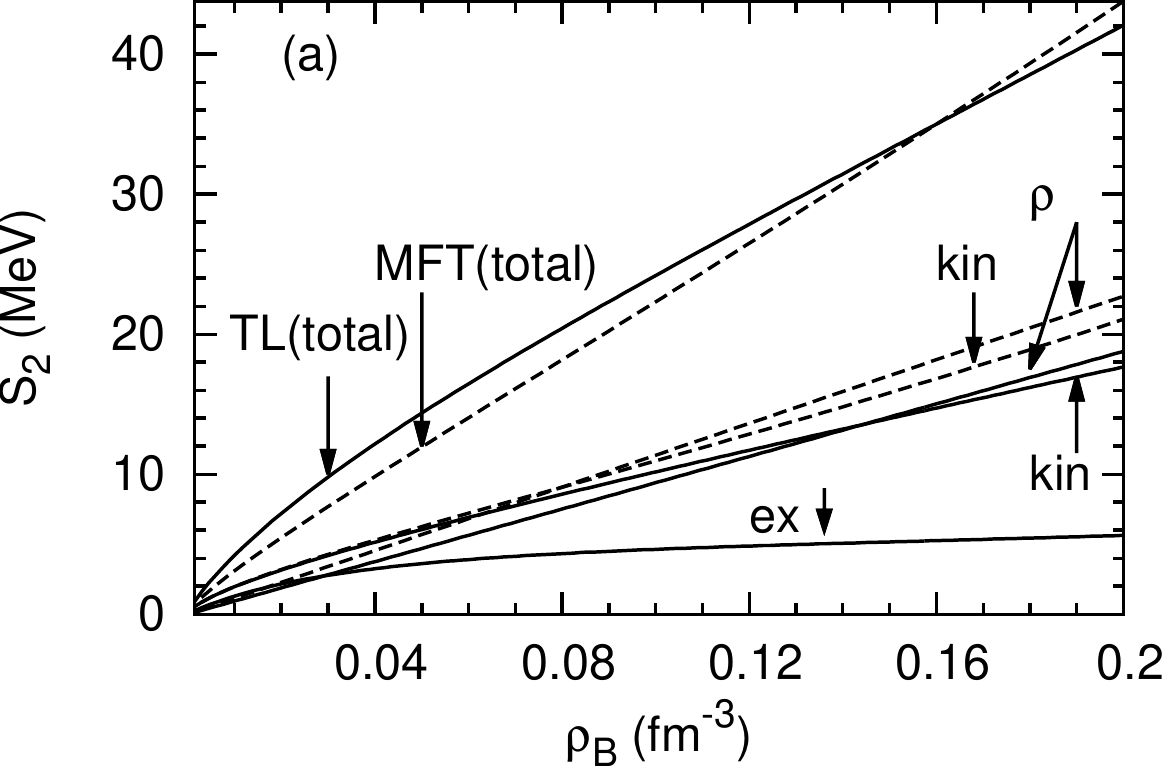} \\
		\hspace{-4mm}	 \includegraphics[angle=0,width=0.51\textwidth]{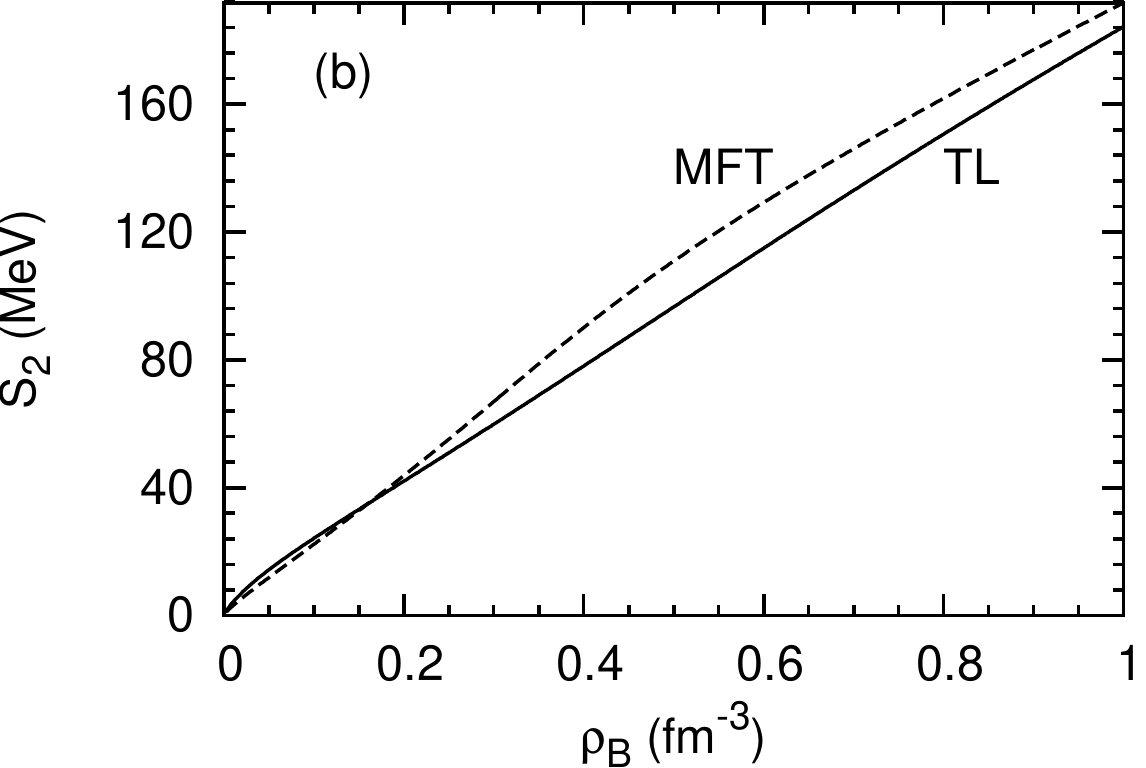}
					\caption{The symmetry energy $S_2$ vs $\rho_B$. In MFT, contributions to $S_2$  arise from 
     both the $\rho$-meson and scalar $\phi$-meson interactions [Eq.~(\ref{eqn:MFTs2})]. The latter piece is termed $S_2^{kin}$ in Eq.~(\ref{eqn:LMFTdecomp}). In the TL calculation, there is an additional  contribution from the exchange diagram [Eq.~(\ref{eqn:deltaS2})].}   \label{fig:zeroT_S2}
\end{figure}

The symmetry energies $S_2$ vs $\rho_B$ for the MFT and TL calculations are  shown in Fig.  \ref{fig:zeroT_S2}. The top panel in this figure shows contributions from terms involving  kinetic,  direct and exchange contributions from the $\rho$-meson to $S_2$ in the two cases. The symmetry energy stiffness parameter $L$ for the TL calculation (85.09 MeV) is smaller than that for MFT (103.62 MeV).  There are three factors all of which reduce  $L$ for  TL: (1) the contribution proportional to $\rho_B$ from $\rho$-meson exchange  is smaller in TL because the $g_\rho$ coupling is smaller than in MFT, (2) the $S_2^{kin}$ is smaller than in MFT because the $M^\ast$ is larger in TL, and  (3) the exchange diagram contribution to symmetry energy has a much weaker density dependence for near nuclear densities.  


\begin{figure}
	\centering
				 \includegraphics[angle=0,width=0.5\textwidth]{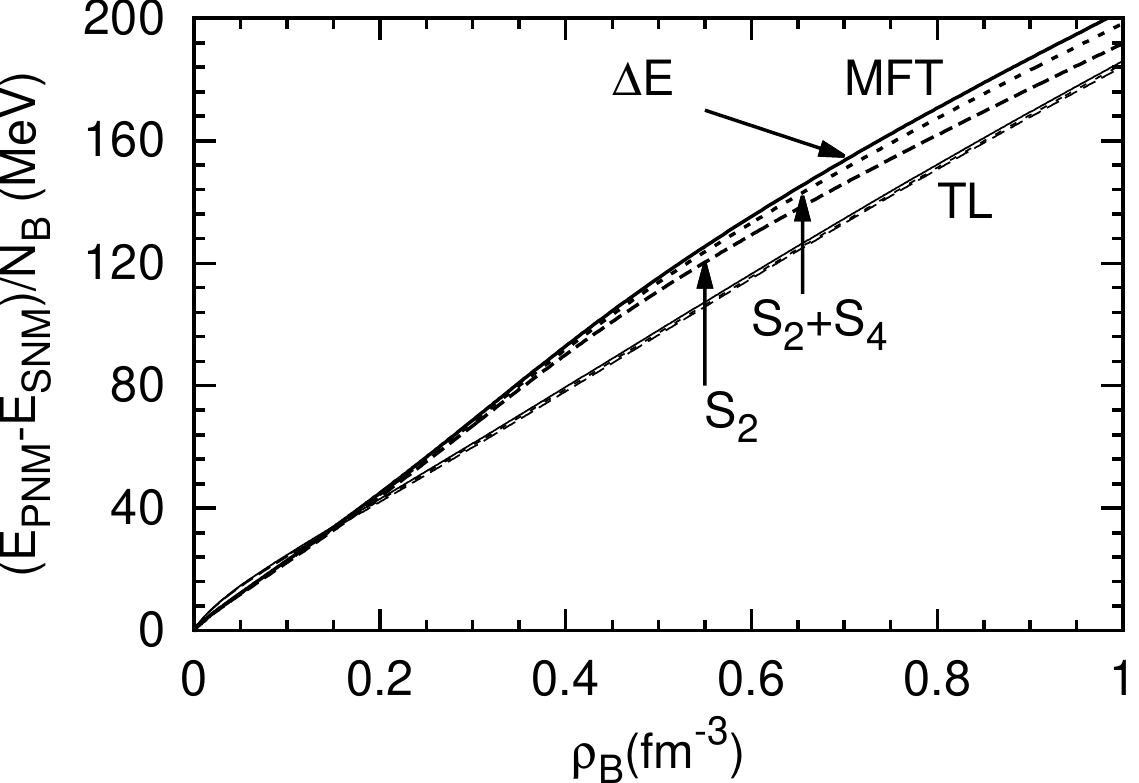}
     \caption{The energy difference between PNM and SNM approximated by $S_2$ and $S_2+S_4$  in MFT and TL calculations. For both cases, the solid curve is the energy difference, the bottom 
     long-dashed curve is $S_2$, and the middle short-dashed curve is $S_2+S_4$.}   \label{fig:zeroT_S2S4}
\end{figure}

The difference between the energy of PNM and SNM receives contributions from beyond quadratic terms in $\alpha=1-2x$, and can be expressed as 
\begin{equation}
\Delta E = (E_{PNM} - E_{SNM})/N_B = \sum_{\ell=2.4,...} S_\ell \alpha^\ell \,,
\end{equation}
where 
\begin{equation}
S_\ell = \frac {1}{\ell !} \left. \frac {\partial^\ell E(\rho_B,\alpha)/N_B}{\partial \alpha^\ell}\right|_{\alpha=0} \,; \quad \ell=2,4,....
\end{equation}
In MFT, we find the coefficient of  the quartic term in $\alpha$ from kinetic sources to be 
\begin{equation}
S_4^{kin} = \frac {1}{27} S_2^{kin} \left[ 1 + \frac {3}{4} \frac {k_F^2}{E_F^{*2}} \left( 1 +  \frac {k_F^2}{E_F^{*2}} \right) \right] 
-\frac {1}{96} \frac{M^*k_F^2}{E_F^{*3}}~ \left. \frac {\partial^2M^*}{\partial x^2}\right|_{x=1/2} \,,
\label{S4kin}
\end{equation}
where $S_2^{kin} = k_F^2/(6E_F^*)$. 
Table~\ref{tab:symenergies} lists the contributions from $S_2^{kin}$and $S_4^{kin}$ as well as that from $\rho$-meson exchange, $S_2^\rho$, to $\Delta E$ at   
the nuclear equilibrium density $\rho_0$. 
The contribution of $S_4^{kin}$ is small relative to $S_2^{kin}$ at near nuclear densities. At supra-nuclear densities,  it helps to bring $S_2+S_4$   close to $\Delta E$ as shown in Fig. \ref{fig:zeroT_S2S4}.

The TL contribution, $S_4=S_4^{kin} + S_4^{ex}$, is not readily amenable for analytical manipulations, but is straightforward to calculate numerically.  In this case, use of Eq.~(\ref{S4kin}) is inappropriate because the MFT piece of the kinetic energy alone does not satisfy $\partial E/\partial M^*$=0. However, its use does not yield significantly different results from the exact numerical calculations up to twice the saturation density but differences become noticeable at higher densities.   
At $\rho_0$, $S_4=0.5$ MeV is of similar magnitude to  $S_4^{kin}$ at the MFT level. 
We found very little difference between the results of $S_2$ and $S_2+S_4$ as functions of $\rho_B$ (see Fig. \ref{fig:zeroT_S2S4}). This feature is attributable to the inherent structure of the exchange terms that have opposite signs relative to the direct terms in the calculation of energies at the densities shown.  
For the structural properties of neutron stars, we report results from the use of  $S_2$ to keep neutron-star matter  charge neutral and in beta-equilibrium for both MFT and TL calculations. Negligible differences were found with the use of $S_2+S_4$.    \\  


\begin{table}
   \caption{Contributions from quadratic and quartic terms 
   of symmetry energies to the  difference $\Delta E$ between the PNM and SNM energies  at the nuclear 
   density $\rho_0=0.16~{\rm fm}^{-3}$.} 
    \begin{ruledtabular} 
   \begin{tabular}{l c c c c c c c c } 
   	Model & $M^*$ & $S_2^{kin}$ & $S_2^\rho$ & $S_2^{ex}$ & $S_2$ & $S_4^{kin}$ &$S_4^{ex} $& $\Delta E$ \\ \hline
   	MFT  & 0.674 & 18.2 & 16.8 & $-$ &  35.0 & 0.63 &   $-$ & 35.9 \\ 
	MFT + TL & 0.785 & 14.8 & 15.1 & 5.1 & 35.0 & 0.6 & $-0.1$ &  35.9 \\
   \end{tabular}      
   \end{ruledtabular}
   \label{tab:symenergies}
\end{table}
%


\begin{figure}
	\centering
		\includegraphics[angle=0,width=0.5\textwidth]{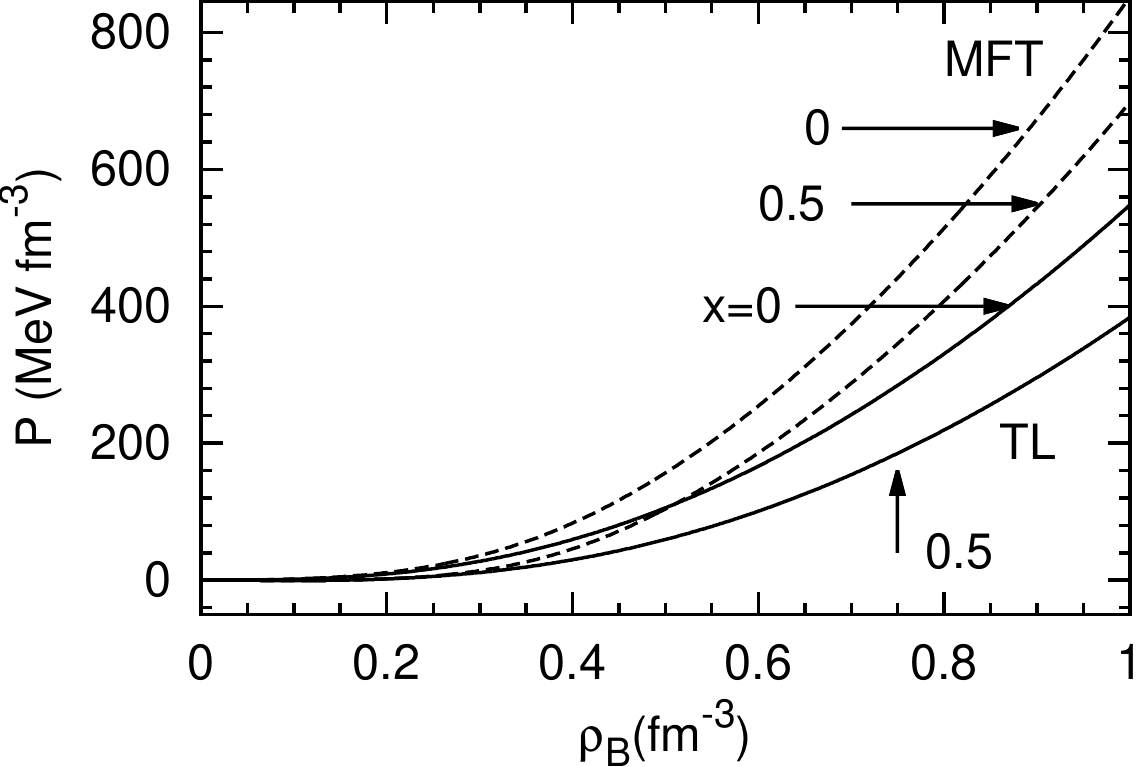}	
     \caption{Pressure vs. density in SNM and PNM.  
     } \label{fig:P_MFT_TL}
\end{figure}

\begin{figure}
	\centering
		\includegraphics[angle=0,width=0.5\textwidth]{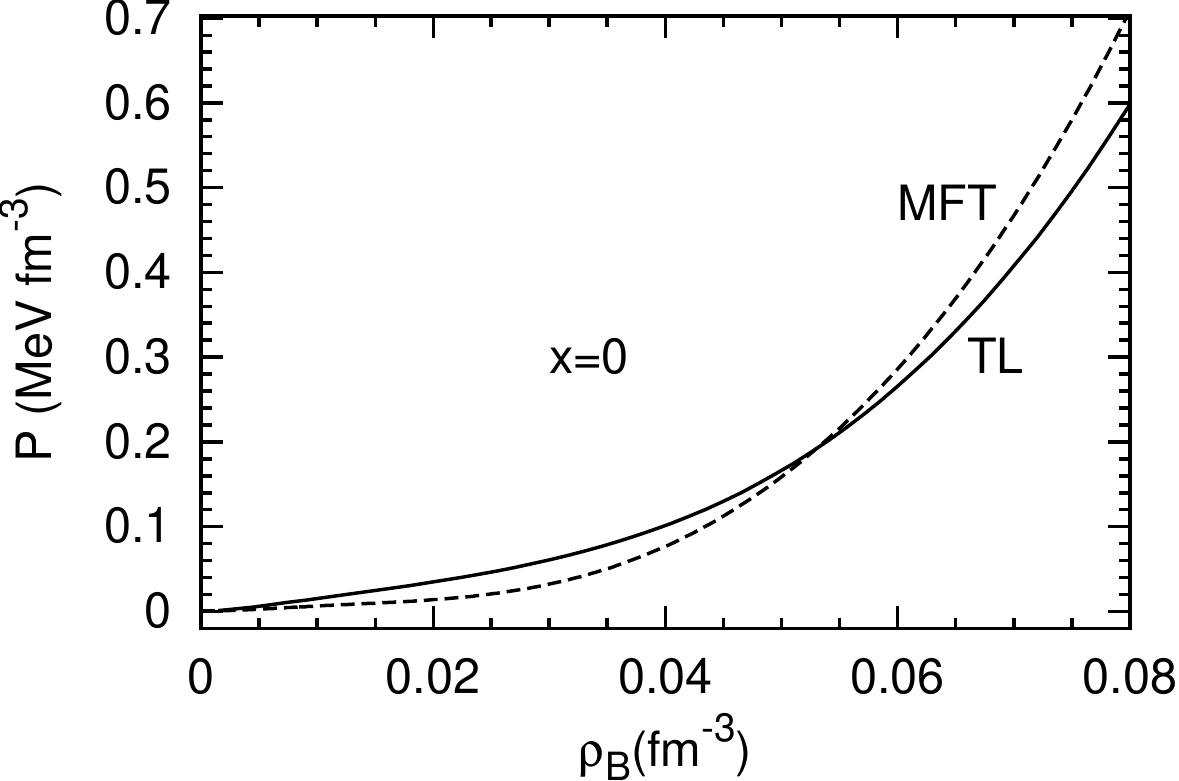}	
     \caption{Same as \fig{fig:P_MFT_TL}, but for PNM in the low density region. } \label{fig:P_MFT_TLv2}
\end{figure}

The pressure  $P$ vs. $\den{B}$ is shown in \fig{fig:P_MFT_TL} for both SNM and PNM.  \fig{fig:P_MFT_TLv2} shows results for PNM at sub-saturation densities.  As expected from the results for energy in \fig{fig:zeroT_BE_mine} and \fig{fig:zeroT_BE_minev2}, the MFT and TL results for pressure are close to each other in SNM for $\den{B} \leq 0.3\, \denunit$; above this region, the MFT pressure becomes larger than that of TL. In PNM,  the situation is somewhat diffrent; for $\den{B}\geq 0.05\, \denunit$, the MFT pressure is larger than that of TL, but at sub-nuclear densities, the trend is reversed.  For neutron star structure, however, these differences below  $\den{B} \leq 0.08\, \denunit$ may not matter because  the appropriate EOS there would be that of inhomogeneous matter containing nuclei. We turn now to address the question of whether these EOS's can support a two solar mass neutron star as required by recent precise determinations \cite{Demorest10,Antoniadis13}. 

\begin{figure}
	\centering
		\includegraphics[angle=0,width=0.5\textwidth]{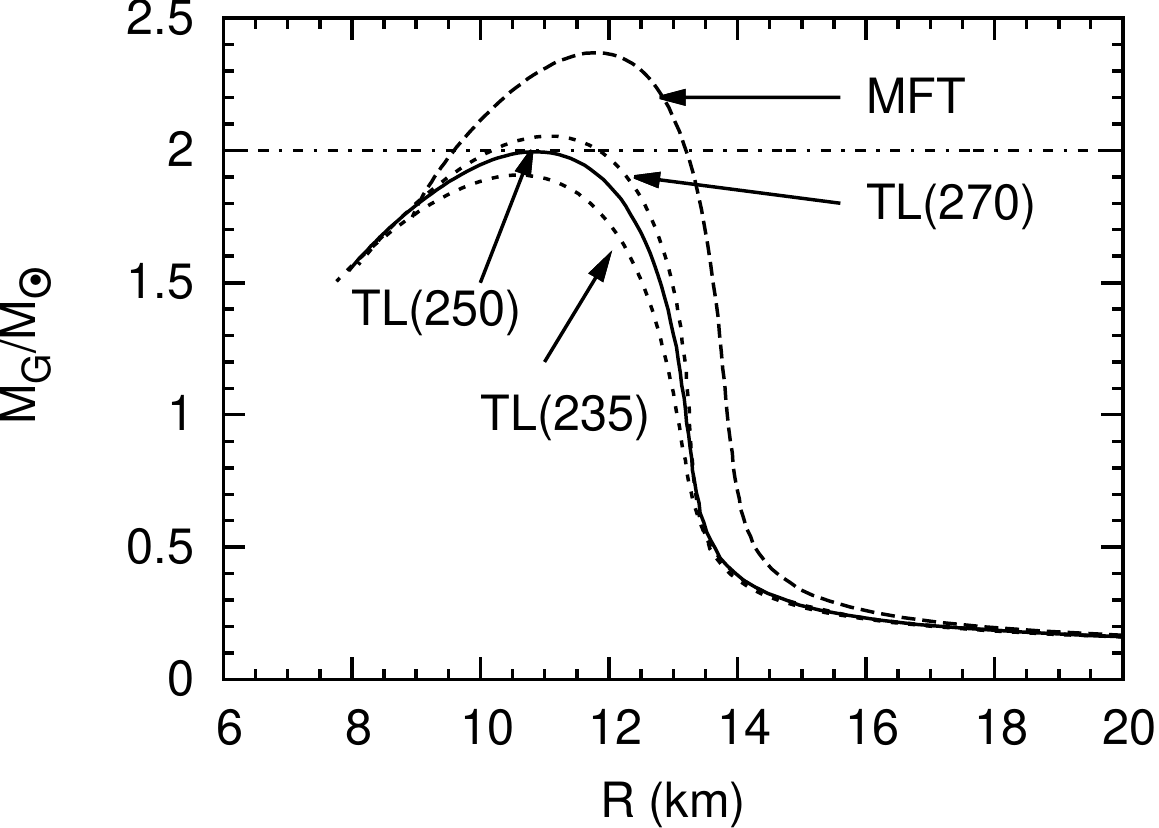}
     \caption{Neutron star mass-radius diagram for the MFT and TL calculations.} \label{fig:neutronstar}
\end{figure}

Constructing EOS's for charge neutral and beta-equilibrated matter for both MFT and TL cases,  mass-radius relations from solutions of the structure equations of Tolman-Oppenheimer-Volkoff  are shown in \fig{fig:neutronstar}. Results for the TL calculations are indicated by values of the compression modulus used (see Table~\ref{tab:two-loop_mine}) used to test  sensitivity. As the EOS's for the TL case are generally softer than that of MFT, lower maximum masses are obtained.  However, two-loop EOS's with compression moduli $\geq 250$ MeV are able to reach the value of $\sim 2M_\odot$. Because of the lower values of the symmetry energy stiffness parameter $L$ in the case of TL calculations, radii that are lower than their MFT counterparts are obtained. The values of radii for  $1.4M_\odot$ stars are significantly lower than that for MFT, and are in the range of values estimated from observations \cite{LattimerPrakash16}.


As mentioned in the introduction, in order to reconcile the nuclear EOS with measurements of collective flow such as the mean transverse momentum vs. rapidity, elliptic flow, {\em etc}., in 
medium energy heavy-ion collision experiments, 
the exchange potential needs to be included in the mean field experienced by nucleons~\cite{Prakash:1988zz,Welke:1988zz, Weber:1992qc,2015PhRvC..92b5801C}. 
A recent  discussion of this topic in the context of non-relativistic models can be found in Ref.~\cite{2015PhRvC..92b5801C}. 
Here we examine the nucleon optical potentials from MFT and TL calculations.  
In the context of a relativistic theory, the Schrodinger-equivalent optical potential, $V_\mathrm{opt}$, is the single particle potential that when used in the non-relativistic Schrodinger equation gives the same scattering results as originally computed in a relativistic theory \cite{Jaminon:1989wj, Weber:1992qc, Danielewicz:1999zn,Gaitanos:2012hg}.  Following Ref.~\cite{Jaminon:1989wj},  one can express the single particle spectrum  in Eq.~(\ref{eqn:especMFT}) for the MFT and Eq.~(\ref{eqn:singlepartcilespec}) for the TL as 
\begin{equation}
\epsilon(p)=\sqrt{p^2+\left(M-\gs\phib\right)^2} +\Sigma_v \left(p\right)\,.
\end{equation}
The optical potential is then expressed as 
\begin{eqnarray}
V_\mathrm{opt}\left(p\right)\equiv \Sigma_v\left(p\right) -\gs \phib +\frac{\left(\gs \phib \right)^2-\Sigma_v\left(p\right)^2}{2 M}+\frac{\Sigma_v\left(p\right)}{M} E_\mathrm{kin} \,, \label{eqn:Voptdef}
\end{eqnarray}
where $E_\mathrm{kin}\equiv \epsilon(p)-M$ is the asymptotic kinetic energy.

\begin{figure}
	\centering
		\includegraphics[angle=0,width=0.5\textwidth]{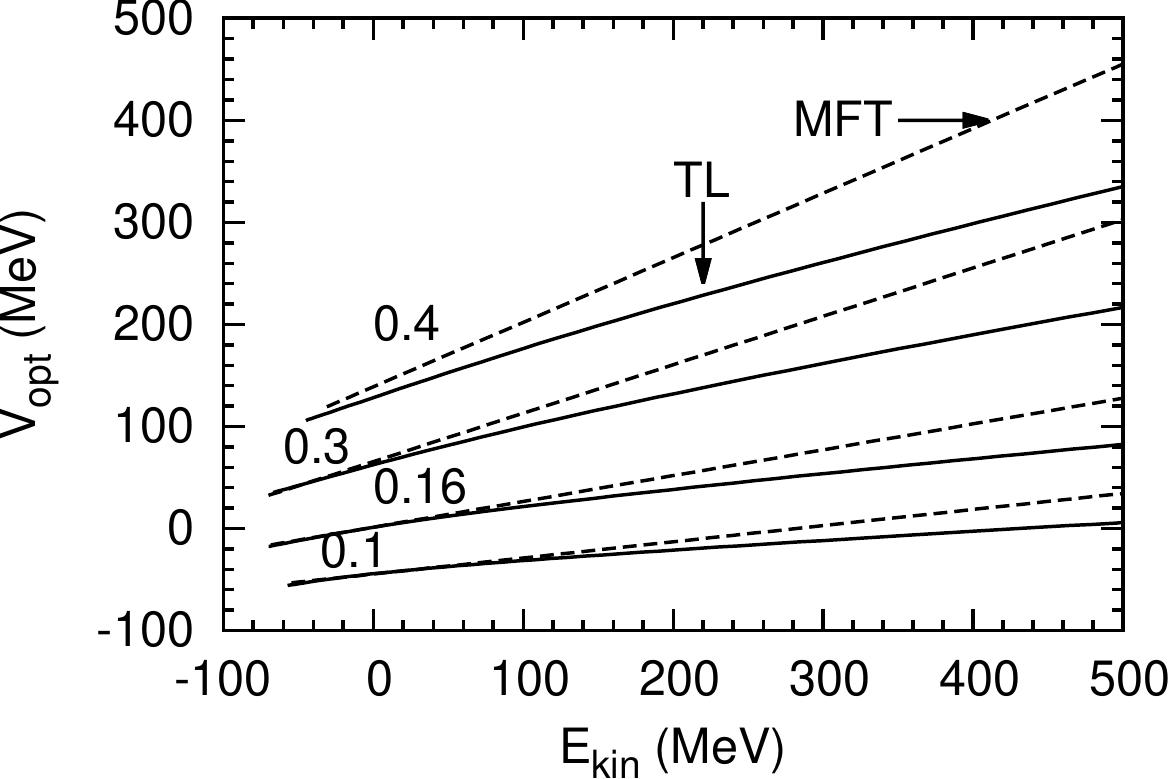}
     \caption{The nucleon's optical potential in SNM  at different densities. For each density, results of MFT lie above those of the TL calculations. For nuclear density, $0.3$, and $0.4\,\denunit$, the potentials are shifted by $50$, $100$, and $150$ MeV, respectively, for clarity. } 
     \label{fig:Vopt}
\end{figure}

In Fig.~\ref{fig:Vopt}, we show $V_\mathrm{opt}\left(p\left(E_\mathrm{kin}\right)\right)$ vs. $E_\mathrm{kin}$ in SNM  for $\rho_B =0.1,\, 0.16,\, 0.3,\,{\rm and}~ 0.4$ $\mathrm{fm}^{-3}$, respectively. For each density, the upper curve is the MFT result whereas the lower curve corresponds to TL calculations. 
Note that results for the three higher densities are shifted  by a constant amount shown in the figures's legend. 
For low $E_\mathrm{kin}$,  the MFT and TL results are close to each other. This trend is, however broken for increasing $E_{\rm kin}$  as in MFT,  $V_\mathrm{opt}$ increases linearly with $E_\mathrm{kin}$ [see Eq.~(\ref{eqn:Voptdef})], much faster than the TL result does. 
Previous microscopic many-body calculations, proton-nucleus scattering measurements, and  phenomenological extractions from heavy-ion collisions all point to a slower increase of $V_\mathrm{opt}$ than in MFT \cite{Danielewicz:1999zn}. Comparing with the results from Ref.~\cite{Danielewicz:1999zn}, the inclusion of  TL contributions makes the $V_\mathrm{opt}$ to agree better with the phenomenological values. Extrapolating the current  TL results to higher $E_\mathrm{kin}$ than shown may be questionable because point nucleon-meson couplings likely oversimplify the physical situation.

\begin{figure}
	\centering
		\includegraphics[angle=0,width=\textwidth]{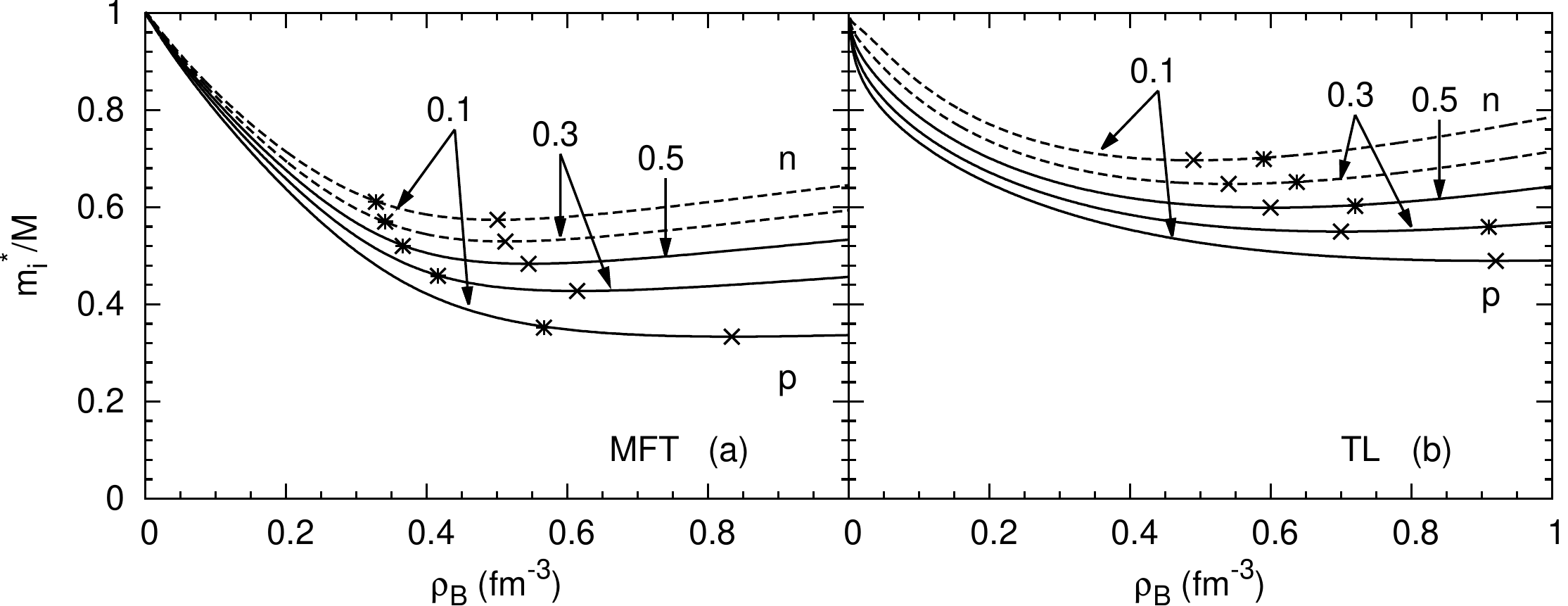}	
     \caption{The neutron and proton effective Landau masses $\mland$ vs. $\den{B}$ for different proton fractions.  
     {The crosses $\times$ indicate the densities at which the first derivatives  become zero, whereas the asterisks $\ast$ mark the densities at which the Dirac  effective mass $\mnstar$ equals either the neutron or the proton Fermi momentum. }} \label{fig:zeroT_mland_splitting_mine}
\end{figure}

The neutron and proton Landau effective masses are shown in  \fig{fig:zeroT_mland_splitting_mine} for different proton fractions $x$. For each $x$, the TL results are larger than those of MFT for the same $\rho_B$. The isospin splitting of the effective masses is such that those for neutron are always larger than for the proton as $x$ decreases from its SNM value of 0.5. This pattern is similar to those of microscopic non-relativistic models and many phenomenological models although exceptions exist in the latter category \cite{Constantinou2014}. For the isospin invariant nucleon-nucleon interactions used in the MFT and TL calculations here,   $\mland_{n}$ at $x$ is the same as  $\mland_{p}$ at $(1-x)$. Beyond the saturation density, the Landau mass for each $x$ exhibits a minimum marked by a cross $\times$ which is characteristic of field-theoretical calculations. For isospin invariant interactions in non-relativistic calculations, this feature would be absent~\cite{Constantinou2014}. To assess the importance of  relativistic kinematics, the density at which the effective Dirac mass $\mnstar$ becomes equal to the proton (neutron) Fermi momentum is marked by asterisk $\ast$ on the proton (neutron) curve. For MFT, the $\times$-density is larger than the $\ast$-density, i.e., the turnover of $\mland$  lies in the relativistic region, whereas for TL the situation is reversed (for $x=0.1$ the $\ast$-density is larger than $1\,\denunit$ and is not shown in the plot).

\begin{figure}
	\centering
		\includegraphics[angle=0,width=\textwidth]{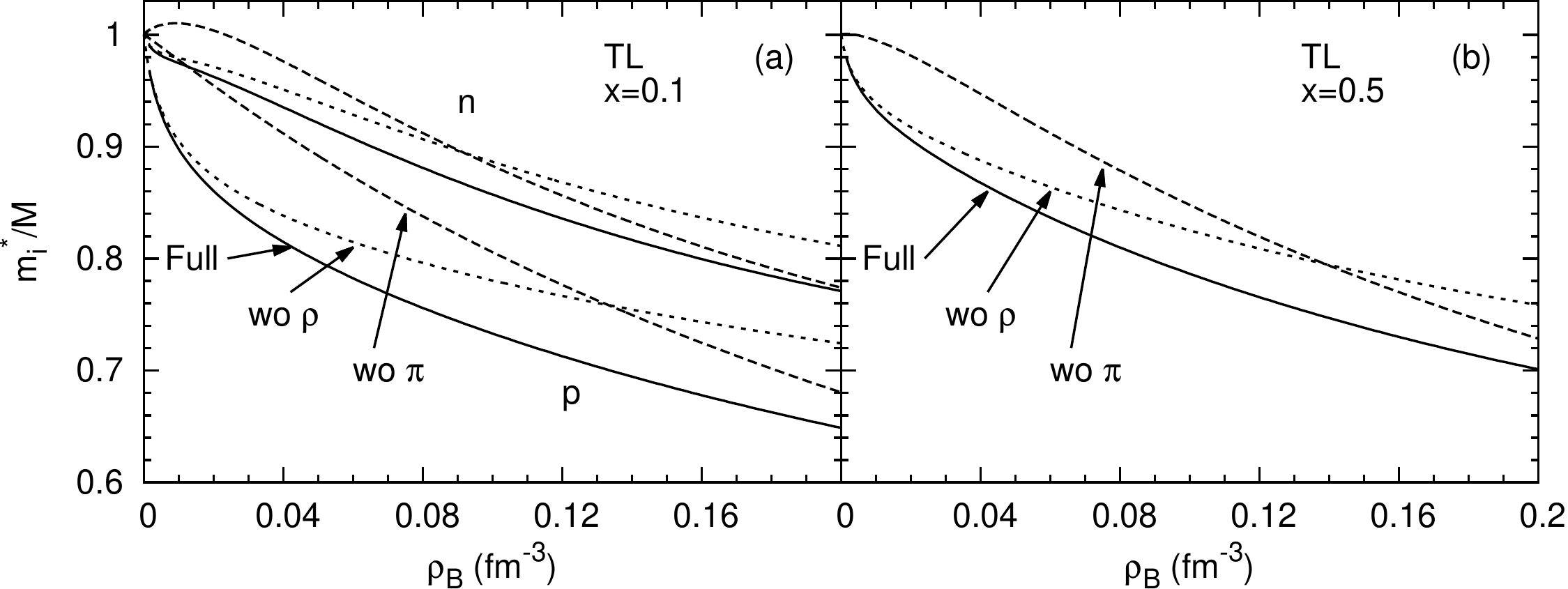}	
     \caption{The neutron and proton Landau effective masses $\mland$ vs. $\den{B}$  for different proton fractions. Here ``wo $\pi$'' and ``wo $\rho$'' are results without including $\pi$ and $\rho$ two-loop contributions in calculating the single particle spectra. } \label{fig:zeroT_mland_wopiworho}
\end{figure}

The behavior $\mland$'s  for the MFT and TL calculations merits examination as 
the differences between $\mland$'s in the TL results at low densities are larger than those in  MFT. The origin of these differences can be traced back to the role of pion exchange in the two-loop contribution.  
The small pion mass sets a distinct low-density scale in the TL results. To illustrate this feature, in  \fig{fig:zeroT_mland_wopiworho} we show the TL results of $\mland_i$  for $x=0.1$ and $0.5$ without including the $\pi$ contribution in the calculation of the single particle spectrum [\eq{eqn:singlespec}].  Results without the contribution from $\rho$ exchange is also shown for comparison. 
For the two proton fractions shown,  and for both the neutron and the proton, excluding the $\pi$ contribution leads to a significant change, i.e., $\partial \mland_i /\partial \den{B}$ is substantially modified, while excluding the $\rho$ contribution changes $\mland_i$ gradually with density. 
In the following section, we will see that $\mland_i$ play central roles in the finite temperature properties of the EOS.

\begin{figure}
	\centering
		\includegraphics[angle=0,width=\textwidth]{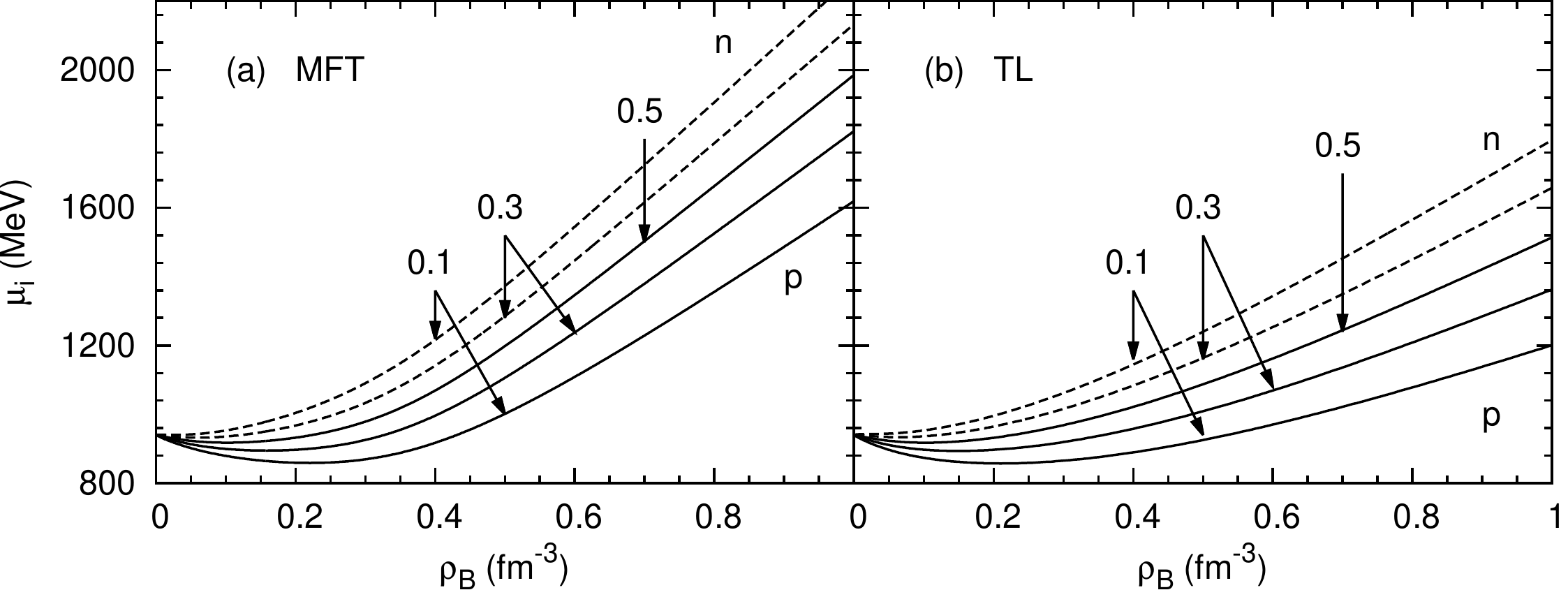}
     \caption{The zero temperature neutron and proton chemical potentials (inclusive of masses) vs density for different proton fractions.} 
     \label{fig:mu_zeroT}
\end{figure}

The neutron and proton chemical potentials for the MFT and TL calculations are shown in \fig{fig:mu_zeroT}.  The results in this figure serve to gauge the role of thermal effects on the chemical potentials (to be presented in subsequent sections).  
As $x$ decreases from its value of 0.5 in SNM, $\mup$ decreases relative to $\mun$ at a given density $\rho_B$. The $\mu_i$ for MFT are larger than those of  TL for  $\rho_B \geq 0.2\sim 0.3\,\denunit$, similar to the comparison in \fig{fig:zeroT_BE_mine}.  

\section{Results at finite temperature}
\label{sec:resultst}

\begin{figure}
	\centering
		\includegraphics[angle=0,width=0.5\textwidth]{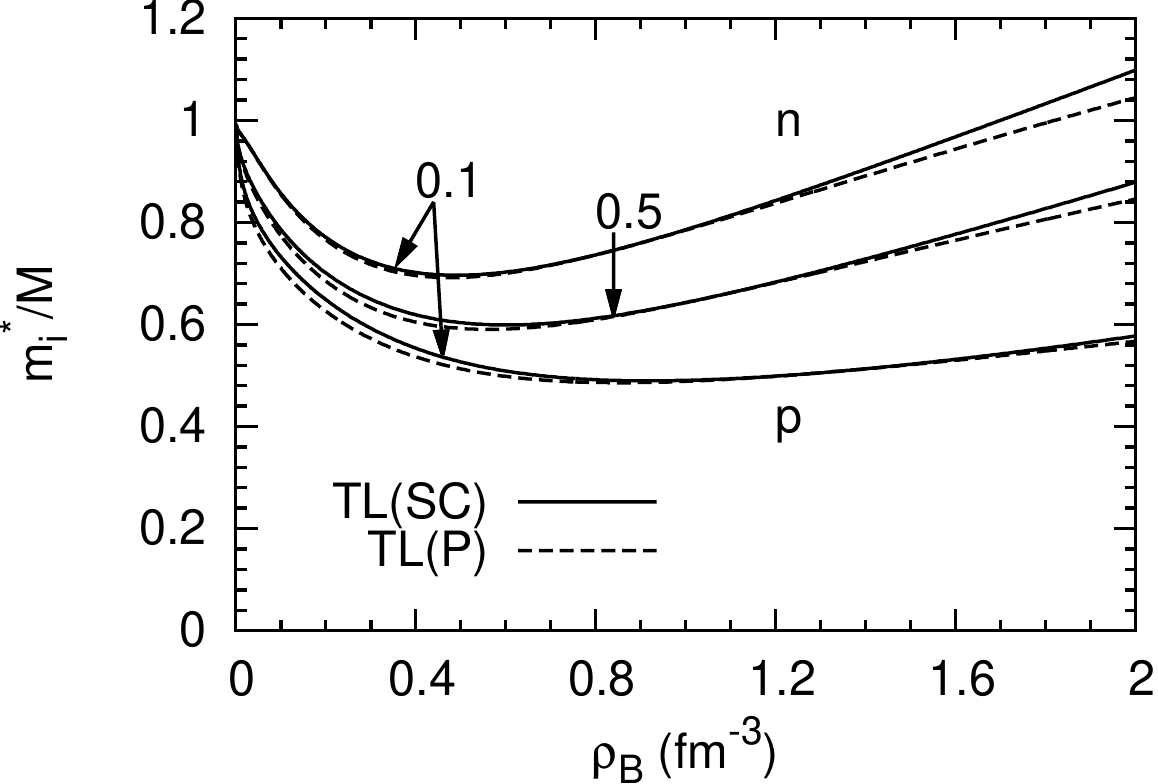}
     \caption{Comparison between the Landau masses $\mland_{p,n}$ (from a self-consistent (SC) calculation) and $m^{\ast '}_{p,n}$ (from the perturbative (P) approach) for proton fractions $x=0.1$ and $0.5$. } \label{fig:mlandSCvsP}
\end{figure}

In this section, we discuss the
thermal components of the state variables defined by  $Q_{\mathrm{th}}\equiv Q(\den{B},x,T)-Q(\den{B},x,T=0)$, where $Q$ stands for any of pressure, energy, chemical potential, {\it etc.}  The full TL and MFT calculations are compared with each other as well as to the results of Fermi-liquid theory (FLT) valid in the limiting situation of degenerate matter in which the temperature is much smaller than the Fermi energy. 
In this limit, the entropy is proportional to the temperature $T$, and the thermal pressure, energy and chemical potentials are proportional to $T^2$. The magnitudes of these state variables are controlled by  the Landau effective masses and their first density derivatives. 
The FLT working formulas are collected in Appendix~\ref{sec:degnondeglimits}. 
As shown in section~\ref{sec:SC}, the TL(SC) and TL(P) calculations yield different Landau effective masses, 
$\mland_{p,n}$ and $m^{\ast '}_{p,n} $. The difference between them is small (below $\approx 3\%$) for $0<\den{B}< 1\,\denunit$, as shown in Figure~\ref{fig:mlandSCvsP}. Discernible differences in the slopes occur only for  $\rho_B \geq 1.5\, \denunit$. 
For contrast, therefore, results of  the thermal properties for both TL calculations are presented below; those labeled TL(P) are based on the perturbative scheme in Sec.~\ref{subsec:finiteTpert}, whereas those termed TL(SC)  refer to the self-consistent approach in Sec.~\ref{sec:SC}.
In the low density region when the thermal de Broglie wavelength becomes much smaller than the inter-particle separation,  $2\pi\den{i}^{1/3}/\sqrt{3 M T} \ll 1$, the non-degenerate situation prevails and the  
single particle distribution is adequately given by the classical Maxwell distribution \cite{Prakash87}. The relevant formulas can be found in Appendix~\ref{sec:degnondeglimits}. 

\begin{figure}
	\centering
		\includegraphics[angle=0,width=\textwidth]{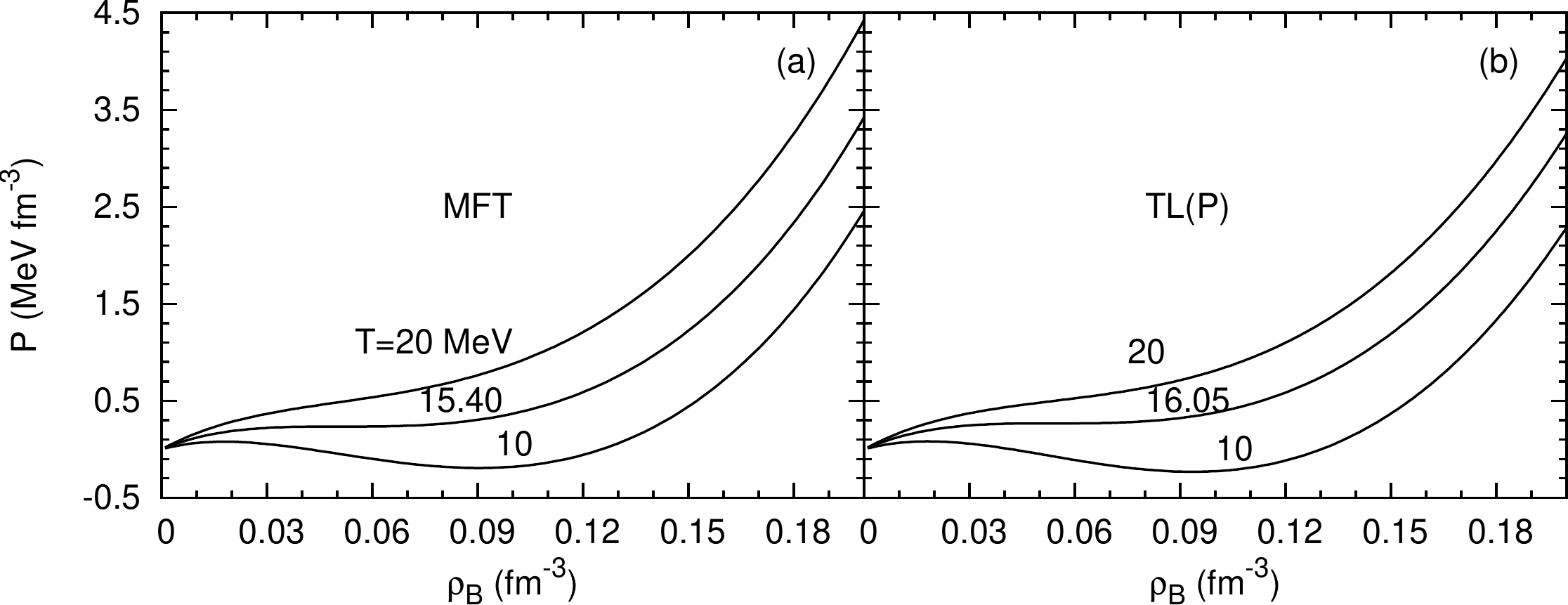}
     \caption{Pressure isotherms at low density for SNM. The middle curve in each panel corresponds to the temperature at which the liquid-gas phase transition occurs. } 
     \label{fig:gas-liquid_trans}
\end{figure}

In \fig{fig:gas-liquid_trans}, we show the total pressure $P$ vs. $\den{B}$ in SNM at select temperatures  for  MFT and TL(P) calculations. For the range of $\rho_B$ and $T$ of relevance here, the TL(SC) and TL(P) pressures closely  overlap as their thermal components are nearly the same (see Fig.~\ref{fig:Pth_T_2050MeV} below). The middle curve in each  panel corresponds to the case for which the relations
\begin{equation}
\left. \frac {dP}{d\rho_B} \right|_{\rho_c,T_c} = \left. \frac {d^2P}{d\rho_B^2} \right|_{\rho_c,T_c} =0 
\end{equation}
are satisfied indicating the occurrence of a liquid-gas phase transition. 
The critical temperature $T_c$, density $\den{c}$, and pressure $P_c$ are $16.05\,\mathrm{MeV}$, $0.055\,\denunit$ and $0.268\,\mathrm{MeV}\,\mathrm{fm}^{-3}$ 
for TL, and $15.40\,\mathrm{MeV}$, $0.051\,\denunit$, and $0.235\,\mathrm{MeV}\,\mathrm{fm}^{-3}$
for MFT, leading to  $P_c/\left(\den{c}\,T_c\right)=0.304$ and $0.299$
 for TL and MFT.  These ratios can be compared to the value $0.375$ for a Van der Wals EOS (see the discussion in Ref.~\cite{Constantinou2014}). Empirical estimates of $T_c$ from nuclear physics experiments lie in the range 15-20 MeV  \cite{Karnaukhov2008,Holt:2013fwa}. 

\begin{figure}
	\centering
		\includegraphics[angle=0,width=\textwidth]{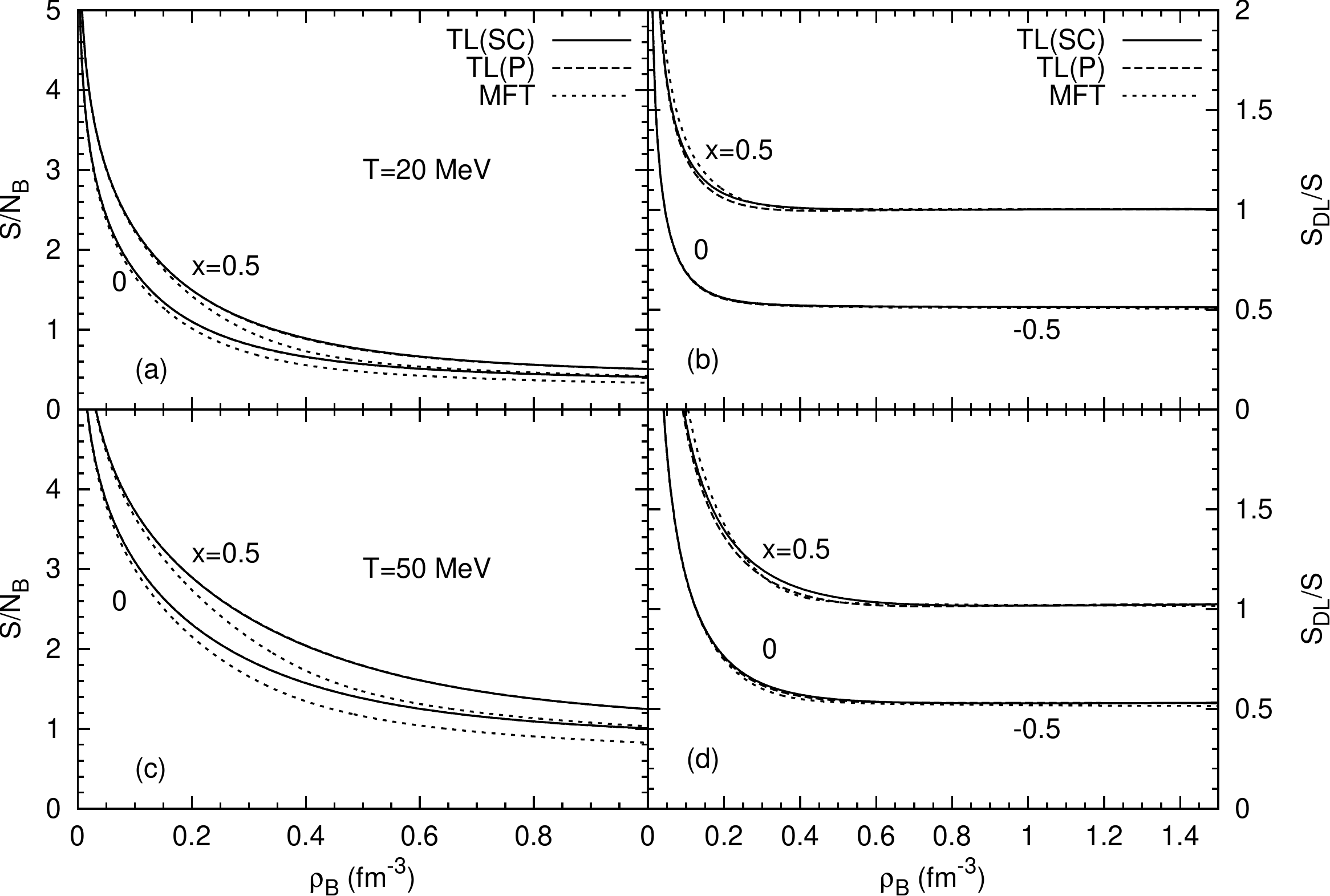}
     \caption{Entropy per nucleon at $T=20$ and $50$ MeV for proton fractions $x=0$ and $0.5$. The left panels compare the TL(P,SC) and MFT results. The right panel shows ratios between the degenerate limits (DL) and the corresponding full results. The ratios for PNM are shifted by -0.5.} \label{fig:S_T_2050MeV}
\end{figure}

Figures \ref{fig:S_T_2050MeV} (a) and (c) show the entropy per baryon at temperatures of $T=20$ and 50 MeV, and for proton fractions $x=0$ (PNM) and $0.5$ (SNM) from MFT and TL calculations.  At sub-nuclear densities, the three calculation results approach their corresponding  classical limits [\eq{eqn:classicalgaslimited}] detailed in Appendix~\ref{sec:degnondeglimits}.  In Ref.~\cite{Constantinou2014},  the MFT results at low density using the same QHD Lagrangian was shown to agree with the non-degenerate limit results including the first order correction from the fugacity expansion. 
In the high density region, the FLT result $S/N_B \propto T\mland_{i}/k_{F,i}^2$ helps to understand the behaviors with respect to $T$ and $\rho_B$. The TL(P,SC) results are about $20$-$30\%$ larger than those of MFT for both proton fractions reflecting the similar behaviors of  the Landau effective masses $\mland_{i}$ shown in \fig{fig:zeroT_mland_splitting_mine} ($m^{\ast '}_{p,n}$ are close to $\mland_{p,n}$ cf. Figure~\ref{fig:mlandSCvsP}). In the density region shown, $0<\den{B}<1\,\denunit$, the TL(P) and TL(SC) curves differ by less than $1\%$ (too small to be seen in the figures), a larger difference occurring for 
$\den{B}>1.2\,\denunit$.
The right panels (b) and (d) show the ratios between the degenerate limit entropy and the full result for each case. Densities below which these ratios differ significantly from unity mark the onset of the semi-degenerate regions before matter enters the non-degenerate regions. 
For all cases shown, the full results  approach their corresponding degenerate limits above 0.4-0.5$\,\denunit$ for $T=20$ MeV. 
 As expected, for results of $T=50$ MeV the densities beyond which the degenerate limit applies,  $\sim 1\,\denunit$, are much larger than for $T=20$ MeV. 
Degenerate limit expressions beyond the leading order FLT results derived in Ref. \cite{cons15} may be used to extend the ranges of partial degeneracy for which an analytical treatment remains valid.

In Figs. \ref{fig:Pth_T_2050MeV}, we show  $P_{\mathrm{th}}$ vs. $\den{B}$ for SNM and PNM at $T=20$ and $50$ MeV, respectively. All results converge to their classical limits [\eq{eqn:classicalgaslimitp}] in the low density region, and also approach their degenerate limits for $\den{B}$ exceeding 0.4-0.5$\, \denunit$ for $T=20$ MeV and  $0.8-0.9\,\denunit$ for $T=50$ MeV. The TL(P,SC) results are systematically larger than the MFT ones at high density owing to differences in their $\mland_{p,n}$'s shown in \fig{fig:zeroT_mland_splitting_mine}. 
Differences between the TL(P) and TL(SC) results at $T=20$ MeV are negligible for $\den{B} < 0.5\,\denunit$, and at most  
$\approx 6\% $ for $ 0.5 < \den{B} < 1 \, \denunit$.  
At $T<20$ MeV, the differences in the degenerate region would be similar to that at $T=20$ MeV 
reflecting the differences in their respective Landau masses in  Fig.~\ref{fig:mlandSCvsP}. 
At $T=50$ MeV, the two results differ by less than $1\%$ for $ 0 < \den{B} < 1 \, \denunit$.  
For both temperatures differences in the pressure are larger than those for the entropy 
because of its dependence on the derivatives of the Landau masses.
Although the qualitative behavior of the degenerate limit results at $T=50$ MeV are similar to the full results, quantitative differences persist unlike at $T=20$ MeV for which the quantitative agreement is better. It is worth pointing out that the TL results do not exhibit the pronounced maximum in the thermal pressure as do those of MFT, but are similar to results of  non-relativistic calculations ({\it e.g}., Ref.~\cite{Constantinou2014,cons15}).

\begin{figure}
	\centering
		\includegraphics[angle=0,width=\textwidth]{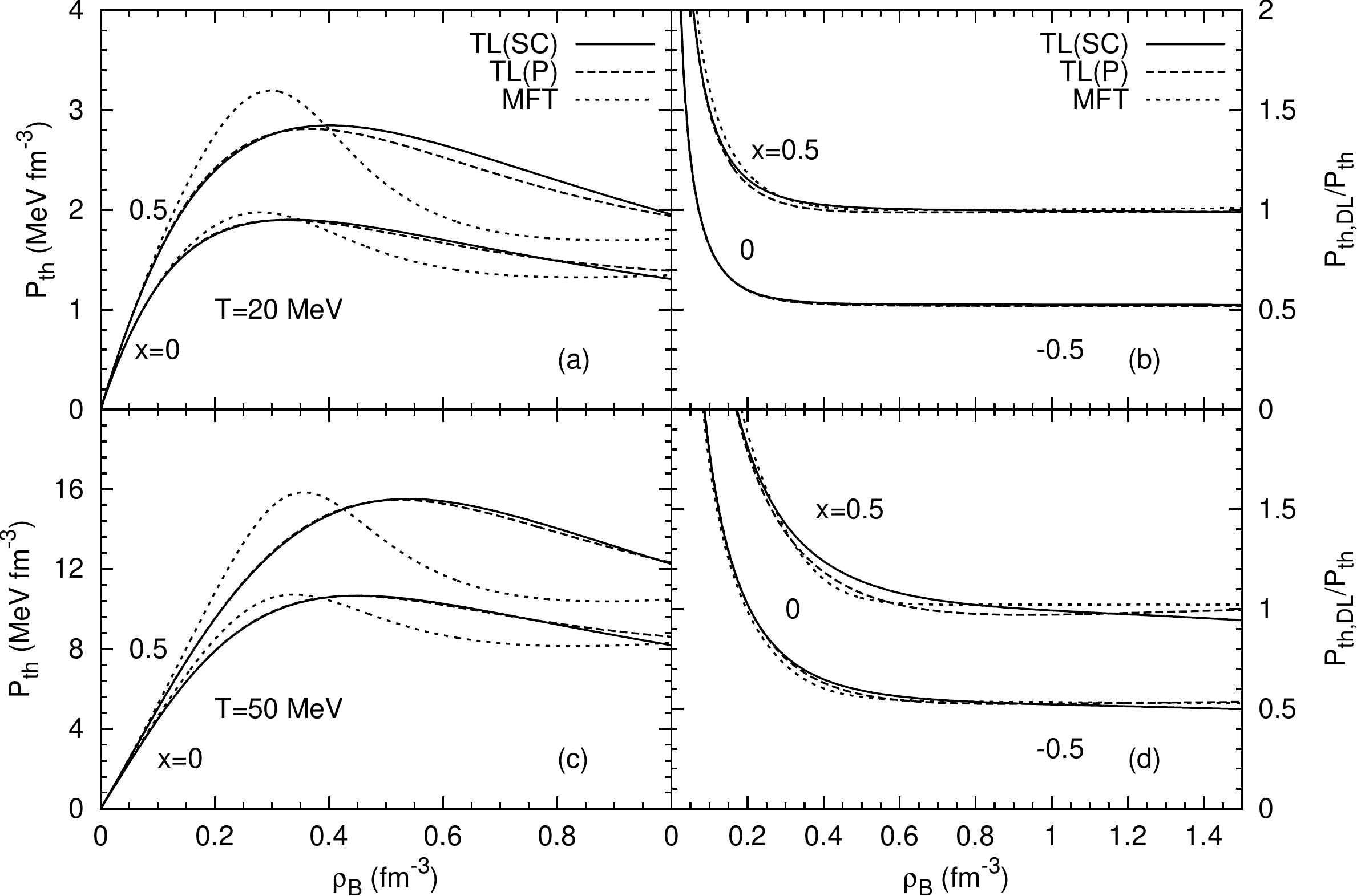}
     \caption{ Thermal pressure at $T=20$ and $50$ MeV for proton fractions $x=0$ and $0.5$. The left panels compare the TL(P,SC) and MFT results. The right panel shows ratios between the results of the full calculations and their degenerate-limit values. The ratios for PNM are  shifted by -$0.5$. } \label{fig:Pth_T_2050MeV}
\end{figure}

\begin{figure}
	\centering
		\includegraphics[angle=0,width=\textwidth]{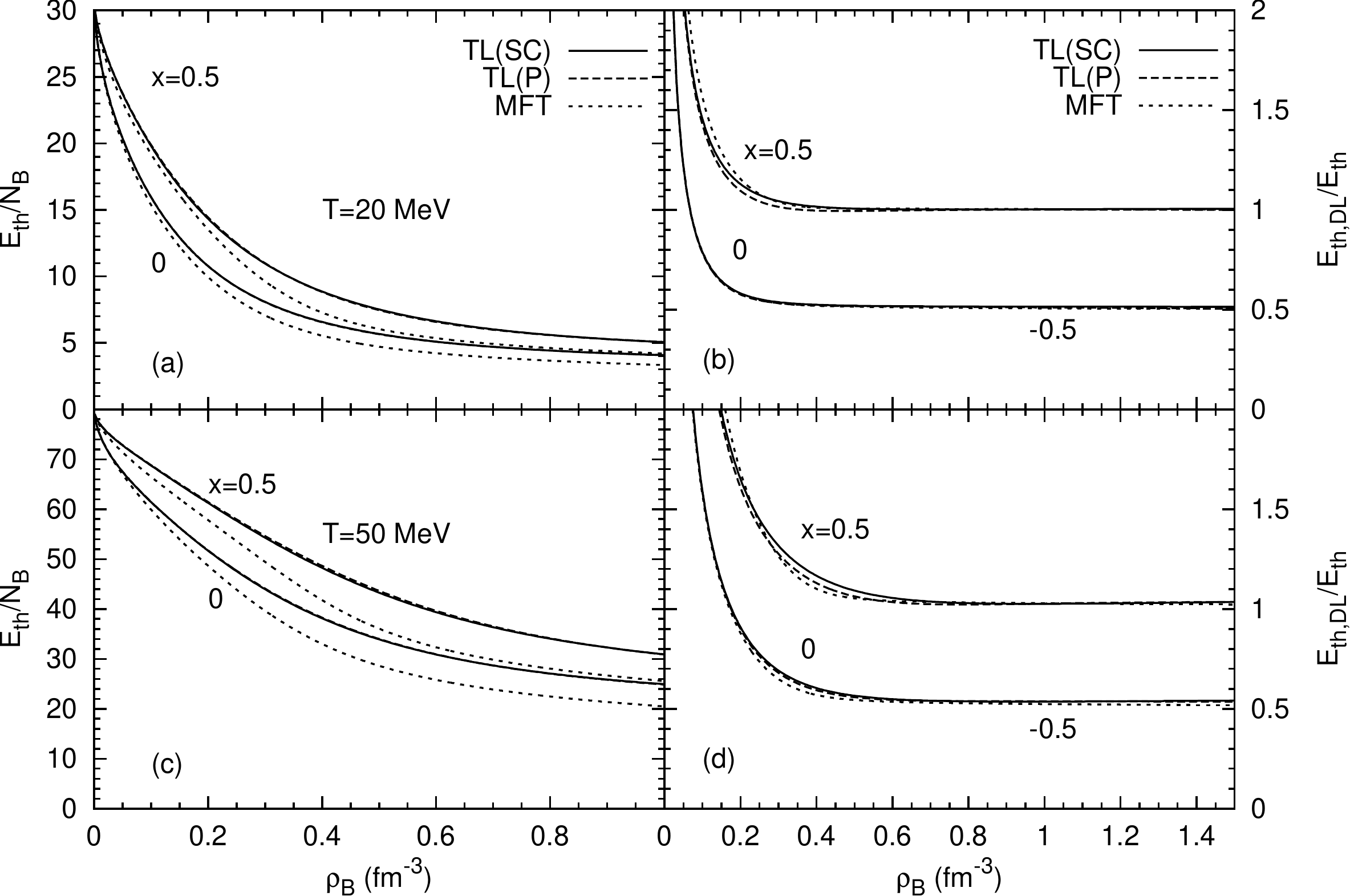}
     \caption{ Thermal energy per nucleon at $T=20$ and $50$ MeV for proton fractions $x=0$ and $0.5$. The left panels compare the TL(P,SC) and MFT results. The right panel shows ratios between the results of the full calculations and their degenerate-limit values. The ratios for  PNM are shifted by -$0.5$. } \label{fig:Eth_T_2050MeV}
\end{figure}

The thermal energies per nucleon $E_{\rm th}/N_B$ at $T=20$ and $50$ MeV are shown in Fig.~\ref{fig:Eth_T_2050MeV}.  In the high density and degenerate limit, the density dependence is the same as for the entropy per baryon, but the temperature dependence is quadratic.  For both SNM and PNM, the relation $(S/N_B)^2 = 4 aE_{\rm th}$ holds, where $a$ is the level density parameter.  Whereas convergence to the low-density classical limit (\eq{eqn:classicalgaslimited}) at both temperatures is good for MFT,  the TL results are influenced by contributions from pion exchange which are required to achieve a similar convergence.   In a different calculation without iso-vector meson exchange (not shown here), the convergence is, however, improved.  For the thermal energy, relativistic corrections (proportional to $T/M^*$) to the classical result of   $1.5\, T$ improve agreement with the full results. Such corrections at the leading $T/M^*$ order are also present for the entropy per baryon (in this case, however, the  leading $\ln\left[\frac{\den{}}{2}\left(\frac{2\pi}{MT}\right)^{3/2}\right] $ term dominates, {\it c.f.} Appendix~\ref{sec:degnondeglimits}).  The small differences between the TL(P) and TL(SC) thermal energies evident in the degenerate region share the same pattern as that for the entropy because of the linear dependence on the Landau masses.

\begin{figure}
	\centering
		\includegraphics[angle=0,width=\textwidth]{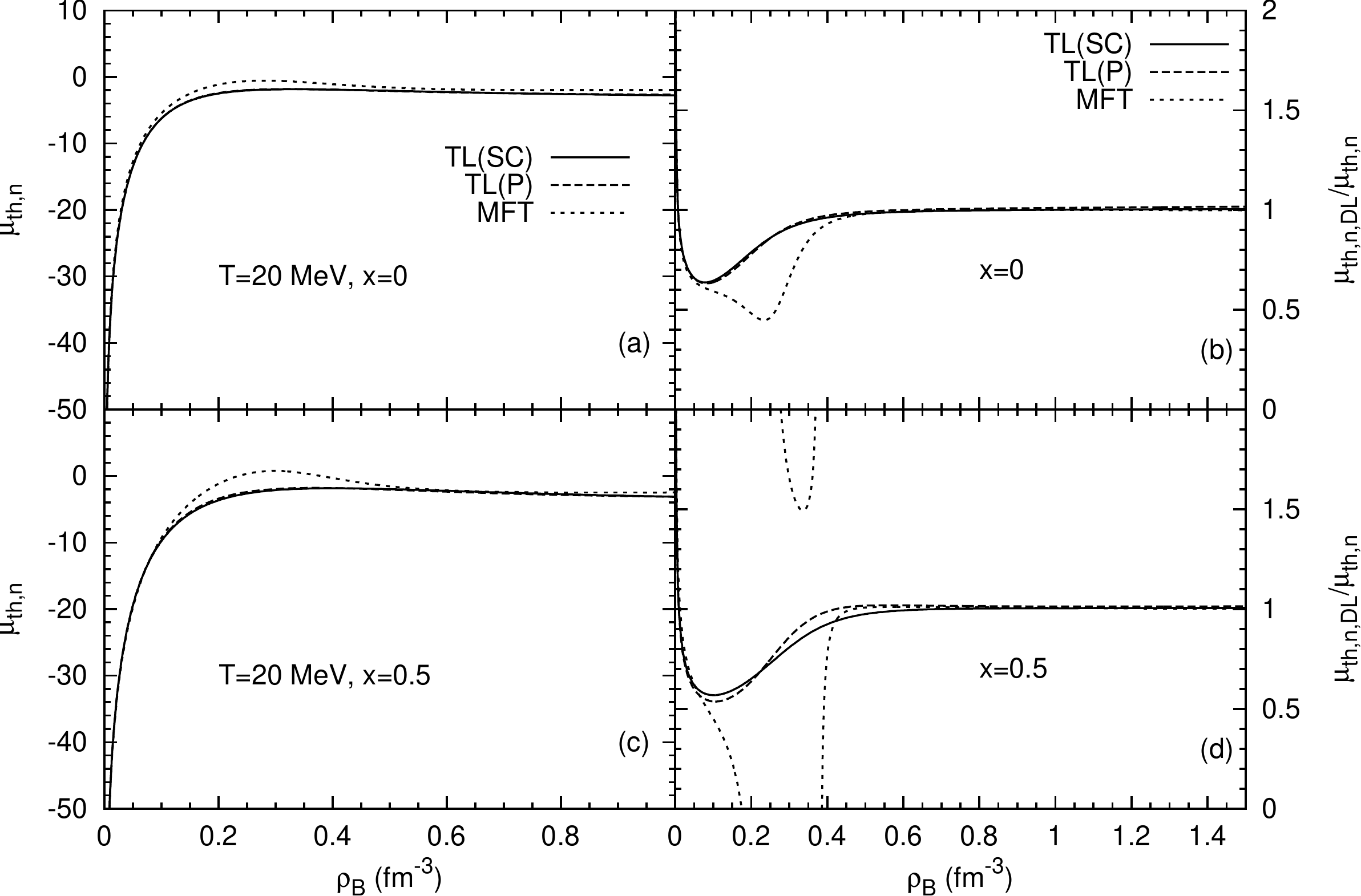}
     \caption{(a) Thermal components of the neutron chemical potentials at $T=20$ MeV at the indicated proton fractions. (b) Ratios of results from full and degenerate-limit calculations.} \label{fig:Muth_T_20MeV}
\end{figure}

\begin{figure}
	\centering
		\includegraphics[angle=0,width=\textwidth]{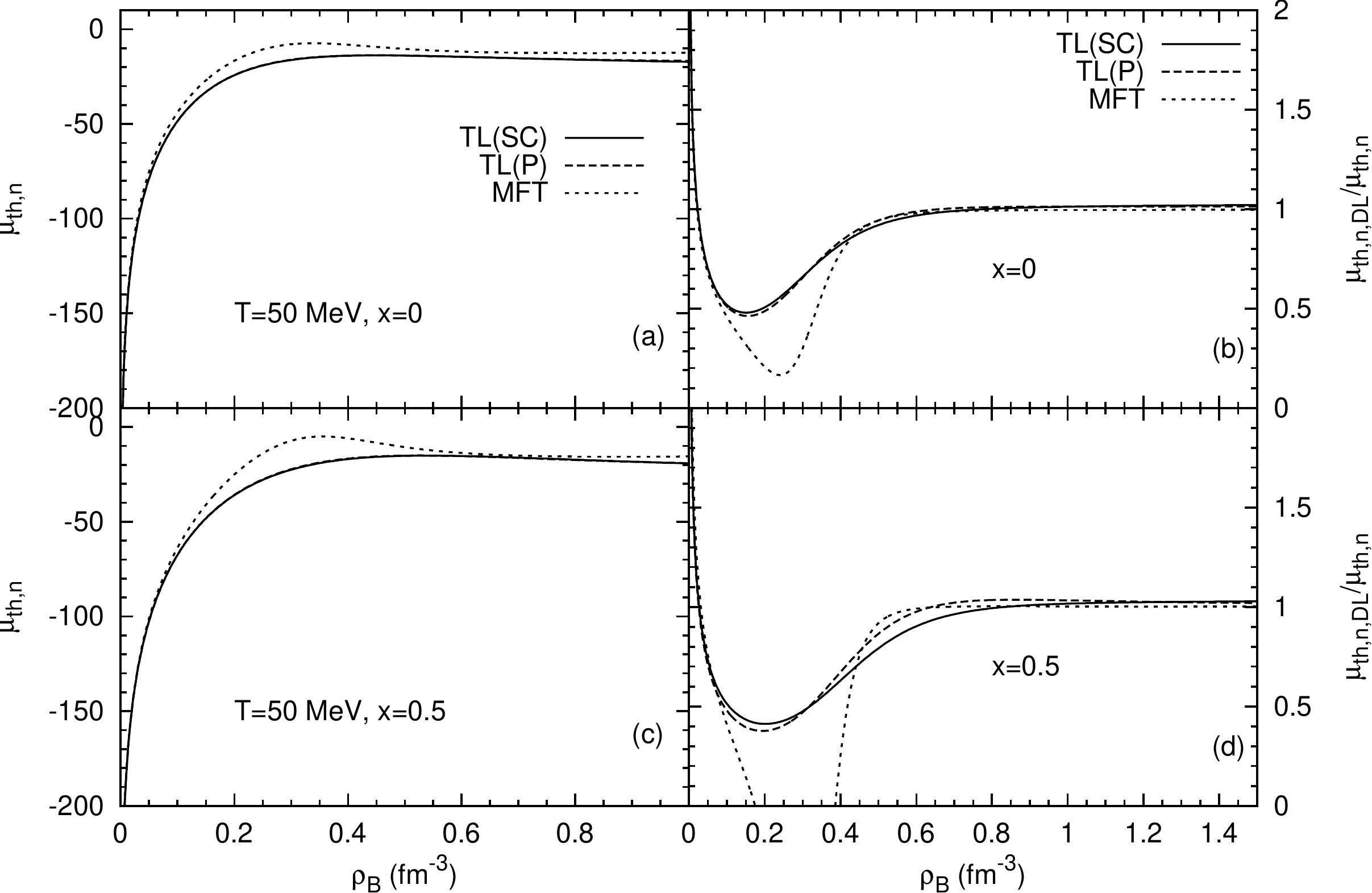}
     \caption{Same as  \fig{fig:Muth_T_20MeV}, but for $T=50$ MeV. } \label{fig:Muth_T_50MeV}
\end{figure}

The thermal components of the neutron chemical potential are shown in Figs.  \ref{fig:Muth_T_20MeV} and \ref{fig:Muth_T_50MeV} from MFT and TL calculations. In the low density region, the results approach to the classical gas limit shown in \eq{eqn:classicalgaslimitmu}. At high density, for both SNM and PNM and temperatures, the TL curves are above the MFT curves, in contrast to the comparisons for the other state variables due to the overall negative sign in Eq.~(\ref{mudeg})  (note also that in 3-dimensions, the chemical potential is always less than the Fermi energy). Differences between the full and degenerate or non-degenerate limit results occur at near nuclear densities for which matter is in the semi-degenerate regime for which an analytical treatment is not possible. This region is indicated by the curves that go off scale in this figure because the exact results approach zero there.    The differences between the TL(P) and TL(SC) chemical potentials resemble those observed for the pressure, because $\mu_p \den{p}+\mu_n \den{n} = \ed - T \sd + P$, which is dominated by the pressure term.


\begin{figure}
	\centering
		\includegraphics[angle=0,width=\textwidth]{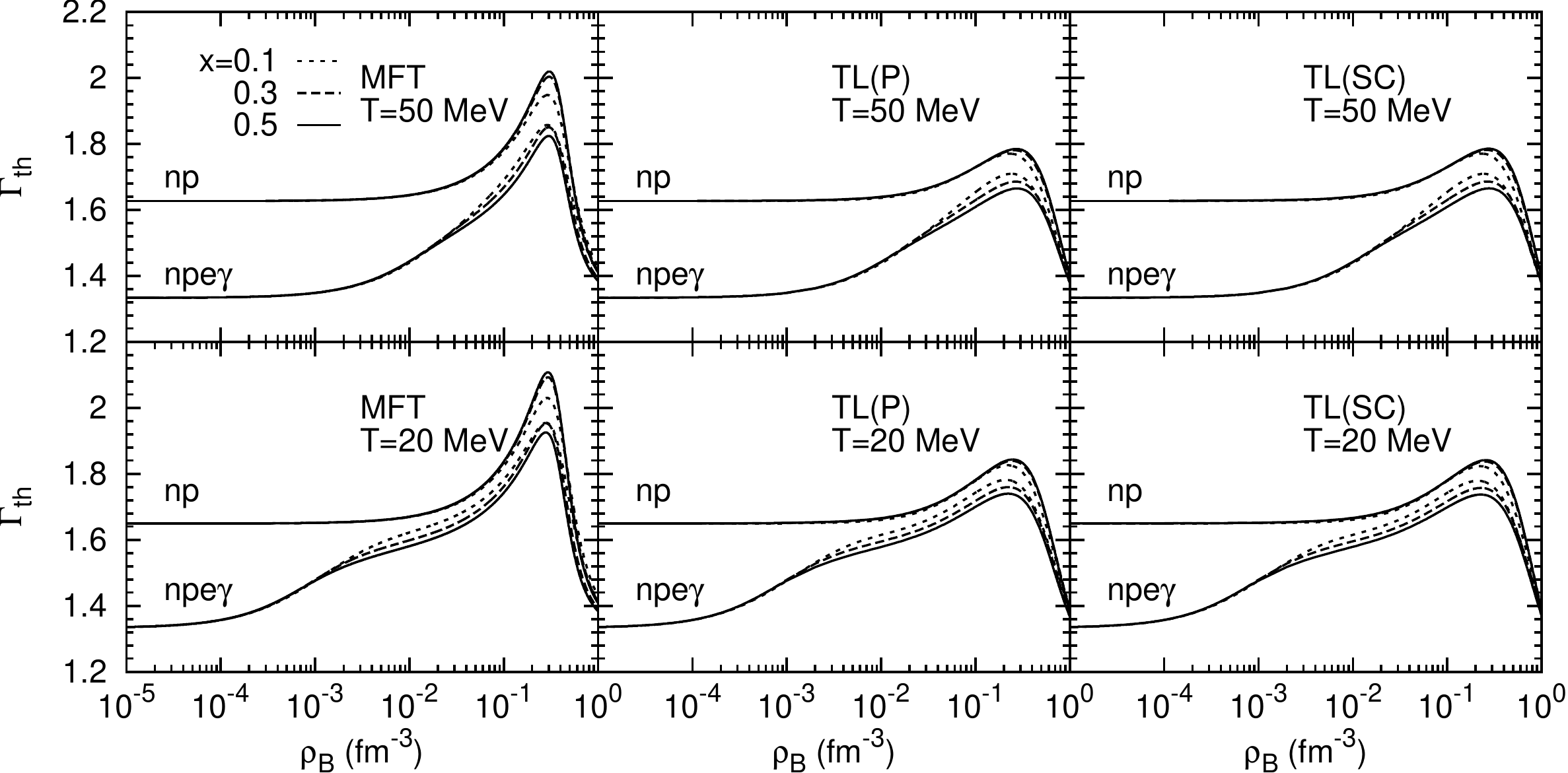}
     \caption{The thermal index $\Gamma_{th}$ at T=$20$ and $50$ MeV, and  proton fractions $0.1$, $0.3$, and $0.5$. The panels from left to right show results of MFT, TL(P), and TL(SC) calculations. The upper curves are for matter  with nucleons only, whereas the lower curves are for nucleons with leptons and photons.} \label{fig:GammathT_20_50MeV}
\end{figure}

In recent astrophysical simulations of core-collapse supernovae, evolution of proto-neutron stars and mergers of compact binaries, 
the thermal index 
\begin{eqnarray}
\Gamma_{\mathrm{th}} \equiv 1+ \frac{P_{\mathrm{th}}}{\ed_{\mathrm{th}}} 
\end{eqnarray}
has been employed to capture thermal effects as  functions of baryon density and lepton fraction $Y_{Le}$  
\cite{Janka1993,Bauswein2010,Hotokezaka2013,Hotokezaka2013,Kaplan2014}). 
In \fig{fig:GammathT_20_50MeV}, results of $\Gamma_{\mathrm{th}}$ are shown for the MFT and TL calculations for matter with only nucleons, and with nucleons, leptons and photons.  
Ideal gas contributions from leptons are included as in Ref.~\cite{Constantinou2014}. Two-loop contributions arising from photon exchange are justifiably neglected owing to the smallness of the fine structure constant~\cite{Kapusta:2006pm}.    
The results in this figure prompt the following observations: 

(1) In the low density region for nucleons only matter,  $\Gamma_{\mathrm{th}}\rightarrow 5/3$ for both TL and MFT calculations for all proton fractions, characteristic of classical non-relativistic gases; 

(2) In the low density region with contributions from leptons and photons,  $\Gamma_{\mathrm{th}}\rightarrow 4/3$ because of the dominant contributions from relativistic leptons;

(3) Around $1\,\denunit$, both MFT and TL with and without leptons yield $\Gamma_{\mathrm{th}} \sim 1.4$.  At higher densities, $\Gamma_{\mathrm{th}}\rightarrow 4/3$ for MFT, because the electron mass and nucleon $M^\ast$'s lose their significance relative to their Fermi momenta; the associated Landau masses $m_i^*$ become proportional to their Fermi momenta leading to $P_{\mathrm{th}}/\ed_{\mathrm{th}}\rightarrow 1/3$.  The density dependence of the nucleon $\mland_i$  in the TL case is more complicated than in MFT and the approach to the asymptotic value of $\Gamma_{\mathrm{th}}$ is postponed to much higher densities than for MFT.

(4) The maximum values of $\Gamma_{\mathrm{th}}$ attained in TL calculations, 1.7-1.8 (for $T=20$ MeV) and 1.6-1.7 (for $T=50$ MeV), are significantly smaller than those in MFT, 1.9-2.1 (for $T=20$ MeV) 1.9-2 (for $T=50$ MeV). 
The dependence on $Y_e$ is weak, but that on $\rho_B$ is more pronounced. In this respect, the TL results here resemble those
of non-relativistic treatments in which finite-range interactions are employed \cite{cons15}. 

\section{Summary and Conclusion}
\label{sec:sumup}

In this paper,  we have studied hot and dense nucleonic matter in the EFT framework of QHD beyond the mean field approximation by including contributions from two-loop diagrams. Based on the same QHD Lagrangian, results of  MFT and TL approximations 
for conditions of relevance to the study of core-collapse supernovae, neutron stars and mergers of compact binaries were calculated and compared. The Lagrangian employed is the same as in Ref. \cite{Lalazissis:1996rd} (termed NL3 in the literature), but with the addition of pseudo-vector nucleon-pion coupling which preserves chiral symmetry. Results at the MFT level 
were checked to be the same as those of NL3 in the literature. The TL calculations add significant density-dependent contributions to the Hartree terms of MFT  from the exchange of iso-scalar, iso-vector and pseudo-scalar mesons.
For both approximations, the same set of five density-independent nucleon-meson couplings  were used. In each case,  the coupling strengths 
were determined utilizing the empirical properties of zero-temperature bulk matter, namely, the binding energy, Landau mass, incompressibility, and symmetry energy all at the equilibrium density of isospin-symmetric nuclear matter.  Our principal findings are summarized below. 

The decrease of the TL Dirac effective mass, $\mnstar$, with density is much slower than in the MFT case. 
This feature, in conjunction with relativistic effects, renders the Landau masses, $m_{n,p}^*$, to also decrease more slowly than in MFT. The overall behavior of   $m_{n,p}^*$ vs $\rho_B$
is qualitatively similar to that of non-relativistic models which consider exchange contributions  from finite-range interactions (see, e.g., Ref.~\cite{Constantinou2014}). 
At $T=0$ in MFT, the $16$ MeV binding energy of SNM results from the competition between the scalar (attraction) and vector (repulsion) meson exchanges, but in the TL calculation part of the repulsion ( $\sim 10$ MeV) comes from the exchange diagrams. 
To achieve saturation at the empirical values, the coupling strengths, 
particularly  $\gs^2$ and $\gv^2$,  exhibit a distinct pattern: $\gs^2 < \gv ^2  $ in MFT, but $\gs^2 > \gv ^2$ in TL; moreover, their magnitudes in the TL case  are about half of those in the MFT. 
The $T=0$ EOS for TL is significantly softer than that of MFT: for SNM,  the difference starts around $\sim 0.4\,\mathrm{fm}^3$, whereas  for PNM,  the dissimilarity begins at a lower density $\sim 0.2\,\mathrm{fm}^3$. 
It is also noteworthy that with minor adjustments of the coupling strengths, the TL energies  in PNM at low and near-nuclear densities agree with those of modern non-relativistic QMC and EFT calculations. The contribution of quartic terms in the neutron-proton asymmetry parameter $x$ to the difference of PNM and SNM energies was found to be small, that in TL calculations being much smaller than in MFT. 

The neutron-star matter EOS's of both MFT and TL calculations  support a 2$M_\odot$ star required by recent precise determinations. In the TL case, a slightly stiffer than the nominal case EOS we studied in detail had better success.   The symmetry energy stiffness parameter $L$ in TL calculations ($\sim$ 83 MeV) is  smaller than that in MFT 
($\sim 103$ MeV) by about 20 MeV. Consequently, TL calculations yield smaller (by about 1 km) neutron star radii than those in MFT. We have verified that even smaller values of $L$ (in the suggested range in Ref.~\cite{LattimerPrakash16}), and hence smaller neutron star radii can be obtained with additional scalar-iso-vector couplings, but at the expense of more involved TL calculations.  Work is in progress to find a minimal set of density-independent couplings 
that yield radii in the range of 11-13 km for  $1.4M_\odot$ stars as indicated by analyses of astrophysical observations  \cite{LattimerPrakash16}.

The density-dependent TL single-particle potentials differ substantially from those of MFT. As is well known, the Schrodinger-equivalent optical potential, $V_\mathrm{opt}(E_\mathrm{kin})$, of MFT increases linearly with $E_\mathrm{kin}$, and is in disagreement with those obtained from analyses of  proton-nucleus scattering and heavy-ion experiments.
The TL results for  $V_\mathrm{opt}$ increase more slowly with energy than those of MFT, and are in  better agreement with extractions from measurements.  The qualitative behaviors with density and energy are similar to those of non-relativistic models in which exchange contributions from finite-range interactions are considered \cite{Constantinou2014}.  The TL calculations of  $V_\mathrm{opt}$ offer a contrast to other modifications of MFT in which either density-dependent couplings and/or momentum cut-off procedures which introduce additional functions and/or parameters are employed~\cite{Gaitanos:2012hg}.

The Landau effective mass plays a key role in low-temperature thermodynamics. 
The MFT and TL  calculations of $\mland_{n,p}$ share a common feature, viz, at a given density $\rho_B$, both 
$m_n^*$ and $m_p^*$ increase with the proton fraction $x$ in matter. 
However,  the isospin splitting at low density in the TL calculation is much larger than in  MFT. 
This difference is caused by pion exchange in the TL calculation, which is absent in MFT.  Moreover,  for given $x$ and  $\rho_B$, the TL $\mland_{n,p}$ are systematically larger than those of MFT.   

Contrasting the thermal properties of nucleonic matter for MFT and TL calculations was another goal of this study. On the formalism level, we demonstrated that our TL finite temperature results recover the zero temperature limit (not a trivial task) and  satisfy thermodynamic consistency. Thermal effects calculated from a self-consistent approach were contrasted with those from a perturbative approach. Except at the highest densities (approaching $\sim 2~{\rm fm^{-3}}$), results from these two methods were found to be consistent with each other. A study of the liquid-gas phase transition (without Coulomb interactions) yielded the critical temperatures  $T_c=16.05~(15.40)$ MeV for the TL (MFT) calculations in agreement with accepted values.  The entropy per particle, thermal components of pressure, energy, and chemical potential were calculated  for $T=20$ and $50$ MeV, and $x=0.1$ and $0.5$ to gauge the range of variation in astrophysical settings.  The results agree with the non-degenerate limits at  low density and high temperature.  At high densities and low temperature for which degenerate conditions prevail, results of FLT reproduce the exact results. Comparisons with the limiting situations revealed the density and temperature ranges for which results of exact calculations are needed.

The thermal index, $\Gamma_{\mathrm{th}}$, increasingly being used in astrophysics simulations, was also computed for 
representative values of $T$ and $x$ with and without  contributions from leptons (electrons and positrons) and photons.  
$\Gamma_{\mathrm{th}}$ varies weakly with $T$ and $x$, but more significantly with $\rho_B$.  
MFT yields a prominent peak at supra-nuclear density with $\Gamma_{\mathrm{th}}$ exceeding (close to) $2$ at $T=20~(50)$ MeV without leptons, whereas in the TL calculations the peaks are suppressed to  values much below 2.   
Contributions from leptons and photons (the latter significant only at high $T$'s) further decrease $\Gamma_{th}$ at all $\rho_B$.   
For  sub-nuclear densities ($<<0.1\, \mathrm{fm}^3$), results of both MFT and TL  approach the non-degenerate limit of $\Gamma_{\mathrm{th}}=5/3$, but the presence of leptons and photons forces $\Gamma_{\mathrm{th}}$ toward 4/3 characteristic of relativistic particles.  At high density, the MFT results can be easily understood as  $\mland_i \rightarrow k_{F,i}$; the dominance of relativity  (with or without leptons)   results in $\Gamma_{\mathrm{th}} \rightarrow 4/3$. 
The $\mland_i$ and $m_i^{\ast '}$ in TL calculations have a more complicated density dependence in the region of interest here.

In conclusion, the TL results are distinctly different from their MFT counterparts for the EOS, the single-particle optical potential, and thermal properties. Based on the  NL3 Lagrangian  with the inclusion of an iso-vector pion-nucleon coupling, the meson exchange diagrams improve results of the single-particle optical potential to be in agreement with extractions from data.  Although the EOS of TL is much softer than that of MFT, results  consistent with current neutron star data are obtained. The TL thermal properties differ significantly from those of MFT, and are in semi-quantitative agreement with results of non-relativistic models that include exchange contributions from finite-range  interactions. 

Our results also suggest some directions for further study. MFT underestimates nuclear level densities for heavy nuclei owing to its low Landau effective mass; the ability to obtain  larger masses in  TL calculations hint at  better predictions. Thus, studies of finite nuclei including TL contributions are worthwhile. As a bonus, one can also learn about the influence of exchange terms on spin-orbit splittings in nuclei, a success enjoyed by MFT.  Unlike the conventional Hatree-Fock method, the TL calculation does not require a self-consistent single particle spectrum; instead, the ground state energy at $T=0$ or the chemical potential at $T\neq 0$ are minimized by adjusting  the meson fields. This procedure could greatly simplify calculations of finite nuclei, results of which can be contrasted with those of the more involved relativistic Hartree-Fock procedure.  

\section*{ACKNOWLEDGEMENTS}
This work was supported 
by the U.S. DOE under grants No. DE-FG02-93ER-40756 (for M.P., and X. Z. while he was at Ohio University where this work was begun) and No.  DE-FG02-97ER-41014  (for X.Z at the University of Washington). The authors are grateful to Brian Muccioli and Constantinos Constantinou for helpful conversations and for several checks of results. X.Z. thanks Jeremy W. Holt for useful discussions.

\appendix

\section{Explicit expressions for the two-loop contribution $\delta\omega_{(1)}$} 
\label{app:simplifiedtwoloop}

Analytical integrations of the angular dependences  render \eqs{eqn:chempot1phi}{eqn:chempot1pi}]   into two-dimensional integrals in the variables  $p=|\vec{p}|$ and  $q=|\vec{q}|$. For the $\phi$ field, we obtain
\begin{eqnarray}
A^{(\phi)}(p,q)&= & p^{2}+q^{2}+{\msstar}^{2}-\left(\estar(p)-\estar(q)\right)^{2} \ , \\
\Theta^{(\phi)}(p,q)&=& \ln\left(\frac{A^{\phi}(p,q)+2pq}{A^{\phi}(p,q)-2pq}\right) \ , \label{eqn:Thetadef}  \\
\delta^{(\phi)}_{d}&=& \frac{1}{{M}^{2}} \frac{1}{(2\pi)^{4}} \int dp dq \frac{pq}{\estar(p)\estar(q)} \Theta^{(\phi)}(p,q)\left[n_{p}(p)n_{p}(q)+n_{n}(p)n_{n}(q)\right] \ , \\
\delta^{(\phi)}_{e}&=& \frac{4}{{M}^{2}} \frac{1}{(2\pi)^{4}} \int dp dq \frac{pq}{\estar(p)\estar(q)} \Theta^{(\phi)}(p,q)\left[n_{p}(p)n_{n}(q)\right] \ .
\end{eqnarray}
Analogous quantities can be defined for the other fields:
$A^{(v,\rho,\pi)}(p,q)$, $\Theta^{(v,\rho,\pi)}(p,q)$, $\delta^{(v,\rho,\pi)}_{d}$, and $\delta^{(v,\rho,\pi)}_{e} $.
Also of use are quantities involving the scalar densities (with the spin degeneracy factor $\gamma_{s}=2$):
\begin{eqnarray}
\den{s,i}&=& \gamma_{s} \int \dvecp{q}\frac{\mnstar}{\estar(q)}n_{i}(q) \ , \notag \\ 
\alpha_{d}&=& \frac{1}{8{\mnstar}^{2}M^{4} \gamma_{s}^{2}} \left(\den{s,p}^{2}+\den{s,n}^{2}\right)  \quad {\rm and} \quad
\alpha_{e} = \frac{1}{2{\mnstar}^{2}M^{4} \gamma_{s}^{2}} \left(\den{s,p}\den{s,n}\right) \ ,
\end{eqnarray}
and 
\begin{eqnarray}
\lambda^{(\phi)}&=& \half \left(\frac{{\mnstar}^{2}}{M^{2}}-\frac{{\msstar}^{2}}{4{M}^{2}}\right) \,, \quad 
\lambda^{(v)} = -\half \left(\frac{{\mnstar}^{2}}{M^{2}}+\frac{m_{v}^{2}}{2{M}^{2}}\right) \ , \notag \\
\lambda^{(\rho)} &=& -\frac{1}{8} \left(\frac{{\mnstar}^{2}}{M^{2}}+\frac{m_{\rho}^{2}}{2{M}^{2}}\right)\,, \quad  \mathrm{and} \quad
\lambda^{(\pi)} = -\frac{1}{8}\frac{m_{\pi}^{2}}{{M}^{2}} \ .
\end{eqnarray}

Utilizing the above expiressions, $\delta\omega_{(1)}$ in \eqs{eqn:chempot1phi}{eqn:chempot1pi} can be written compactly as 
\begin{eqnarray}
\frac{\delta\omega_{(1,\phi)}}{{M}^{4}} &=& \gamma_{s}\gs^{2} \left(\alpha_{d}+\lambda^{(\phi)}\delta^{(\phi)}_{d}\right) \ , \notag \\
\frac{\delta\omega_{(1,v)}}{{M}^{4}} &=& \gamma_{s}\gv^{2} \left(2 \alpha_{d}+\lambda^{(v)}\delta^{(v)}_{d}\right) \ , \notag \\
\frac{\delta\omega_{(1,\rho)}}{{M}^{4}} &=& \gamma_{s}\grho^{2} \left[\half \left(\alpha_{d}+\alpha_{e}\right) +\lambda^{(\rho)}\left(\delta^{(\rho)}_{d}+\delta^{(\rho)}_{e}\right)\right] \ , \notag \\
\frac{\delta\omega_{(1,\pi)}}{{M}^{4}} &=& \gamma_{s}\left(\frac{g_{A}}{f_{\pi}}\right)^{2} {\mnstar}^{2} \left[\alpha_{d}+\alpha_{e} +\lambda^{(\pi)}\left(\delta^{(\pi)}_{d}+\delta^{(\pi)}_{e}\right)\right] \ . \label{eqn:grandchempv2}
\end{eqnarray}

\section{Non-relativstic (low density) limit expressions at  zero temperature} 
\label{subsec:TLhnlimit}

For  $\frac{k_F}{\mnstar} \ll 1$, non-relativistic conditions prevail. The expressions derived here help us to understand the behavior of the TL contributions at low densities, and also serve as checks of numerical calculations of the exact expressions at zero temperature. 
The emerging structure is the same as in previous non-relativistic studies that include exchange interactions \cite{Prakash:1988zz,Welke:1988zz}. 
The various TL contributions can be expressed in terms of the generic function 
\begin{eqnarray}
F(k_{F,i} ,\, k_{F,j}; m) \equiv
\int \dvecp{p} \dvecp{q}   \frac{\theta(k_{F,i}-|\vec{p}|)\, \theta(k_{F,j}-|\vec{q}|)}{(\vec{p}-\vec{q})^2+m^2} \,,
\end{eqnarray} 
where $i$ and $j$ stand for neutron or proton, and $m$ is the mass of the meson through which nucleon-nucleon interactions are occurring.  Explicitly, the TL contributions in the non-relativistic limit are
\begin{eqnarray}
\delta\ed_{(1,\phi)}&=&\frac{\gamma_s}{2} \gs^2 \left[F(k_{F,p} ,\, k_{F,p}; \msstar)+ F(k_{F,n} ,\, k_{F,n}; \msstar)\right] \notag \\ 
\delta\ed_{(1,V)}&=&-\frac{\gamma_s}{2} \gv^2 \left[F(k_{F,p} ,\, k_{F,p}; m_v)+ F(k_{F,n} ,\, k_{F,n}; m_v)\right] \notag \\ 
\delta\ed_{(1,\rho)}&=&-\frac{\gamma_s}{2} \grho^2 \left[ \frac{1}{4}\left(F(k_{F,p} ,\, k_{F,p}; m_\rho)+ F(k_{F,n} ,\, k_{F,n}; m_\rho)\right)+F(k_{F,p} ,\, k_{F,n}; m_\rho)\right] \notag \\ 
\delta\ed_{(1,\pi)}&=&\frac{\gamma_s}{2} \left(\frac{g_A}{f_{\pi}}\right)^2 m_{\pi}^2 \bigg[\frac{1}{4m_{\pi}^2} \left(\frac{\rho_{p}^2+\rho_{n}^2}{4}+\rho_p \rho_n\right) \notag \\
&&-\frac{1}{4}\left(F(k_{F,p} ,\, k_{F,p}; m_\pi)+ F(k_{F,n} ,\, k_{F,n}; m_\pi)\right)-F(k_{F,p} ,\, k_{F,n}; m_\pi)\bigg]
\end{eqnarray} 
The integrations can be done analytically with the result
\begin{eqnarray}
&& (2\pi)^4 \, F(k_{F,i} ,\, k_{F,j}; m) \notag \\ 
& =& \frac{2}{3} (k_{F,i}^3 +k_{F,j}^3) \left[k_{F,i}+k_{F,j}-m \arctan{\left(\frac{k_{F,i}+k_{F,j}}{m}\right)} \right] \notag \\ 
&&-\frac{2}{3} (k_{F,i}^3 -k_{F,j}^3) \left[k_{F,i}-k_{F,j}-m \arctan{\left(\frac{k_{F,i}-k_{F,j}}{m}\right)} \right] \notag \\ 
&&+ \ln{\left(\frac{m^2+(k_{F,i}+k_{F,j})^2}{m^2+(k_{F,i}-k_{F,j})^2}\right)} \left[\frac{m^4}{24}+\frac{k_{F,i}^2+k_{F,j}^2}{4} m^2 -\frac{\left(k_{F,i}^2-k_{F,j}^2\right)^2}{8}\right] \notag \\
&&-\frac{5}{6} k_{F,i} k_{F,j}\left(k_{F,i}^2+k_{F,j}^2\right) -\frac{m^2}{6} k_{F,i} k_{F,j} \ . 
\end{eqnarray}

\section{Single-particle spectrum}
\label{sec:spectrum}

From the zero-temperature energy density functional,  the single-particle spectrum is obtained from \cite{Prakash87,Constantinou2014, Constantinouthesis} :
\begin{eqnarray}
\gamma_s\,\espec{i}{p}=\frac{\partial \ed \left[n_{p,n};\phib[n_{p,n}], \vb[n_{p,n}], \rhob[n_{p,n}] \right] }{\partial n_{i}(p)} =\left. \frac{\partial \ed\left[n_{p,n};\phib, \vb, \rhob \right] }{\partial n_{i}(p)} \right\vert_{bg} \,,
\end{eqnarray}
where $\espec{i}{k}$, $\phib,\, \vb,$ and $\rhob$ are functionals of $n_{p,n}(k)$. Because the meson field expectation values  minimize $\ed$,  their functional derivatives are zero.  The single-particle spectrum at the MFT level is  
\begin{eqnarray}
\epsilon_{(0),i}(p)&=&\sqrt{p^{2}+{\mnstar}^{2}} + \gv \vb + t_3 \half \grho \rhob  \ , \label{eqn:especMFT}
\end{eqnarray}
with $t_3=+1$ for proton and $-1$ for neutron. After including the two-loop contributions,  
\begin{eqnarray}
\epsilon_{(1),i}(p) &=& \epsilon_{(0),i}(p) + \delta\epsilon_{(1),i}(p) \,, \\
\delta  \epsilon_{(1,\phi,v),i}(p) &=&  \frac{\partial \delta\ed_{(1,\phi,v)}}{\partial n_{i}(p)} =\frac{(-)}{4\estar(p)} \int d\tau_{\vec{q}} n_{i}(q)\left[\gs^{2}\ f_{s}\ D(\msstar)+\gv^{2}\ f_{v}\ D(m_{v}) \right]  \ , \\
\delta  \epsilon_{(1,\rho,\pi),p}(p) &=&  \frac{\partial \delta\ed_{(1,\rho,\pi)}}{\partial n_{p}(p)} =\frac{(-)}{4\estar(p)} \int d\tau_{\vec{q}} \frac{1}{4} \left[n_{p}(q)+2n_{n}(q)\right] \times   \notag \\
&&\quad\quad\quad \quad\quad\quad \left[\grho^{2} \ f_{v}\ D(m_{\rho})+\left(\frac{g_{A}}{f_{\pi}} \right)^{2}  {\mnstar}^{2}\ f_{pv}\ D(m_{\pi}) \right] \ , \\
\delta  \epsilon_{(1,\rho,\pi),n}(p) &=&\delta  \epsilon_{(1,\rho,\pi),p}(p)\big(n_p\leftrightarrow n_n\big) \ , \label{eqn:especTL}
\end{eqnarray}
and $n_{i}(q)=\theta \left(k_{F,i}-q\right)$. 
From these results, we can compute $\mland_{i}(\den{p,n})$, which controls the thermodynamics at low temperature and high density (degenerate limit).  
Through  a procedure similar to that in Appendix~\ref{app:simplifiedtwoloop}, we obtain a compact expression for $\epsilon_{(1),i}(p)$ with the help of the functions
\begin{eqnarray}
\Lambda_{i}&\equiv& \frac{\den{s,i}}{4 \gamma_{s} {\mnstar}M}  \ , \\
\Delta^{(\phi,v,\rho,\pi)}_{i}(p) &\equiv& \frac{1}{(2\pi)^{2}} \int dq \frac{q}{\estar(q)} \Theta^{(\phi,v,\rho,\pi)}(p,q) n_{i}(q) \ , \\
\psibar{\Lambda}_{i}&\equiv& \left(\gs^{2}+2 \gv^{2}\right) \Lambda_{i}+\left(\half \grho^{2}+\left(\frac{g_{A}\mnstar}{f_{\pi}}\right)^{2}\right) \left(\Lambda_{i}+2\Lambda_{j}\right)  \ , \\
\psibar{\Delta}_{i}(p)&\equiv& \bigg[\gs^{2} \lambda^{(\phi)}\Delta_{i}^{(\phi)}(p)+\gv^{2} \lambda^{(v)}\Delta_{i}^{(v)}(p)+\grho^{2} \lambda^{(\rho)}\left(\Delta_{i}^{(\rho)}(p)+2 \Delta_{j}^{(\rho)}(p)\right) \notag  \\ 
&&{}+\left(\frac{g_{A}\mnstar}{f_{\pi}}\right)^{2}  \lambda^{(\pi)}\left(\Delta_{i}^{(\pi)}(p)+2 \Delta_{j}^{(\pi)}(p)\right)\bigg] \ . \label{eqn:LambdaDeltabardef}
\end{eqnarray}
with $\Theta^{(\phi,v,\rho,\pi)}(p,q) $ as in \eq{eqn:Thetadef}. The lower indices $``i"$ and $``j"$ are for protons and neutrons, but the two are always different in the current discussion. The spectrum is then
\begin{eqnarray}
\epsilon_{(1),i}(p) &=& \estar(p) + \gv \vb + t_3 \half \grho \rhob + \psibar{\Lambda}_{i} \frac{M}{\estar(p)} +\frac{{M}^{2}}{p \estar(p)} \psibar{\Delta}_{i}(p)  \ . \label{eqn:singlepartcilespec}
\end{eqnarray} 
From \eq{eqn:mlanddef}, the Landau effective mass becomes
\begin{eqnarray}
\mland_{i}=\estar_{F,i} \left[1-\frac{M \psibar{\Lambda}_{i}}{{\estar_{F,i}}^{2}}-\frac{{M}^{2}}{k_{F,i}} \left(\frac{1}{k_{F,i}^{2}}+\frac{1}{{\estar_{F,i}}^{2}}\right)\psibar{\Delta}_{i}(k_{F,i}) +\frac{{M}^{2}}{k_{F,i}^{2}} \left( \frac{d \psibar{\Delta}_{i}}{d p}\right)_{p=k_{F,i}} \right]^{-1} \label{eqn:mlandeq} \ ,
\end{eqnarray}
where
\begin{eqnarray}
\frac{d \Delta^{(\phi)}_{i}(p)}{d p}=\frac{1}{\pi^{2}} \int dq~ \frac{q^{2}}{{\estar(q)}} \frac{A^{(\phi)}(p,q)-2p^{2}\frac{\estar(q)}{\estar(p)}}{\left(A^{(\phi)}(p,q)\right)^{2}-4p^{2}q^{2}} \ n_{i}(q) \ .
\end{eqnarray}

\section{Degenerate and non-degenerate limit expressions}
\label{sec:degnondeglimits}

In the absence of collective effective effects close to the Fermi surface, Landau's Fermi Liquid Theory (FLT) \cite{ll9,flt,Prakash87} 
enables the calculation of the degenerate-limit thermal properties for a general single-particle spectrum. 
To leading order in temperature effects, the explicit forms of the entropy density, thermal 
energy, thermal pressure, and thermal chemical potential are  
\begin{eqnarray}
s &=& 2T\sum_i a_i \rho_i\,, \label{sdeg} \qquad
\frac{E_{th} }{N_B}= \frac{T^2}{\rho_B}\sum_i a_i \rho_i  \label{edeg}\\
P_{th} &=& \frac{2T^2}{3}\sum_i a_i \rho_i\left(1-\frac{3}{2} \sum_j   \frac{\rho_j}{m_i^*}\frac{\partial m_i^*}{\partial \rho_j}\right) \,, \label{pdeg} \\
\mu_{i,th} &=& -T^2\left(\frac{a_i}{3}+\sum_j \frac{\rho_ja_j}{m_j^*}\frac{\partial m_j^*}{\partial \rho_i}\right) \,, \label{mudeg}
\end{eqnarray}
where ${\displaystyle{ a_i =\frac{\pi^2}{2}\frac{m_i^*}{k_{F,i}^2} }}$ is the level density parameter. Expressions for next-to-leading order in temperature effects have been recently worked out in Ref. \cite{cons15}.   

In the non-degenerate limit when the particle's de Broglie wave length is much smaller than the inter-particle distance, {\it i.e.,}  $2\pi\den{i}^{1/3}/\sqrt{3 M T} \ll 1$, the Fermi-Dirac distribution approaches the classical Maxwell distribution. In this case, the thermal state variables can be expended in terms of fugacity.  Here we only collect the relevant formulas for the leading order terms, which are the same as those of the classical gas state variables. 
Assuming both protons and neutrons are in this limit, 
\begin{eqnarray}
S/N_B&\simeq& \frac {1}{\rho_B} \sum_i \den{i} \bigg\{\frac{5}{2}-\ln\left[\frac{\den{i}}{2}\left(\frac{2\pi}{M^*T}\right)^{3/2}\right] 
+ \frac {\rho_i}{8} \left (\frac {\pi}{M^*T}\right)^{3/2}  -  \frac {15T}{4M^*}
\bigg\} \ ,  \label{eqn:classicalgaslimitsd}\\
P_{\mathrm{th}} &\simeq&  \sum_i\left\{ \rho_i T \left[ 1 + \frac {\rho_i}{4} \left(\frac {\pi}{M^*T} \right)^{3/2} \right]   - P_{Fi}^*\right\} 
- \delta V \quad 
\label{eqn:classicalgaslimitp} \\
\frac{E_{\mathrm{th}}}{N_B} &\simeq& \frac {1}{\rho_B} \sum_i \left\{ \frac{3}{2} T\rho_i 
\left[ 1 + \frac {\rho_i}{4} \left( \frac {\pi}{M^*T} \right)^{3/2} + \frac {5T}{4M^*} \right]  - {\cal T}_{F_i}^*\rho_i \right\} + \frac {\delta V}{\rho_B} + M^\ast \ ,  \label{eqn:classicalgaslimited}  \\ 
\mu_{\mathrm{th},i} &\simeq&  T \left\{ \ln\left[\frac{\den{i}}{2}\left(\frac{2\pi}{M^*T}\right)^{3/2}\right] 
+ \frac {\rho_i}{2} \left(\frac {\pi}{M^*T}  \right)^{3/2} - \frac {15T}{8M^*} \right\} + \left(M^* - E_{F_i}^* \right)
\label{eqn:classicalgaslimitmu}
\end{eqnarray}
where $P_{F_i}^*$ and ${\cal T}_{F_i}^*$ are the Fermi pressure and energy of species $i$ at $T=0$. 
Above, the next-to-leading order expressions  in terms of  the fugacity and $T/\mnstar$ ({\it i.e.,} relativistic  corrections) for MFT are from Refs.  \cite{Prakash87,Constantinouthesis}, where expressions for $\delta V$ arising from the $T$-dependence of $M^*$ 
can also be found.  Contributions from exchange terms are not included above, but are expected to be negligibly small at very low densities as they are proportional to $\rho_i^2$.

\clearpage

\bibliographystyle{h-physrev3}

\bibliography{./now}

\end{document}